\documentclass[11pt,a4paper]{article}                           

\usepackage{amsmath}
\usepackage{amssymb}
\usepackage[usenames]{color} 
\usepackage{multirow}
\usepackage{graphicx}
\usepackage{epstopdf}
\usepackage{array}
\usepackage{hhline}
\usepackage{cite}
\usepackage{microtype,colortbl}
\usepackage[bf]{caption}
\usepackage{color}
\usepackage[usenames,dvipsnames]{xcolor}
\usepackage{pstool}
\usepackage{tikz}
\usepackage{ytableau}
\usepackage{latexsym,stmaryrd}
\usepackage[bookmarks=false]{hyperref}
 \usepackage{lscape}

 \usepackage{tensor}
 \usepackage{dsfont}

\usepackage[margin=10pt,font=small,labelfont=bf]{caption}
\usepackage[margin=0pt,font=small,labelfont=normalfont,skip=22pt]{subcaption}

\usepackage{ifpdf}
\def\hybrid{\topmargin -20pt    \oddsidemargin 0pt
        \headheight 0pt \headsep 0pt
        \textwidth 6.25in       
        \textheight 9 in       
        \marginparwidth .875in
        \parskip 5pt plus 1pt 
          \jot = 1.5ex
   }
\hybrid
\numberwithin{equation}{section}
\numberwithin{table}{section}

\newcommand{\beq}{\begin{equation}}  \newcommand{\eeq}{\end{equation}}
\newcommand{\bal}{\begin{aligned}}   \newcommand{\eal}{\end{aligned}}
\newcommand{\bea}{\begin{eqnarray}}  \newcommand{\eea}{\end{eqnarray}}

\def\ov{\overline}

\newcommand{\bmat}{\left(\begin{array}}
\newcommand{\emat}{\end{array}\right)}

\newcommand{\bbC}{\mathbb{C}}
\newcommand{\bbR}{\mathbb{R}}

\newcommand{\nn}{\nonumber}

\newcommand{\cO}{\mathcal{O}}

\newcommand{\cE}{\mathcal{E}}

\newcommand{\cC}{\mathcal{C}}

\newcommand{\cK}{\mathcal{K}}
\newcommand{\cN}{\mathcal{N}}

\newcommand{\cV}{\mathcal{V}}

\usepackage{cancel}

\newcommand{\be}{\begin{equation}}
\newcommand{\ee}{\end{equation}}

\DeclareMathOperator{\im}{im}

\definecolor{Gray}{gray}{0.95}

\newcommand{\eq}[1]{\begin{equation}
                     \begin{split} #1 \end{split}
                     \end{equation}}
\newcommand{\op}{\hspace{1pt}}

\begin{document}


\vspace*{1.5cm}

\begin{center}
{\LARGE
Moduli Stabilization in Asymptotic Flux Compactifications
\\}
\end{center}

\vspace{0.5cm}

\begin{center}
  Thomas W.~Grimm\footnote{t.w.grimm@uu.nl},
  Erik Plauschinn\footnote{e.plauschinn@uu.nl}, 
  Damian van de Heisteeg\footnote{d.t.e.vandeheisteeg@uu.nl}
\end{center}

\vspace{0.4cm}

\begin{center} 
\textit{
Institute for Theoretical Physics, Utrecht University \\
Princetonplein 5, 3584CC Utrecht \\
The Netherlands \\
}
\end{center} 

\vspace{2cm}


\begin{abstract}
\noindent
We present a novel  strategy to systematically study complex-structure moduli stabilization 
in Type IIB and F-theory flux compactifications. In particular, we determine vacua in any 
asymptotic regime of the complex-structure moduli space by exploiting powerful tools 
of asymptotic Hodge theory. 
In a leading approximation the moduli dependence of the vacuum conditions are shown to be polynomial with a 
dependence given by sl(2)-weights of the fluxes. This simple algebraic dependence can be 
extracted in any asymptotic regime, even though in nearly all asymptotic regimes essential exponential 
corrections have to be present for consistency. 
We give a pedagogical introduction to the sl(2)-approximation as well as 
a detailed step-by-step procedure for constructing the corresponding Hodge star operator.  To exemplify the construction, we 
present a detailed analysis of several Calabi-Yau three- and fourfold examples.
For these examples we illustrate that the vacua in the sl(2)-approximation match the vacua obtained with 
all polynomial and essential exponential corrections rather well, and we determine 
the behaviour of the tadpole contribution of the fluxes. 
Finally, we discuss the structure of vacuum loci and their relations to several swampland conjectures. In particular, 
we comment on the realization of the so-called linear scenario in view of the tadpole conjecture.

\end{abstract}


\clearpage

\tableofcontents

\newpage
\section{Introduction}

In order to obtain phenomenologically interesting effective theories 
from string theory, it is vital to ensure that neutral scalar fields have a 
sufficiently large mass in order to not contradict fifth force bounds.
The most prominent way to achieve this is by considering string theory 
solutions that include background fluxes \cite{Grana:2005jc,Douglas:2006es,Blumenhagen:2006ci}. Such fluxes can, 
for example, fix the complex-structure of the compactifications geometry 
and thus give potential complex-structure deformation moduli a mass. The 
most prominent and best understood scenario where this happens in a controlled 
way are Type IIB orientifold compactifications with O3/O7-planes and their more 
general F-theory counterparts. In these settings the precise way 
how moduli stabilization occurs has been studied for many examples by explicitly 
constructing the compact Calabi-Yau geometry, specifying a flux, and deriving 
the scalar potential for the complex-structure moduli. However, 
it is difficult to draw general conclusions from 
such specific examples, which makes it hard 
to interpret the findings in these special cases in a broader context. 

In recent years the swampland program, which aims at identifying general principles 
that every effective theory consistent with quantum gravity has to satisfy, has forced 
us to develop novel approaches to study general properties of effective theories arising 
from string theory. In particular, the well-defined Type IIB and F-theory settings, are 
a perfect testing ground for many swampland conjectures. This does require, however, 
to move beyond studying specific examples and rather uncover the universal 
structures appearing in all such compactifications. While this is a challenging task in general, 
it has recently become clear that it becomes tractable if we focus on the asymptotic 
regimes of the field space and apply the powerful tools of asymptotic Hodge theory. 
Following \cite{Grimm:2019ixq} we will call the such scenarios  \textit{asymptotic flux compactifications} in 
the following. To introduce them in more detail, we recall that the relevant compactification geometries for Type IIB 
and F-theory settings without fluxes are Calabi-Yau three- and fourfolds, respectively. 
These manifolds admit a deformation space known as the complex-structure 
moduli space, which parametrizes the allowed choices of complex-structures on these geometries.     
It is crucial to realize that this moduli space is not compact, but rather has boundaries and thus 
associated asymptotic regions. Asymptotic Hodge theory gives a detailed understanding 
of how the Hodge decomposition of forms behaves in these regions and provides a 
dictionary for how data associated to an individual boundary component determines the 
asymptotic form of the flux-induced scalar potential. 

In this work we suggest that the mathematical results of asymptotic Hodge theory provide 
a systematic algorithm to perform moduli stabilization near any boundary of moduli space. Finding 
flux vacua requires to solve a self-duality condition on the fluxes involving the Hodge star. The moduli 
dependence of the Hodge star is, in general, a very complicated transcendental function that depends 
on many details of the compactification geometry. However, in the asymptotic regime there exists an approximation 
scheme to extract the moduli dependence in essentially three steps: (1) the sl(2)-approximation, (2) the nilpotent orbit 
approximation, (3) the fully corrected result. In steps (1) and (2) a certain set of corrections is consistently neglected, turning the extremely hard problem of finding flux vacua into a tractable algebraic problem. It is important to stress that it is a very non-trivial fact that such approximations exist in all asymptotic regimes. In particular, 
this is remarkable because it was shown in \cite{Bastian:2021eom} that in almost all asymptotic regimes exponential corrections are essential 
when deriving the Hodge star from the period integrals of the unique $(3,0)$-form or $(4,0)$-form 
of the Calabi-Yau manifold. The determination of the approximated Hodge star is non-trivial, 
but can be done explicitly for any given example following the algorithms that we discuss 
in detail in this work. 

The most involved computations are needed to determine the sl(2)-approximation. The crucial 
result of \cite{CKS} is that in any boundary regime in which the moduli satisfy a certain
hierarchy a number of $n$ commuting sl$(2,\bbC)$ algebras and a boundary Hodge decomposition 
can be determined. This data in turn determines a split of the flux space and the asymptotic form 
of the Hodge star operator used, for example, in the flux vacuum conditions. We collect some intuitive arguments why such structures emerge 
at the boundary and discuss the step-by-step procedure to construct the sl(2)s and the boundary 
Hodge decomposition. This algorithm has already been given in \cite{Grimm:2018cpv} and can be extracted from the 
original works \cite{CKS, Kato}. The approach is iterative in the number of moduli that admit a hierarchy and 
uses solely linear algebra, even though in a somewhat intricate way. Therefore, it can be easily 
implemented on a computer and any given example can be evaluate very rapidly. We discuss 
one example in detail and present all the asymptotic data needed in the sl(2)-approximation. Remarkably,
the sl(2)-approximation can also be used abstractly, i.e.~one can parametrise unknown coefficients 
and study the general form of the asymptotic flux scalar potential and vacuum conditions by 
considering all possible sl(2)-representations. This 
more abstract approach can be viewed as an avenue to study general aspects of flux vacua and 
eventually provide a classification of possibilities at least in the regime where the sl(2)-approximation is 
valid. Initial steps in this program have been carried out for two-moduli Calabi-Yau fourfolds in \cite{Grimm:2019ixq}.  
Independent of whether one uses the sl(2)-approximation concretely for specific examples or abstractly in a classification, 
we stress that with its help the search for flux vacua is turned into a rather simple algebraic problem. 
It can therefore be viewed as providing the computationally ideal starting point when
looking for vacua of the fully corrected potential.

In order to make use of any results obtained in the sl(2)-approximation, it is crucial to show how closely it approximates 
the full answer. In particular, the second approximation step, the nilpotent orbit approximation, will differ from the 
sl(2)-approximation by subleading polynomial corrections. The presence of these corrections 
complicates the field dependence of the Hodge star  significantly and the search for vacua becomes much more involved
and quickly computationally intractable. We show by studying a number of explicit Calabi-Yau threefold examples, that 
the sl(2)-approximation actually provides a rather good approximation, even for a mild hierarchy among the moduli. 
We show this not only for a large complex-structure regime, but also at so-called conifold-large complex structure 
boundaries that are not straightforwardly 
accessible via large complex-structure/large volume mirror symmetry. The periods of the latter setting 
have been derived in \cite{Blumenhagen:2020ire,Demirtas:2020ffz} and we will see explicitly that the general sl(2)-approximation can be  
evaluated from this information. Our findings will support the claim that the successive approximation scheme 
is powerful not only in specific asymptotic regimes, but rather provides a general systematics 
 in all asymptotic regimes in Type IIB  flux compactifications. 
 
 In the final part of this work we show that our strategy is equally applicable to Calabi-Yau fourfolds used in F-theory flux compactifications. 
We hereby focus on discussing the large complex-structure regime, for which the necessary data that determines the sl(2)-approximation 
can be given in terms of the intersection numbers and Chern classes of the mirror Calabi-Yau fourfold. In order to provide an example computation we 
discuss an explicit Calabi-Yau fourfold that realizes the so-called linear scenario recently proposed in \cite{Marchesano:2021gyv}.\footnote{A related Type IIA version of such 
a scenario has been first proposed in \cite{Palti:2008mg}.} We show that this scenario admits a flat direction in the sl(2)-approximation 
that is subsequently lifted in the nilpotent orbit approximation. Furthermore, we will argue that this does not necessarily imply that 
the tadpole conjecture of \cite{Bena:2020xrh} is violated by adapting a strategy recently suggested in \cite{Plauschinn:2021hkp}. The conjecture claims that in high-dimensional 
moduli spaces not all fields can be stabilized by fluxes without overshooting the tadpole bound. If true, it would have far reaching 
consequences for string phenomenology. The desire to provide evidence for this conjecture, or to disprove it, can be viewed 
as another motivation for implementing a systematic approximation scheme. Remarkably, our constructions can be carried out 
for examples with many moduli without much effort and therefore give an exciting opportunity to explore moduli stabilization 
on higher-dimensional moduli spaces \cite{toappear}.

This paper is organized as follows. In section \ref{Mod-stab-rev} we briefly recall some basics about 
Type IIB orientifold compactifications, the conditions on flux vacua, and the form of the Hodge star as a function of the 
complex-structure moduli. We give a detailed introduction to the sl(2)-approximation and its derivation 
in section \ref{sec_type_ii_boundary}. The idea is to first collect some intuitive ideas how to see the emergence of an 
sl(2) and thereafter delve in the technical details of the general construction. For concreteness we 
will also exemplify the construction on one explicit example. In section \ref{sec-mod-stab} we then introduce 
the moduli stabilization scheme with the successive approximation steps. We show in a number of 
examples how close the vacua determined in the sl(2)-approximation are to the vacua obtained with 
Hodge star that has all polynomial corrections. Finally, in section \ref{F-theory} we extend the discussion 
to Calabi-Yau fourfolds and F-theory. For simplicity, in this case we only discuss the large complex-structure 
regime in detail. We determine the sl(2)-approximation for a concrete fourfold example and 
show how moduli can be stabilized using the linear scenario of \cite{Marchesano:2021gyv}. This will allow us to comment 
on the compatibility of this scenario with the tadpole conjecture. 


\section{Moduli stabilization for type IIB orientifolds}\label{Mod-stab-rev} 

In this section we briefly review moduli stabilization for type IIB orientifolds 
with O3- and O7-planes, focusing on the complex-structure and axio-dilaton 
sectors. (For more extensive reviews we refer the reader to \cite{Grana:2005jc,Douglas:2006es,Blumenhagen:2006ci}, 
and for recent systematic analyses of moduli-stabilization scenarios to 
\cite{
Grimm:2019ixq,
AbdusSalam:2020ywo,
Marchesano:2020uqz,
Marchesano:2021gyv,
Cicoli:2021tzt
}).
This part is meant to establish our notation and conventions
but contains no new results, and the reader familiar with 
the topic can safely skip to section~\ref{sec_type_ii_boundary}.


\subsubsection*{Orientifold compactifications}

We consider compactifications of type IIB string theory on Calabi-Yau threefolds $Y_3$,
subject to an orientifold projection which  contains a holomorphic involution $\sigma$.
This involution is chosen to act on the K\"ahler 
form and on the holomorphic three-form of $Y_3$ 
as $\sigma^* J= + J$ and $\sigma^*\Omega = -\Omega$,
and $\sigma$ splits the cohomology groups of $Y_3$ 
into even and odd eigen\-spaces 
as $H^{p,q}(Y_3) = H^{p,q}_+(Y_3) \oplus H^{p,q}_-(Y_3) $. 
Relevant for our purpose is the orientifold-odd third cohomology of $Y_3$, for which we 
can choose an integral symplectic basis as
\eq{
\label{basis_01}
  \{ \alpha_I , \beta^I \} \in H^3_-(Y_3)\,, \hspace{50pt} I = 0, \ldots, h^{2,1}_- \,.
}
For this paper we assume that $h^{2,1}_+=0$, and we therefore 
set $h^{2,1}_-=h^{2,1}$ in the following. 
The only non-vanishing pairings of the basis forms in \eqref{basis_01} are given by
\eq{
\label{basis_02}
  \int_{Y_3} \alpha_I \wedge \beta^J = \delta_I{}^J \,.
}


\subsubsection*{Moduli}

The effective four-dimensional theory obtained after compactification 
preserves $\mathcal N=1$ supersymmetry and 
contains  scalar fields which parametrize
deformations of $Y_3$.  Of interest to us are the axio-dilaton 
$\tau$ and the complex-structure moduli $t^i$, which we define as
\eq{
  \tau = c + i\op s\,, \hspace{80pt} t^i = x^i + i \op y^i \,, \hspace{30pt} i = 1,\ldots, h^{2,1} \,,
}
and our conventions are such $s>0$ and $y^i>0$. 
The K\"ahler moduli  are not relevant for our discussion. 
The K\"ahler potential describing the dynamics of the moduli fields is given by 
\eq{
  \label{kpot}
  K =  - \log\bigl[ -i(\tau-\bar \tau) \bigr] - \log \left[ +i\int_{Y_3} \Omega \wedge \bar \Omega \right]
  - 2\log \mathcal V\,,
}
where $\Omega$ depends on the complex-structure moduli $t^i$ and 
the volume $\mathcal V$ of the threefold depends on the K\"ahler moduli. 
The holomorphic three-form $\Omega$ of 
the Calabi-Yau orientifold will play an important role in our subsequent discussion, and 
it can be expanded in the symplectic basis \eqref{basis_01} 
as follows
\eq{
  \label{def_123}
  \Omega = X^I \alpha_I - \mathcal F_I \beta^I \,, \hspace{60pt} I = 0, \ldots, h^{2,1}\,.
}
Using the intersections \eqref{basis_02} we can then express the K\"ahler potential for 
the complex-structure moduli $ K^{\rm cs} = - \log \left[ +i\int_{Y_3} \Omega \wedge \bar \Omega \right]$
in terms of a period vector $\Pi$ and a symplectic pairing $\eta$ as
\eq{
\label{kpot_100}
  K^{\rm cs} = - \log \Bigl[ - i\, \bar\Pi^T \eta\op \Pi \Bigr], 
  \hspace{50pt}
  \arraycolsep2pt 
  \Pi = \left( \begin{array}{r}X\\ -\mathcal F\end{array}\right)\,,
  \hspace{20pt}
    \eta =  \left( \begin{array}{cc}
  0   & +\mathds 1
  \\[1pt]
  - \mathds 1 & 0
  \end{array}\right) .
}


\subsubsection*{Fluxes}

In order to generate a potential for the axio-dilaton and the complex-structure moduli
we consider NS-NS and R-R three-form fluxes $H_3$ and $F_3$ along the internal space
$Y_3$. These can be expanded into the integral symplectic basis \eqref{basis_01} as
\eq{
\label{tad_02}
  H_3 = h^I \alpha_I - h_I \beta^I \,, \hspace{50pt}
  F_3 = f^I \alpha_I - f_I \beta^I \,, 
}
where the expansion coefficients $h^I, h_I, f^I, f_I$ are integers. 
These fluxes generate a scalar potential in the effective four-dimensional  theory,
which is encoded in the  superpotential \cite{Gukov:1999ya}
\eq{
\label{pot_03}
 W = \int_{Y_3} \Omega \wedge G_3 \,, \hspace{50pt} 
 G_3 = F_3 - H_3 \op \tau \,.
}
The fluxes furthermore contribute to the tadpole cancellation conditions. In particular, 
the fluxes \eqref{tad_02} appear in the D3-brane tadpole condition in the combination 
(which in our conventions is positive)
\eq{
\label{tad_01}
  N_{\rm flux} = \int_{Y_3} F_3 \wedge H_3 = h_I f^I - h^If_I   \,.
}


\subsubsection*{Scalar potential}

The effective four-dimensional theory resulting from compactifying type II string theory on 
Calabi-Yau orientifolds can be described in terms of $\mathcal N=1$ supergravity. 
After taking into account the no-scale property of the K\"ahler potential, 
the corresponding  F-term potential takes the form
\eq{
\label{pot_04}
 V_F = e^K \op F_{A\vphantom{\ov B}}\op G^{A\ov B}\op \ov F_{\ov B}\,,
}
where $A,B$ label the axio-dilaton and the complex-structure moduli,
where $F_A = \partial_A W + (\partial_A K) W$ denotes the F-terms and where $G^{A\ov B}$ is the inverse of 
the K\"ahler metric computed from \eqref{kpot}.
The global minimum of \eqref{pot_04} is of Minkowski-type and corresponds to vanishing F-terms, that is
$F_{A} = 0$.
As discussed  in \cite{Giddings:2001yu,Grimm:2004uq}, these conditions can equivalently be expressed as an imaginary
self-duality condition for the flux $G_3$ which was given in \eqref{pot_03}. In particular, with $\star$ denoting the Hodge-star operator 
of $Y_3$ the F-term conditions
are equivalent to 
\eq{
  \label{eom_001}
  \star G_3 = i \op G_3 \,.
}
Note that the value of the superpotential in the minimum $W_0$ can be zero or non-zero. The first possibility 
corresponds to supersymmetric vacua (including the K\"ahler moduli sector) while the second possibility 
is important for the KKLT and Large Volume Scenarios \cite{Kachru:2003aw,Balasubramanian:2005zx}.
In particular, for KKLT and small $W_0$ is needed which has been studied recently in \cite{Demirtas:2019sip,Demirtas:2020ffz,Blumenhagen:2020ire,Honma:2021klo,Demirtas:2021nlu, Demirtas:2021ote,Broeckel:2021uty,Bastian:2021hpc}.


\subsubsection*{Hodge-star operator}

Let us become more concrete about the Hodge-star operator acting 
on the third cohomology.
This operator can be determined from a matrix $\mathcal{N}_{IJ}$,
which  is computed from the periods
 $\Pi= (X^I,  - \mathcal{F}_I)$ as $\mathcal{N}_{IJ} = ( \mathcal{F}_I  ,  \ D_{\bar{i}} \bar{\mathcal{F}}_I ) (X^J , \ D_{\bar{j}}  \bar{X}^J )^{-1}$.
However, in a frame for which a prepotential $\mathcal F$ exists, 
one can use its second derivatives $\mathcal F_{IJ} = \partial_I\partial_J \mathcal F$
to determine  $\mathcal N = \mathcal R + i\op \mathcal I$  as
\eq{
\label{pm}
{\cal N}_{IJ}=\overline{\mathcal {F}}_{IJ}+2i \, \frac{
{\rm Im}(\mathcal F_{IM}) X^M \, {\rm Im}(\mathcal F_{JN}) X^N}{
           X^P \,{\rm Im}(\mathcal F_{PQ}) X^Q}  \,.
}
For the integral symplectic basis  $\{\alpha_I,\beta^I\}$ introduced in 
\eqref{basis_01} the action of the  Hodge-star operator then takes the form 
\eq{
  \label{hodgestar01}
  \star \binom{\alpha}{\beta} = 
  C \op \binom{\alpha}{\beta} \,,
  \hspace{40pt}
  C= \left( \begin{array}{cc}
  \mathcal R\op \mathcal I^{-1} &  -\mathcal I - \mathcal R\op \mathcal I^{-1} \mathcal R
  \\
  \mathcal I^{-1} & - \mathcal I^{-1} \mathcal R
  \end{array}\right) \,,
}
while for elements in the Dolbeault cohomology the Hodge-star operator acts
as
\eq{\label{C_def}
  C \op \omega^{p,q} = i^{p-q} \op \omega^{p,q} \,,
  \hspace{40pt}
  \omega^{p,q} \in H^{p,q}(Y_3) \,.
}
With the help of the relation \eqref{hodgestar01}, we can now become more concrete about the self-duality condition \eqref{eom_001}.
First, we compute  $\mathcal M= \eta\op C$ and 
recall the symplectic matrix $\eta$ from \eqref{kpot_100} as
\eq{
\label{matrices_001}
  \mathcal M =    
  \left( \begin{array}{cc}
   -\mathcal I - \mathcal R \mathcal I^{-1} \mathcal R
   &
  - \mathcal R \mathcal I^{-1} 
  \\
  - \mathcal I^{-1} \mathcal R & - \mathcal I^{-1} 
  \end{array}\right) ,
  \hspace{50pt}
  \eta =  \left( \begin{array}{cc}
  0   & +\mathds 1
  \\
  - \mathds 1 & 0
  \end{array}\right) .
}  
Denoting the flux vectors by $\mathsf H_3 = (h^I,-h_I)^T$ and $\mathsf F_3 = (f^I,-f_I)^T$,
the self-duality condition \eqref{eom_001} can be written in matrix form notation as
\eq{
  \label{eom_003}
 \mathsf F_3 = \bigl( \eta\op \mathcal M \op s + \mathds 1\op  c \bigr) \op \mathsf H_3 \,.
}


\section{The sl(2)-approximation to the Hodge star}
\label{sec_type_ii_boundary}

In this section we study the asymptotic regimes in complex-structure moduli space in more detail. 
In particular, in view of moduli stabilization we are interested in the boundary behaviour of the Hodge-star matrix 
$C$ defined in \eqref{hodgestar01}. We will show how its leading form $C_{\rm sl(2)}$, which we call the  
sl(2)-approximated Hodge star, can be computed systematically in every asymptotic regime. 
Our approach relies on techniques coming from asymptotic Hodge theory, originally developed in the mathematical works \cite{Schmid, CKS}. For applications of the underlying technology in the swampland program we refer to \cite{Grimm:2018ohb,Grimm:2018cpv,Corvilain:2018lgw,Font:2019cxq,Grimm:2019wtx,Grimm:2019ixq,Grimm:2019bey,Gendler:2020dfp,Lanza:2020qmt,Grimm:2020cda,Bastian:2020egp,Calderon-Infante:2020dhm,Grimm:2020ouv,Grimm:2021ikg,Lanza:2021qsu,Castellano:2021yye,Palti:2021ubp,Bastian:2021hpc}.


\subsection{Sketch of the general idea}
\label{sec_hodge_gen}

Let us first introduce some of the main ideas for studying the boundary behaviour 
of the Hodge-star matrix.  For simplicity, in this section we consider only one modulus 
sent to the boundary but discuss the general case in section~\ref{sec_tech_details} below.
Much of this following discussion can also be found in \cite{Grimm:2020cda,Grimm:2021ikg}, but we will provide here a 
slightly different angle on the construction.


\subsubsection*{Boundaries and monodromy symmetries}

In complex-structure moduli space one can naturally associate to 
each boundary a discrete symmetry, known as a monodromy symmetry. 
Sending only a single modulus to the boundary we can  choose 
local coordinates such that $ z=0$ corresponds to the boundary locus, but
a more useful parametrization is given by
\eq{\label{coordinates}
   t =  x + i\op  y = \frac{1}{2\pi \op i} \log  z \,,
}
where the boundary corresponds to the limit $ t\to i\op \infty$.  
The monodromy symmetry is realized by 
encircling the boundary 
as $z\to e^{2\pi \op i} z$, which corresponds to a shift of the coordinate $x$ of the form $x\to x+ 1$. But, even though the effective theory is invariant under this shift, 
certain quantities transform non-trivially. A prominent example is the 
period vector $\Pi$ of the holomorphic three-form $\Omega$ shown in \eqref{def_123}, which behaves as
\eq{
  \label{monodromy_002}
  \Pi
  \hspace{5pt} \xrightarrow{\hspace{5pt}x\to x+1\hspace{5pt}}  \hspace{5pt}
  \Pi' = T\, \Pi\,,
}
where matrix notation is understood. Here, $T$ denotes an integer-valued monodromy matrix which in 
order for the K\"ahler potential \eqref{kpot_100} to be invariant has to satisfy 
\eq{
  T^T \eta \, T =  \eta\,,
  \hspace{50pt} T \in \mbox{Sp}(2h^{2,1}+2,\mathbb Z) \,.
}
Although not obvious, it turns out that for Calabi-Yau threefolds $Y_3$ the monodromy matrices
$T$ can always be made unipotent \cite{Landman}, that is $(T-\mathds 1)^{m+1}= 0$ for some $m\geq 0$.\footnote{This might require 
sending $z \rightarrow z^n$ and amounts to removing a possible semi-simple part of a general monodromy matrix $T$. } 
Furthermore, to each $T$ we can associate  a so-called log-monodromy matrix defined as 
$N = \log T$, which is an element of the Lie algebra $\mathfrak{sp}(2h^{2,1}+2,\mathbb R)$
and therefore satisfies
\eq{
  \label{def_log_mo}
  N = \log T\,,
  \hspace{50pt}
  ( \eta \op N) - ( \eta \op N)^T = 0 \,, 
  \hspace{50pt} N \in \mathfrak{sp}(2h^{2,1} +2,\mathbb R) \,.
}
Since the monodromy matrices $T$ are unipotent, it follows that the log-monodromy matrices
$N$ are nilpotent, that is $N^{m+1}=0$ for some $m\geq 0$. 
Note that 
the symmetry $N$ might induce an approximate 
continuous shift symmetry of the moduli space metric near certain boundaries. This is 
familiar, for example, from the large complex-structure regime. In order to simplify the naming 
we will refer to $x$ as being the axion and $y$ is being the saxion, even if no continuous symmetry 
is restored in the limit.


\subsubsection*{Nilpotent orbit theorem}

The nilpotent orbit theorem allows us to descirbe the moduli dependence of 
differential forms close to the boundary. 
An example is
the 
holomorphic three-form $\Omega$, and one finds that in the limit  $t\to i\op\infty$
the period vector $\Pi$ defined in \eqref{kpot_100} can be expressed as
\eq{
  \label{npot_001}
  \Pi(t) = e^{t\op N} e^{\Gamma(z)} \op \mathbf{a}_0 \,,
}
where $\Gamma(z)\in\mathfrak{sp}(2h^{2,1} +2,\mathbb C)$ is a matrix which depends
holomorphically on $z = e^{2\pi \op i t}$. The $(2h^{2,1} +2)$-dimensional 
vector $\mathbf a_0$ is a reference point 
which is independent
of $t$ but which in general depends holomorphically on all moduli not sent to the boundary.
We have therefore expressed the dependence of the period vector $\Pi$ on $t$ near the boundary 
in a simple form. Expanding the second exponential in \eqref{npot_001}  we find a natural 
split of $\Pi(t)$ as 
\beq \label{Pi-exp}
   \Pi(t) = \Pi_{\rm poly} +  \Pi_{\rm exp} = e^{t\op N}  \op \mathbf{a}_0 +  \cO(e^{2\pi i t})
\eeq
where we have collected all polynomial terms in $ \Pi_{\rm poly} =e^{t\op N}  \op \mathbf{a}_0 $ while 
all exponentially suppressed terms reside in $ \Pi_{\rm exp}$. It will be crucial below that 
the nilpotent orbit theorem implies that an expansion of the form \eqref{Pi-exp} 
also occurs for all its holomorphic derivatives.

Let us furthermore recall that for a Calabi-Yau threefold $\Omega$ is an element of 
$H^{3,0}(  Y_3)$. Taking a holomorphic (covariant) derivative of $\Omega$
lowers its holomorphic degree by one, and hence the fourth (covariant) derivative of $\Omega$ has to vanish. 
Combining this observation with \eqref{npot_001} and ignoring some technical subtleties, 
we find the necessary condition $N^4 = 0$. The highest non-vanishing power $m$ of $N$ depends 
on the boundary under consideration. We find the four choices
\eq{
  \label{npot_021}
  N^{m}\neq 0\, , \ N^{m+1}=0 \ , \hspace{50pt}\mbox{for}\quad  0\leq m \leq 3 \,,
}  
where $m=0$ implies that there is no unipotent monodromy associated to the boundary. 

\subsubsection*{Essential exponential corrections}

As stressed in \eqref{npot_021} the nilpotency order of $N$ does not have to be 
maximal, i.e.~$m=3$. This implies that the polynomial $\Pi_{\rm poly} =e^{t\op N}  \op \mathbf{a}_0 $ 
appearing in the expansion \eqref{Pi-exp} can be of any degree smaller or equal three and is given 
by the highest $k$ with non-vanishing $N^k a_0$. For a Calabi-Yau threefold the full middle cohomology 
can be obtained by taking holomorphic derivatives of $\Omega$. However, if 
$k<3$ the derivative of $\Pi_{\rm poly}$ with respect to $t$ cannot generate all three-forms. 
This implies that some of the exponential corrections in $ \Pi_{\rm exp}$ have to be present. Such corrections 
are thus essential near any boundary with $k<3$ and were termed  essential exponential corrections or essential instantons in \cite{Bastian:2021eom}.
A more precise statement can be made by introducing the expansion 
\beq \label{Pi-exp2}
    \Pi(t) =  e^{t\op N}  \op \big(\mathbf{a}_0 + e^{2\pi i t} \mathbf{a}_1 + e^{4\pi i t} \mathbf{a}_2 + \ldots \big)\ .
\eeq
Denoting by $k_i$ the lowest integer
such that $N^{k_i} \mathbf{a}_i = 0$, one finds that the term $\mathbf{a}_{i+1}$ has to be included whenever 
$k_0+ \ldots + k_i < 3 $. Essential instanton corrections are thus needed at almost all boundaries. 
We stress that essential instanton corrections will be taken into account in the rest of this work. 
While they can be constructed systematically as shown in \cite{Bastian:2021eom}, we will use there presence in a 
more indirect way in the following.


\subsubsection*{Nilpotent matrices and $\mathfrak{sl}(2)$}

As we have seen above, $N$ is a nilpotent matrix which belongs to a 
symplectic Lie algebra $\mathfrak{sp}(2h^{2,1}+2,\mathbb C)$. 
The classification of nilpotent elements of semi-simple Lie algebras is a
well-studied mathematical problem, and in the following we want  
to outline the main ideas of this classification. The classifications for $N \in \mathfrak{sp}(2h^{2,1}+2,\mathbb R)$ 
is slightly more involved and we will only quote the result in the following. 
For a clearer presentation, let us for this paragraph denote the nilpotent Lie algebra element 
by $N^-$ and let us use $m$ instead of $h^{2,1}+1$. 
\begin{itemize}

\item Let $N^-\in\mathfrak{sp}(2m,\mathbb K)$  be a nilpotent element of the Lie algebra $\mathfrak{sp}(2m,\mathbb C)$. Here 
 $\mathbb{K}$ can be either $\bbR$ or $\bbC$.  
A theorem by Jacobson and Morozov  states that one can always find elements 
$N^0,N^+\in \mathfrak{sp}(2m,\mathbb K)$ 
such that the algebra generated by $\{ N^-, N^+, N^0\}$ is isomorphic to $\mathfrak{sl}(2,\mathbb K)$. 
Recall that  $\mathfrak{sl}(2,\mathbb K)$ is generated by a triple $\{ n^-, n^+, n^0\}$ that satisfies the commutation relations
\eq{
\label{comm}
  [n^0,n^+]=+2\op n^+\,, \hspace{40pt}
  [n^0,n^- ]=-2\op n^- \,, \hspace{40pt}
  [n^+,n^-]= n^0\,.
}
Here, $n^+$ is a raising operator, $n^-$ is a lowering operator and $n^0$ is the weight operator, and we 
note that $\mathfrak{sl}(2,\mathbb K)$ is closely related to the Lie algebra $\mathfrak{su}(2)$. 

\item The triple $\{ n^-, n^+, n^0\} \in \mathfrak{sl}(2,\mathbb K)$ is represented by $2m\times 2m$ dimensional matrices 
$\{N^-,N^+,N^0\}$ acting on a $2m$-dimensional vector space $V$ over $\mathbb{K}$. 
However, this representation of  $\mathfrak{sl}(2,\mathbb K)$
is in general reducible and can therefore be expressed as a direct sum of irreducible 
representations of  $\mathfrak{sl}(2,\mathbb K)$. 
Concretely, this means that up to conjugation we can write each $N^-,N^+,N^0$ as
\eq{
  \label{nil_001}
  N^-,N^+,N^0=\left[\hspace{4pt}
  \arraycolsep0pt
  \begin{array}{ccc@{\hspace{6pt}}c@{\hspace{4pt}}}
  \setlength{\fboxsep}{6pt}
  \fbox{\hspace{0pt}$\ast$\hspace{0pt}}
  \\  
  &\setlength{\fboxsep}{14pt}
  \fbox{\hspace{0pt}$\ast$\hspace{0pt}}
\\
  &&\setlength{\fboxsep}{6pt}
  \fbox{\hspace{0pt}$\ast$\hspace{0pt}}
  \\
  &&& \ddots
  \end{array}
  \right] ,
}
where each block corresponds to an irreducible $\mathfrak{sl}(2,\mathbb K)$ representation
of $\{ n^-,n^+,n^0\}$ of a certain dimension. Note that typically some of these blocks 
correspond to the one-dimensional representation of $\mathfrak{sl}(2,\mathbb K)$ which 
is simply a zero. 
Also, the dimensions of the blocks 
have to add up to $2m$.

\item The classification of all possible nilpotent matrices $N^-$ becomes a combinatorial problem of how these can 
be decomposed in irreducible pieces.  Considering $N^- \in \mathfrak{sp}(2m,\mathbb C)$ it is well-known that 
Young diagrams classify irreducible representations. In the case of $N^- \in \mathfrak{sp}(2m,\mathbb R)$ this  problem is solved using 
so-called signed Young diagrams as explained, for example, in \cite{BrosnanPearlsteinRobles, Kerr2017,Grimm:2018cpv}. 

\end{itemize}
Let us now apply the above discussion to our situation. The nilpotent 
$\mathfrak{sp}(2h^{2,1} +2)$ matrix $N$ introduced in \eqref{def_log_mo}
can be identified with the representation $N^-$ of $n^-$ acting on the vector space $V = H^{3}(Y_3,\mathbb K)$.
Therefore, there exists a matrix $S\in \mbox{Sp}(2h^{2,1} +2)$
such that $S^{-1} N \op S$ has a block-diagonal form where
each block $\nu$ is a matrix representation $\mathsf N^-_{[\nu]}$ of the $\mathfrak{sl}(2)$ lowering operator $n^-$,
that is
\eq{
  S^{-1} N \op S = \left[\hspace{4pt}
  \arraycolsep0pt
  \begin{array}{ccc@{\hspace{6pt}}c@{\hspace{4pt}}}
  \setlength{\fboxsep}{6pt}
  \fbox{$\mathsf N^-_{[1]}$\hspace{-1pt}}
  \\  
  &\setlength{\fboxsep}{14pt}
  \fbox{$\mathsf N^-_{[2]}$\hspace{-2pt}}
\\
  &&\setlength{\fboxsep}{6pt}
  \fbox{$\mathsf N^-_{[3]}$\hspace{-1pt}}
  \\
  &&& \ddots
  \end{array}
  \right] .
}
Let us finally note that the weight operators $\mathsf N^0_{[\nu]}$ in each irreducible block 
are determined only up to conjugation. This freedom will be fixed in
asymptotic Hodge theory by picking a certain  Deligne splitting to be discussed below. 
Furthermore, we have suppressed in this discussion that the nilpotent matrix $N$ and the associated triple $\{N^-=N,N^+,N^0\}$ 
has to be compatible with the Hodge decomposition in the limit $t \rightarrow i \infty$. This  
     aspect will be also relevant in the discussion of the Hodge star and will be more central in section \ref{sec_tech_details}.


\subsubsection*{Weight-space decomposition}

We now want to get a better understanding of how a nilpotent matrix 
acts on  vectors, for instance how in \eqref{npot_001} the matrix $N$ acts  on $\mathbf a_0$.
This leads us to the  weight-space decomposition
under the action of $\mathfrak{sl}(2,\mathbb K)$, where again we can consider $\mathbb{K}$ being either $\bbR$ 
or $\bbC$. We start with a \textit{single irreducible} $\mathsf n$-dimensional 
representation $\{\mathsf N^-,\mathsf N^+,\mathsf N^0\}$
of $\mathfrak{sl}(2,\mathbb K)$, which corresponds to one particular block in \eqref{nil_001}. 
(We suppress the subscript $[\nu]$ for now.)
As is known from Lie-algebra representation theory, the vector space $\mathsf V$ on which 
the $\mathsf n\times \mathsf n$-dimensional 
matrices $\{\mathsf N^-,\mathsf N^+,\mathsf N^0\}$ are acting can be decomposed into one-dimensional weight spaces as
\eq{
  \label{wd_001}
  \mathsf V = \mathsf V_{d} \oplus \mathsf V_{d-2} \oplus \ldots \oplus \mathsf V_{-d}\,,
  \hspace{50pt}
  \mathsf V_{\ell} = \{ v\in \mathsf V : \mathsf N^0 \op v = \ell \op v \} \,,
}
where the eigenvalues $\ell$ of $\mathsf N^0$ are the weights and $d=\mathsf n-1$ is the highest weight. 
The raising and lowering operators $\mathsf N^+$ and $\mathsf N^-$ then map between these 
spaces as $\mathsf N^+: \mathsf V_{\ell} \to \mathsf V_{\ell+2}$ and $\mathsf N^-: \mathsf V_{\ell} \to \mathsf V_{\ell-2}$.
From this decomposition we find that $\mathsf N^-$ satisfies
\eq{
   (\mathsf N^-)^{d+1} = (\mathsf N^+)^{d+1}  = 0 \,.
}
So far we have focussed on a single block of the decomposition \eqref{nil_001}, but 
we can now combine these building blocks as follows. The triple of 
$2m\times 2m$ dimensional matrices $\{N^-,N^+,N^0\}$ is a  sum of 
triples $\{\mathsf N^-_{[\nu ]},\mathsf N^+_{[\nu ]},\mathsf N^0_{[\nu]}\}$, which 
is therefore acting on a $2m$-dimensional vector space that can be decomposed as
\eq{
  \label{wd_005}
  V = \mathsf V_{[1]} \oplus \mathsf V_{[2]} \oplus \ldots\,.
}
The nilpotent matrix $N=N^-$ satisfies \eqref{npot_021}, and therefore
the largest allowed highest weight of the subspaces satisfies $d^{(i)}\leq 3$. In other words, 
in the decomposition \eqref{nil_001} at most four-dimensional irreducible representations of 
$\mathfrak{sl}(2,\mathbb K)$ can appear.


\subsubsection*{Hodge star}

The nilpotent orbit theorem can be used to determine the periods \eqref{npot_001} of the 
holomorphic three-form near the 
boundary. From this expression one can, in principle, determine the Hodge-star operator in that limit
for every asymptotic regime. 
However,  this approximation can be still too involved to 
be of practical use for moduli stabilization, since it will generally 
contain many sub-leading polynomial and exponential corrections.  
We are therefore going to perform 
other approximations as follows:
\begin{itemize}

\item Let us focus again on one particular block in the decomposition \eqref{nil_001}, and consider the 
weight-space decomposition shown in \eqref{wd_001}. For a given triple $\{\mathsf N^-,\mathsf N^+,\mathsf N^0\}$ we now introduce a real operator $\mathsf C_{\infty} $
satisfying the relations $\mathsf C^{-1}_{\infty} \mathsf N^+  \mathsf C_{\infty}=\mathsf N^-$, 
$\mathsf C^{-1}_{\infty} \mathsf N^0\op \mathsf C_{\infty}=-\mathsf N^0$ and the requirement $\mathsf C_{\infty}^2=-\mathds 1$.
This operator maps the subspaces $\mathsf V_{\ell} $ as
\eq{
  \label{wd_304}
  \mathsf C_{\infty} : \mathsf V_{\ell} \to \mathsf V_{-\ell} \,.
}
While these conditions significantly constrain $\mathsf C_\infty$ they do not 
fix it completely. In order to fix $\mathsf C_\infty$ we require that it corresponds to 
the Hodge star operator acting on this representation after being appropriately extended to the boundary.
To make this more precise, we combine the  $\mathsf C_\infty$
of the individual irreducible representations of 
$\mathfrak{sl}(2,\mathbb C)$ into a $C_\infty \in \mathfrak{sp}(2m,\bbR)$ acting 
on the full vector space $V$ given in \eqref{wd_005}.
$C_{\infty}$ is then obtained from the full Hodge star operator $C$ 
via the limiting procedure 
\begin{equation}\label{Cinfty_limit}
   C_\infty = \lim_{y \to \infty} e \, C \, e^{-1} \,,
   \hspace{70pt}
   e= \exp \left[\tfrac{1}{2} \log y \, N^0 \right] ,
\end{equation}
where $y=\mbox{Im}\, t$ is send to the boundary and where $N^0$ denotes the weight operator 
in the triple $\{ N, N^+,N^0\}$. Let us note that the simple expression \eqref{Cinfty_limit} is 
somewhat deceiving, since the construction of an appropriate $N^0$ requires to check for compatibility 
of the choice with the Hodge decomposition. In practice, as we will see in section \ref{sec_tech_details}, we will construct 
$C_\infty$ and the triple $\{ N, N^+,N^0\}$ at the same time. 

\item The operator $C_{\infty}$ in \eqref{wd_304} does not contain any dependence on 
the complex-structure variable 
$y=\mbox{Im}\, t$ which is send to the boundary. We therefore introduce 
a so-called sl(2)-approximated Hodge star operator by
\eq{
  \label{wd_305}
  C_{\rm sl(2)} : \mathsf V_{\ell} \to  \: \mathsf V_{-\ell} \,,\qquad C_{\rm sl(2)} v  = y^\ell C_{\infty} v \ ,
}
where $v  \in V_{\ell}$ and the power in the variable $y$ corresponds to the weight of the weight-space $\mathsf V_{\ell}$. 
Using $e$ defined in \eqref{Cinfty_limit} we can make \eqref{wd_305} more precise as follows
\eq{
  \label{wd_306}
   C_{\rm sl(2)} =  e^{-1} \op C_{\infty} \, e \,, 
}
which produces precisely the type of mapping shown in \eqref{wd_305}. Again, the action 
on the full vector space $V$ is obtained by combining the action in each subspace. 
Note that in \eqref{wd_306} we set the axion  to zero, which can be 
re-installed by replacing $C_{\rm sl(2)} \to e^{+ x N} C_{\rm sl(2)} e^{- x N}$. 
Finally, the relation to the Hodge-star matrix \eqref{matrices_001} is
\eq{
  \mathcal M_{\rm sl(2)} = \eta\op    C_{\rm sl(2)}  \,.
}

\item So far much of the above discussion was possible on the real vector space $V=H^3(Y_3,\bbR)$. 
As soon as one aims to talk about the Hodge decomposition and the compatibility with 
the construction of the $\mathfrak{sl}(2,\mathbb C)$ representation one is forced to work over $\bbC$.
Let us denote by  $Q$ the 
operator acting on elements of $H^{q,3-q}$ with eigenvalue $q-\frac{3}{2}$. Since $\overline{H^{p,q}} = H^{q,p}$ the operator 
$Q$ is imaginary and we have $Q \in i\, \mathfrak{sp}(2m,\bbR)$. It follows from \eqref{C_def} that the Hodge star operator $C$
can be written in terms of $Q$ as $C = e^{\pi i Q} = (-1)^{Q}$. In analogy
to \eqref{Cinfty_limit} one can then extract the information about 
the boundary Hodge decomposition by evaluating 
\beq
     Q_\infty  = \lim_{y \to \infty} e \, Q \, e^{-1} \,.
\eeq
As we will explain in the next subsection, also $Q_\infty$ should actually be constructed together with the triple $\{N,N^+,N^0 \}$, since these operators are linked through non-trivial compatibility conditions. To display these compatibility relations 
it is useful to introduce a complex triple $\{L_{-1},L_0,L_{+1} \}$ by setting 
\beq
   L_{\pm 1} = \frac{1}{2}(N^+ + N^- \mp i N^0)\ , \qquad L_0 = i(N^- - N^+)\ .  
\eeq 
The algebra satisfied by $\{L_{-1},L_0,L_{+1},Q_\infty \}$ then reads
\beq
   [L_0,L_{\pm 1}] = \pm 2 L_{\pm 1}\ , \qquad [L_1,L_{-1}] = L_0\ , \qquad [Q_\infty, L_\alpha] = \alpha L_\alpha\ . 
\eeq
Note that this is the algebra of $\mathfrak{sl}(2,\bbC) \oplus \mathfrak{u}(1)$ if one introduces 
the operator $\hat Q = Q_\infty - \frac{1}{2} L_0$ as the generator of the $\mathfrak{u}(1)$.

\end{itemize}
For our purpose of moduli stabilization, the sl(2)-approximation \eqref{wd_306} of the Hodge star operator 
is particularly useful as it one the one hand contains non-trivial information about the boundary behaviour
and on the other is still manageable for practical purposes. In particular, the field dependence in the 
sl(2)-approximation is polynomial and thus leads to algebraic equations when used, for example, in 
the vacuum condition \eqref{eom_001}. It is, however, important to stress that even in the sl(2)-approximation 
of the Hodge star we automatically include essential exponential corrections to the periods as discussed 
around \eqref{Pi-exp2}. 
The intuitive understanding that exponential corrections can lead to a polynomially 
growing Hodge star arises by noting that after taking a holomorphic derivative of $\Pi$ there is the freedom to rescale 
the result and remove an overall exponential factor. In the computation of the Hodge star, one indeed checks that exponential 
terms in the leading terms precisely cancel and ensure the leading polynomial behavior with coefficients set by, in general, several $\mathbf{a}_i$.


\subsubsection*{Summary of main steps}

Let us finally summarize the necessary steps to construct the  sl(2)-approximated 
Hodge-star operator shown in equation \eqref{wd_306}:
\begin{enumerate}

\item One has to choose a modulus $y = \mbox{Im}\,t$ which approaches the boundary of moduli space. 
Associated to this boundary, the corresponding axion $x=\mbox{Re}\,t$ admits a 
discrete symmetry corresponding to the monodromy transformation of the period vector shown in \eqref{monodromy_002}.

\item The associated monodromy matrix $T$ can be made unipotent, and induces a 
nilpotent log-monodromy matrix $N=\log T$. For Calabi-Yau threefolds each boundary 
has an $0\leq m\leq 3$ with $N^m \neq 0$ and $ N^{m+1}=0$.

\item The log-monodromy matrix $N$ can be interpreted as a lowering operator in an 
$\mathfrak{sl}(2,\mathbb C)$ triple $\{ N^- = N, N^+, N^0\}$. 
Then, the weight operator $N^0$ needs to be constructed which is used in the definition of the 
sl(2)-approximated Hodge-star operator shown in \eqref{wd_306}. 

\item The $\mathfrak{sl}(2,\mathbb C)$ triple $\{ N^- = N, N^+, N^0\}$ has 
to be compatible with the Hodge decomposition extended to the boundary. 
The latter can be encoded by an operator $Q_\infty$ that is constructed 
jointly with the triple. 

\end{enumerate}
Let us emphasize that the above steps apply when sending a single modulus to the boundary. 
Additional complications arise when two or more moduli $t^i$ are considered. 
More concretely, even though the corresponding log-monodromy matrices 
$N_i$ can be shown to commute, when including the associated weight operators $N^0_i$ these operators generically do not all commute 
with each other and hence one cannot construct a consistent weight-space decomposition immediately. 
How to deal with this situation will be explained in the next section.


\subsection{Constructing the multi-moduli sl(2)-approximation}
\label{sec_tech_details}

In this subsection we present a more rigorous approach for constructing 
the sl(2)-approximated Hodge star operator, including the situation 
with multiple moduli sent to the boundary. 
We first introduce the main concepts from asymptotic Hodge theory, 
and then describe an algorithm for obtaining the sl(2)-approximation of the Hodge-star operator. 
This algorithm can be extracted from \cite{CKS} and has also been discussed in \cite{Grimm:2018cpv}.


\subsubsection*{Approximations}

Let us start by introducing two of approximations which can be made near boundaries in complex-structure moduli space. We use the coordinates $t^ i=x^i+i\op y^i$ (with $i=1,\ldots, n$) introduced in \eqref{coordinates}, for which the boundary is located at $y^i = \infty$.
\begin{itemize}

\item As a first approximation we  consider the regime $y^i \gg 1$, which allows us to drop exponential corrections 
of order $\mathcal O(e^{2\pi i \op t^i})$ in the Hodge-star matrix \eqref{matrices_001}. We refer to this near-boundary region as the \textit{asymptotic regime}. While this already simplifies the Hodge star to a polynomial expression, it will still depend rather non-trivially on the relative ratios between the saxions $y^i$. 

\item For a second approximation we  assume relative hierarchies between the 
coordinates $y^i$, which corresponds to the sl(2)-approximation. This means we consider an ordering of the saxions as $y^1 \gg y^2 \gg \ldots \gg y^n\gg 1$, and we refer to this regime  as the \textit{strict asymptotic regime}. This regime can be specified more accurately by constraining the ratios $y_i/y_{i-1}$ by the inequalities 
\beq \label{strict-regime}
 \frac{y_1}{y_2} > \lambda\ ,\quad\frac{y_2}{y_3} > \lambda\ , \ \ldots\ , \ \frac{y_{n-1}}{y_n} > \lambda\ , \ y_n > \lambda\ ,
\eeq
where $\lambda \geq 1$ is a real number. Note that the choice of a strict asymptotic regime specifies an ordering. This implies, in particular, that 
a considered regime includes the ordered limit in which one takes first 
$y^1$ to the boundary, after that $y^2$ and so on. 
Furthermore, we emphasize that different orderings typically give rise to different asymptotic expressions 
for the considered coupling functions in the effective theories. The expansion parameter for these asymptotic expansions 
will be $1/\lambda$. Keeping only the leading terms then becomes more accurate for larger $\lambda$.

\end{itemize}


\subsubsection*{Main concepts I: (pure) Hodge structures and nilpotent orbits}

In order to explain how to construct the sl(2)-approximation, we introduce the relevant notions from asymptotic Hodge theory. For definiteness, let us consider a Calabi-Yau threefold and recall the familiar decomposition of the three-form cohomology $H^3(Y_3, \mathbb{C})$ into $(p,q)$-forms
\begin{equation}
H^3(Y_3, \mathbb{C} ) = \bigoplus_{p=0}^3 H^{3-p,p} 
= H^{3,0} \oplus H^{2,1} \oplus H^{1,2} \oplus H^{0,3}\, ,
\end{equation}
where $\bar{H}^{p,q} = H^{q,p}$. In asymptotic Hodge theory this $(p,q)$-decomposition is  reformulated in terms of the Hodge filtration $F^p$ with $0 \leq p \leq 3$, which groups all three-forms with at least $p$ holomorphic indices together. 
More concretely, for a Calabi-Yau threefold we have 
\eq{
  \arraycolsep2pt
  \begin{array}{lcl}
  F^3 &=& H^{3,0}\,,
  \\[4pt]
  F^2 &=& H^{3,0} \oplus H^{2,1}\,,
  \\[4pt]
  F^1 &=& H^{3,0} \oplus H^{2,1} \oplus H^{1,2} \,,
  \\[4pt]
  F^0 &=& H^{3,0} \oplus H^{2,1} \oplus H^{1,2} \oplus H^{0,3} \,,
  \end{array}
  \hspace{50pt}
  0 \subset F^3 \subset F^2 \subset F^1 \subset F^0\,,
}
and the two formulations are related by 
\begin{equation}\label{def_Hpq}
H^{p,q} = F^p \cap \bar{F}^q \, , \hspace{50pt} F^p = \bigoplus_{k=p}^3 H^{k,3-k}\, .
\end{equation}
The Hodge star operator $\star$ can be evaluated on the elements of the Hodge decomposition 
and will be denoted by $C$ when acting on cohomology classes. It acts as 
\beq \label{Weil-operator}
     C \omega = i^{p-q} \omega \ , \qquad \omega \in H^{p,q}\ . 
\eeq

The Hodge structure and the Hodge filtration vary non-trivially over the complex-structure moduli space. We have already briefly discussed 
that the $(3,0)$-form $\Omega$, as a representative of $F^3 = H^{3,0}$, has a holomorphic dependence on the coordinates. In fact, one 
can find holomorphically varying sections for spanning all spaces $F^p$. While in general the moduli dependence of these sections is very complicated, we 
can provide more information of form of the $F^p$ in the asymptotic regime $y^i \gg 1$ and constrain the moduli dependence on the $t^i$, $i=1,...,n$. 
It is a key result of the nilpotent orbit theorem by Schmid \cite{Schmid} that there is always a nilpotent orbit $ F^p_{\rm nil}$ that approximates 
the $F^p$. The idea is that, similar to the expansion of the $\Omega$ periods in \eqref{Pi-exp}, we can drop exponential corrections proportional $z^i = e^{2\pi i\op t^i}$ 
near the boundary for all $(p,q)$-forms. In practice this means we approximate the Hodge filtration $F^p$ in the following way
\begin{equation} \label{nilp-form}
F^p \simeq F^p_{\rm nil} = e^{t^i N_i} F^p_0\, ,
\end{equation}
where  $F^p_0$ is the limiting filtration which is obtained through
\begin{equation}\label{Fp0}
F^p_0 \equiv \lim_{t^i \to i \infty} e^{-t^i N_i} F^p(t)\, .
\end{equation}
Here we have introduced multiple log-monodromy matices $N_i$ which can be shown to commute  $[ N_i , N_j ] = 0$ if one focuses on one 
boundary located at $z_1=z_2=...=z_n=0$.

Let us note that the vector spaces $F^p_0$ can be thought of as setting the leading terms in the holomorphic expansion in $z^i = e^{2\pi i\op t^i}$
around $z^i=0$, which are constant with respect to changes in $t^i$. In other words, 
in the asymptotic regime the dependence  of the Hodge filtration $F^p$ 
on the complex-structure moduli $t^i$ is captured entirely through the factor $e^{t^i N_i}$. Note that this means that 
$F^3_{\rm nil}$ is represented by specifying the periods $e^{t^i N_i} \mathbf{a}_0$, i.e.~the polynomial term in \eqref{Pi-exp}.
In other words, referring back to \eqref{Pi-exp2}, the one-dimensional space $F^3_{\rm nil}$ does not contain the information 
about the essential exponential corrections and the derivatives of $F^3_{\rm nil}$ with respect to $t^i$ will, in general, 
not span the full spaces $F^p_{\rm nil}$, $p<3$. Nevertheless this information is encoded in the full nilpotent orbit 
when considering all $F^p_{\rm nil}$, i.e.~looking at $p=0,...,3$. In fact, it was shown in \cite{Bastian:2021eom} how the $F^p_{\rm nil}$ can 
be used to reconstruct the essential exponential corrections. In this process one repeatedly uses the fact $F^p_{\rm nil}$  
are vector spaces and that overall exponential factors can be simply ignored. In other words, the nilpotent orbit 
form \eqref{nilp-form} states that these spaces are spanned by polynomial expressions up to overall rescalings in each direction. 
Importantly, the Hodge star $C$ acting as in \eqref{Weil-operator} only depends on the $(p,q)$-splitting of $H^{3}(Y_3, \mathbb{C} )$, i.e.~its split into vector 
spaces. This implies that it is equally independent of overall rescalings in each direction.


\subsubsection*{Main concepts II: mixed Hodge structures}

As we have discussed above, we can associate 
a (pure) Hodge structure $H^{p,q}$ to the Hodge filtration $F^p$, and their relation has been given in equation \eqref{def_Hpq}. 
A natural question then is whether a similar structure underlies the limiting filtration $F^p_0$ -- and indeed, the relevant 
framework in this case is that of a mixed Hodge structure. 
Concretely, this means we consider another splitting of the three-form cohomology $H^3(Y_3, \mathbb{C})$ known as the Deligne splitting $I^{p,q}$, which is  more refined than the $(p,q)$-form decomposition $H^{p,q}$. 
We discuss the following three aspects thereof: \label{page_deligne}
\begin{itemize}

\item The Deligne splitting requires us to introduce another set of vector spaces based on the log-monodromy matrices $N_i$. 
For definiteness, say we are interested in a limit involving the first $k$ saxions $y^1,\ldots, y^k \to \infty$
for which we define
$N=N_{(k)} = N_1 + \ldots + N_k.$\footnote{In fact, one can consider any element $N=c_1 N_1 + \ldots +c_k N_k$ in the linear cone $c_i>0$. The resulting vector spaces $W_\ell(N)$ are independent of the choice of $c_i$.}
The so-called monodromy weight filtration for $N$ is then given by
\begin{equation}\label{Wfiltr}
W_{\ell}(N)= \sum_{j \geq \max(-1,\ell-3)} \ker N^{j+1} \cap \im N^{j-\ell+3}\, .
\end{equation}
It is instructive to compare these vector spaces with the weight decomposition of $H^3(Y_3, \mathbb{C})$ 
under an $\mathfrak{sl}(2)$-triple. Namely, the log-monodromy matrix $N$ is nilpotent
and can therefore be considered as the lowering operator of a $\mathfrak{sl}(2)$-triple. 
The corresponding weight operator allows for a decomposition of 
$H^3(Y_3, \mathbb{C})$ into weight eigenspaces $V_{\ell}$ similarly as in \eqref{wd_001},
and the monodromy-weight filtration is then given by 
\eq{
W_\ell = \bigoplus_{m}^{\ell-3} V_m \,.
}

\item After having introduced the monodromy weight filtration $W_\ell(N)$, we can now define the Deligne splitting $I^{p,q}$. 
It is given by the following intersection of vector spaces 
\begin{equation}\label{Ipq}
I^{p,q} = F_{0}^{p} \cap W_{p+q} \cap \bigg( \bar{F}_{0}^{q}\cap W_{p+q}+\sum_{j\geq 1} \bar{F}_{0}^{q-j}\cap W_{p+q-j-1} \bigg)\, ,
\end{equation}
where in the case of a Calabi-Yau threefold we have $p,q=0,\ldots,3$. These vector spaces 
can be arranged into a diagram, which we have shown in figure~\ref{fig_deligne}.
The limiting filtration $F^p_0$ as well as the monodromy weight filtration $W_\ell$ can be recovered from the Deligne splitting 
via the relations
\begin{equation}
F^p_0 = \bigoplus_{r \geq p} \bigoplus_s I^{r,s}\, , \hspace{50pt} 
W_\ell = \bigoplus_{p+q = \ell} I^{p,q}\,.
\end{equation}

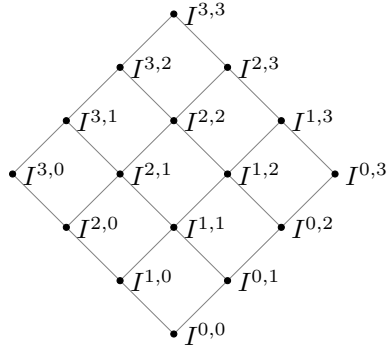
\begin{figure}[t]
\centering
\begin{tikzpicture}[baseline={([yshift=-.5ex]current bounding box.center)},scale=1,cm={cos(45),sin(45),-sin(45),cos(45),(15,0)}]
  \draw[step = 1, gray, ultra thin] (0, 0) grid (3, 3);
  \draw[fill] (3, 0) circle[radius=0.04] node[right]{\small $I^{0,3}$};
  \draw[fill] (2, 0) circle[radius=0.04] node[right]{\small $I^{0,2}$};
  \draw[fill] (1, 0) circle[radius=0.04] node[right]{\small $I^{0,1}$};
  \draw[fill] (0, 0) circle[radius=0.04] node[right]{\small $I^{0,0}$};
  \draw[fill] (3, 1) circle[radius=0.04] node[right]{\small $I^{1,3}$};
  \draw[fill] (2, 1) circle[radius=0.04] node[right]{\small $I^{1,2}$};
  \draw[fill] (1, 1) circle[radius=0.04] node[right]{\small $I^{1,1}$};
  \draw[fill] (0, 1) circle[radius=0.04] node[right]{\small $I^{1,0}$};
  \draw[fill] (3, 2) circle[radius=0.04] node[right]{\small $I^{2,3}$};
  \draw[fill] (2, 2) circle[radius=0.04] node[right]{\small $I^{2,2}$};
  \draw[fill] (1, 2) circle[radius=0.04] node[right]{\small $I^{2,1}$};
  \draw[fill] (0, 2) circle[radius=0.04] node[right]{\small $I^{2,0}$};
  \draw[fill] (3, 3) circle[radius=0.04] node[right]{\small $I^{3,3}$};
  \draw[fill] (2, 3) circle[radius=0.04] node[right]{\small $I^{3,2}$};
  \draw[fill] (1, 3) circle[radius=0.04] node[right]{\small $I^{3,1}$};
  \draw[fill] (0, 3) circle[radius=0.04] node[right]{\small $I^{3,0}$};
\end{tikzpicture}
\caption{Arrangement of the Deligne splitting $I^{p,q}$ for Calabi-Yau threefolds.
\label{fig_deligne}}
\end{figure}

\item Let us note that the expression \eqref{Ipq}  is somewhat involved, but for particular 
cases it reduces to a simpler form. 
More concretely, in general the vector spaces $I^{p,q}$ are related under complex conjugation only up to lower-positioned elements, that is 
\begin{equation}\label{Ipqbar}
\bar{I}^{p,q} = I^{q,p} \mod \bigoplus_{r<q,s<p} I^{r,s}\, .
\end{equation}
The special case $\bar{I}^{p,q} = I^{q,p}$ is called $\mathbb{R}$-split, 
for which the Deligne splitting \eqref{Ipq} reduces to\footnote{A simple check of this relation is to see how one can recover the pure Hodge structure \eqref{def_Hpq} corresponding to a log-monodromy matrix $N=0$. In this case the monodromy weight filtration \eqref{Wfiltr} is given by $W_\ell = 0$ for $\ell < 3$ and $W_\ell = H^3(Y_3)$ for $\ell \geq 3$, and hence the Deligne splitting reduces to $I^{p,q} = F_{0}^{p} \cap \bar{F}_{0}^{q}$ for $p+q=3$ while all others vanish.\label{foot_check}} 
\begin{equation}\label{Ipq_Rsplit}
I^{p,q}_{\text{$\mathbb{R}$-split}}=   F_{0}^{p} \cap \bar{F}_{0}^{q} \cap W_{p+q} \, .
\end{equation}

\end{itemize}


\subsubsection*{Algorithm I: rotation to the $\mathfrak{sl}(2)$-splitting}

Having introduced some of the main concepts of asymptotic Hodge theory, we now turn to the algorithmic part of determining 
the sl(2)-approximation of the Hodge-star operator.
A crucial observation relevant for the sl(2)-orbit theorem \cite{Schmid, CKS} is that any Deligne splitting can be rotated to a special $\mathbb{R}$-split, known as the $\mathfrak{sl}(2)$-split. This rotation is implemented by rotating the limiting filtration $F^p_0$ and consists of two steps: first we use a real phase operator $\delta$ to rotate to an $\mathbb{R}$-split, and then we rotate to the $\mathfrak{sl}(2)$-split via another real operator $\zeta$
\eq{
  \label{notation_8492}
  \fbox{$I^{p,q}$\vphantom{$\mathfrak{sl}(2)$}} 
  \hspace{20pt}\xrightarrow{\hspace{10pt}\delta\hspace{10pt}}\hspace{20pt}
  \fbox{$\hat I^{p,q}$ $=$ $\mathbb R$-split $I^{p,q}$\vphantom{$\mathfrak{sl}(2)$}} 
  \hspace{20pt}\xrightarrow{\hspace{10pt}\zeta\hspace{10pt}}\hspace{20pt}
  \fbox{$\tilde I^{p,q}$ $=$ $\mathfrak{sl}(2)$-split $I^{p,q}$} 
}

Let us begin by performing the rotation to the $\mathbb{R}$-split. We introduce a grading operator $\cN^0$ to ascertain how the complex conjugation rule $\bar{I}^{p,q} = I^{q,p}$ is modified by \eqref{Ipqbar}. This grading 
operator will serve as a starting point to construct a triple with $N= N_1 + \ldots + N_k$ as the lowering operator. To be precise, we the grading operator $\cN^0$ acts on an element of $I^{p,q}$ as follows
\begin{equation}\label{Hcomplex}
\cN^0 \, \omega_{p,q} = (p+q-3)\,  \omega_{p,q}\, , \hspace{40pt} \omega_{p,q} \in I^{p,q}\, .
\end{equation}
Clearly this means that when the Deligne splitting is not $\mathbb{R}$-split,  i.e.~$\bar{I}^{p,q} \neq I^{q,p}$, then $\cN^0$ is not a real operator since $I^{p,q}$ and $I^{q,p}$ are part of the same eigenspace of $\cN^0$. In fact, we can use the way that $\cN^0$ transforms under complex conjugation to determine how the $F^p_0$ should be rotated. We can achieve this by writing the transformation rule conveniently as 
\begin{equation}\label{Hconjugate}
\bar{\cN}^0 = e^{-2 i \delta} \cN^0 \op e^{+2i \delta}\, , 
\hspace{50pt}
\delta \in \mathfrak{sp}(2h^{2,1} +2,\bbR)\,,
\end{equation}
where $\delta$ denotes the phase operator of the rotation. Note that $\delta$ commutes with all 
log-monodromy matrices $N_i$. This operator can be decomposed with respect to its action on $I^{p,q}$ as follows
\begin{equation}\label{delta_decomp}
\delta = \sum_{p,q \geq 1} \delta_{-p, -q}\, , \hspace{50pt} \delta_{-p,-q} \, I^{r,s} = I^{r-p, s-q}\, .
\end{equation}
The fact that $\delta$ only has components with $p,q \geq 1$ follows from the fact that under complex conjugation the $I^{p,q}$ are related only modulo lower-positioned elements, as can be seen from \eqref{Ipqbar}. One can then proceed and solve \eqref{Hconjugate} for the components $\delta_{-p,-q}$ of the phase operator as\footnote{Here we used the identity $e^{X} Y e^{-X} = Y + [X, Y] +\frac{1}{2!} [X, [X, Y] ]+ \ldots$, and that $[H, \delta_{-p,-q} ] = -(p+q) \delta_{-p,-q}$.}
\begin{equation}\label{delta}
\begin{aligned}
\delta_{-1, -1} &= \frac{i}{4} (\bar{\cN}^0- \cN^0)_{-1,-1} \, , \quad \delta_{-1, -2} = \frac{i}{6} (\bar{\cN}^0-\cN^0)_{-1,-2} \,, \quad \delta_{-1, -3} = \frac{i}{8} (\bar{\cN}^0-\cN^0)_{-1,-3} \, , \\
\delta_{-2,-2} &= \frac{i}{8} (\bar{\cN}^0-\cN^0)_{-2,-2} \, , \quad \delta_{-2, -3} = \frac{i}{10} (\bar{\cN}^0-\cN^0)_{-2,-3} - \frac{i}{5} [\delta_{-1, -2}, \delta_{-1,-1}]\, , \\
\delta_{-3, -3} &= \frac{i}{12} (\bar{\cN}^0-\cN^0)_{-3,-3} - \frac{i}{3} [\delta_{-2,-2}, \delta_{-1,-1}]\, ,
\end{aligned}
\end{equation}
and the other components follow by complex conjugation.  The rotation of the limiting filtration $F^p_0$ to the $\mathbb{R}$-split is then straightforwardly given by
(as shown in \eqref{notation_8492}, we use a hat to distinguish the $\mathbb{R}$-split case from the non-$\mathbb{R}$-split case) 
\begin{equation}\label{FpR}
\hat{F}^p_0 \equiv e^{-i \delta} F^p_0\, .
\end{equation}

Next we want to rotate from the $\mathbb{R}$-split to the $\mathfrak{sl}(2)$-split. We parametrize this rotation by an algebra element $\zeta \in \mathfrak{sp}(2h^{2,1} +2,\bbR)$. For a recent discussion in the physics literature on how $\zeta$ is fixed we refer to  \cite{Grimm:2020cda}.  For our purposes let us simply write down the obtained result, which tells us that $\zeta$ can be expressed componentwise in terms of $\delta$ as\footnote{These relations apply for Calabi-Yau threefolds, and their extended version for Calabi-Yau fourfolds is given in \eqref{fourfold_eta} and \eqref{fourfold_zeta}.}
\begin{equation}\label{zeta}
\begin{aligned}
\zeta_{-1,-2} &= -\frac{i}{2} \delta_{-1,-2}\, , \qquad &\zeta_{-1,-3} &= - \frac{3i}{4} \delta_{-1,-3}\, , \\
 \zeta_{-2,-3} &= -\frac{3i}{8} \delta_{-2,-3} - \frac{1}{8} [ \delta_{-1,-1}, \delta_{-1,-2}]\, , \qquad &\zeta_{-3,-3} &= -\frac{1}{8} [ \delta_{-1,-1}, \delta_{-2,-2}]\, ,
\end{aligned}
\end{equation}
and all other components either vanish or are fixed by complex conjugation. Let us emphasize that these $(p,q)$-decompositions of the operators are computed with respect to the $\mathbb{R}$-split $\hat{I}^{p,q}$ obtained from \eqref{FpR} and not $I^{p,q}$. In particular, this means that one cannot use the components $\delta_{-p,-q}$ following from \eqref{delta} directly, since these were evaluated with respect to the starting Deligne splitting $I^{p,q}$. One should rather compute the $\mathbb{R}$-split $\hat{I}^{p,q}$ first explicitly by using \eqref{Ipq} for $\hat{F}^p_0$, and subsequently decompose $\delta$ with respect to $\hat{I}^{p,q}$ in order to determine $\zeta$. The $\mathfrak{sl}(2)$-split is then obtained by applying $\zeta$ on the limiting filtration in the following way
\begin{equation}\label{Fptilde}
\tilde{F}^p_0= e^{\zeta} \hat{F}^p_0= e^{\zeta}  e^{-i \delta} F^p_0\, ,
\end{equation}
where we used a tilde to distinguish the $\mathfrak{sl}(2)$-split from the other two cases. The $\mathfrak{sl}(2)$-split $\tilde{I}^{p,q}$ is then straightforwardly computed by using \eqref{Ipq} for $\tilde{F}^p_0$, similar to how the $\mathbb{R}$-split $\hat{I}^{p,q}$ was obtained.


\subsubsection*{Algorithm II: iterating through the saxion hierarchies}
We now iterate the above procedure through all saxion hierarchies in order to obtain the sl(2)-approximation in the regime $y^1 \gg y^2 \gg \ldots \gg y^n \gg 1$ specified by \eqref{strict-regime}. We start from the lowest hierarchy where all saxions are taken to be large $y^1, \ldots, y^n \to \infty$, and move one saxion at a time up to the highest hierarchy $y^1 \to \infty$. A flowchart illustrating how this iteration runs has been provided in figure \ref{algorithm}. Our construction follows the same steps as \cite{CKS, Kato, Grimm:2018cpv}, in particular, we point out that \cite{Grimm:2018cpv} already contains some examples that have been worked out explicitly. 
\tikzstyle{block} = [draw,, rectangle, 
    minimum height=3em, minimum width=2em, align=center]
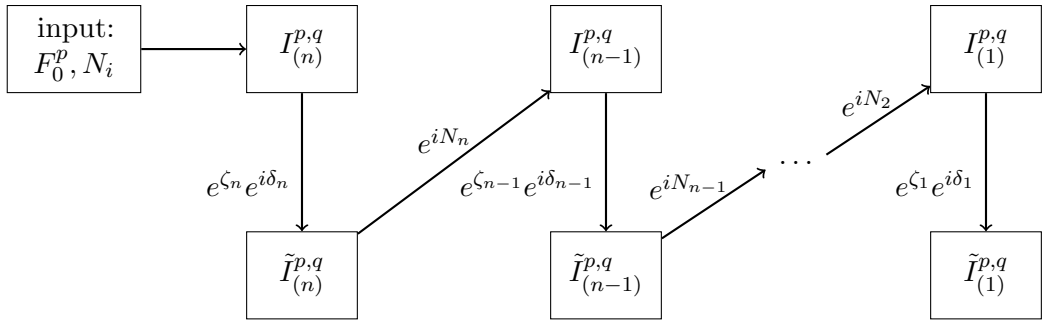
\begin{figure}[h!]
\centering
\begin{tikzpicture}
\node[block, text width=1.5cm] (aa) at (-3,3) {input: \\
$F^p_0, N_i$};
\node[block, text width=1.2cm] (a) at (0,3) {$I^{p,q}_{(n)}$};
\node[block, text width=1.2cm] (b) at (0,0) {$\tilde{I}^{p,q}_{(n)}$};
\node[block, text width=1.2cm] (c) at (4,3) {$I^{p,q}_{(n-1)}$};
\node[block, text width=1.2cm] (d) at (4,0) {$\tilde{I}^{p,q}_{(n-1)}$};

\node (e) at (6.5,1.5) { $\quad \ldots \quad $};

\node[block, text width=1.2cm] (f) at (9,3) {$I^{p,q}_{(1)}$};
\node[block, text width=1.2cm] (g) at (9,0) {$\tilde{I}^{p,q}_{(1)}$};

\draw[thick, ->] (aa) -- (a) ;
\draw[thick, ->] (a) -- (b) node[pos=0.65, left] {$e^{\zeta_n}e^{i\delta_n}$};
\draw[thick, ->] (b) -- (c) node[pos=0.65, left] {$e^{i N_n}$};
\draw[thick, ->] (c) -- (d) node[pos=0.65, left] {$e^{\zeta_{n-1}}e^{i\delta_{n-1}}$};
\draw[thick, ->] (d) -- (6.1, 1.4) node[pos=0.75, left] {$e^{i N_{n-1}}$};

\draw[thick, ->] (6.9,1.6) -- (f) node[pos=0.75, left] {$e^{i N_{2}}$};
\draw[thick, ->] (f) -- (g) node[pos=0.65, left]  {$e^{\zeta_{1}}e^{i\delta_{1}}$};
\end{tikzpicture}
\caption{Flowchart illustrating  the algorithm for constructing the sl(2)-approximation. This figure focuses on the construction of the $\mathfrak{sl}(2)$-split Deligne splittings $\tilde{I}^{p,q}_{(k)}$.  We labelled each arrow by the rotation that has to be applied on the limiting filtration $F^p_0$ according to \eqref{recursion}. In particular, each downward arrow corresponds to an iteration of the $\mathfrak{sl}(2)$-splitting algorithm for the Deligne splitting $I^{p,q}_{(k)}$, where $\delta_k, \zeta_k$ are determined by \eqref{delta} and \eqref{zeta}. }\label{algorithm}
\end{figure}

\begin{enumerate}

\item Our starting data from the periods is given by the limiting filtration $F^p_0$ obtained from \eqref{Fp0}, together with the log-monodromy matrices $N_i$. We begin from the lowest hierarchy where all saxions are taken to be large. This means we consider the monodromy weight filtration \eqref{Wfiltr} of $N_{(n)}=N_1 + \cdots + N_n$. By using \eqref{Ipq} we subsequently compute the Deligne splitting  $I^{p,q}_{(n)}$, where we included a subscript to indicate that it involves all $n$ limiting coordinates. We can then apply the algorithm laid out above to compute rotation operators $\delta_n, \zeta_n$ by using \eqref{delta} and \eqref{zeta} in order to obtain the $\mathfrak{sl}(2)$-split $\tilde{I}^{p,q}_{(n)}$ through \eqref{Fptilde}.

\item The next step is to consider the Deligne splitting for the limit $y^1, \ldots, y^{n-1} \to \infty$. Similar to the previous step we can compute the monodromy weight filtration \eqref{Wfiltr} for $N_{(n-1)}=N_1 + \cdots + N_{n-1}$. However, the limiting filtration we should consider requires slightly more work. Let us denote the limiting filtration of the $\mathfrak{sl}(2)$-split obtained in the previous step by $\tilde{F}^p_{(n)}$. Rather than taking this limiting filtration, we should apply $N_{n}$ in the following way
\begin{equation}
F_{(n-1)}^p = e^{i N_n} \tilde{F}^p_{(n)}\, .
\end{equation}
Roughly speaking this can be understood as fixing the saxion $y^n$ to a value, i.e.~$y^n = 1$, while keeping the other saxions large. We then take this limiting filtration and compute the Deligne splitting $I^{p,q}_{(n-1)}$ at hierarchy $n-1$. 

\item From here on the construction continues in the same manner as for the previous step, where we start from 
$I^{p,q}_{(n-1)}$.
We first determine the rotation matrices $\delta_{n-1}, \zeta_{n-1}$ to obtain the $\mathfrak{sl}(2)$-split $\tilde{I}^{p,q}_{(n-1)}$, and then move on to the next hierarchy by applying $ N_{n-1}$.

\end{enumerate}
To summarize, the full iterative process can therefore be described by the two recursive identities
\begin{equation}\label{recursion}
\tilde{F}^p_{(k)} = e^{\zeta_k} e^{i \delta_k} F^p_{(k)}\, , \hspace{50pt} F^p_{(k-1)} = e^{i N_{k}} \tilde{F}^{p}_{(k)}\, .
\end{equation}


\subsubsection*{Algorithm III: constructing the sl(2) triples and sl(2)-approximated Hodge star}

We now discuss the final steps in constructing the sl(2)-approximated Hodge star $C_{\rm sl(2)}$. The above iteration of the $\mathfrak{sl}(2)$-splitting algorithm already provided us the necessary data in the form of the Deligne splittings $\tilde{I}^{p,q}_{(k)}$. The remaining task is now to read off the relevant quantities for constructing $C_{\rm sl(2)}$ from this data.
The strategy is to first determine a set of mutually communting sl(2)-triples 
\beq \label{commuting_triples}
    \{N^-_k,N^+_k,N^0_k \}\ , \qquad k=1,...,n\ ,
 \eeq 
one associated to each $N^k$. Subsequently we will read off the charge operator $Q_\infty$ and the boundary Hodge star $C_\infty$ and then derive 
$C_{\rm sl(2)}$  in generalization of \eqref{wd_306}. 

We begin by determining the weight operators $N^{0}_{(k)}=N^0_1+\ldots N^0_k $. Their action on the 
$\mathfrak{sl}(2)$-split $\tilde{I}^{p,q}_{(k)}$ is given as grading operators that multiply elements  as
\begin{equation}\label{def_weight}
N^0_{(k)}\,  \omega_{p,q} = (p+q-3)\,  \omega_{p,q} \, , \hspace{50pt} \omega_{p,q} \in \tilde{I}^{p,q}_{(k)} \, .
\end{equation}
Since the iteration of the $\mathfrak{sl}(2)$-splitting algorithm provides us with explicit expressions for the vector spaces $\tilde{I}^{p,q}_{(k)}$, this property suffices to write down the grading operators $N_{(k)}$ explicitly. The weight operators associated with the individual saxions are then determined via
\begin{equation}
N^0_k =N^0_{(k)} - N^0_{(k-1)}\, ,
\hspace{50pt}
N^0_{(0)}=0\,.
\end{equation}
Next we determine the lowering operators $N_k^-$ of the commuting $\mathfrak{sl}(2)$-triples. The idea is to construct these lowering operators out of the log-monodromy matrices $N_k$. However, while the $N_k$ commute with each other, generally they do not yet commute with the weight operators $N^0_k$ of the other $\mathfrak{sl}(2)$-triples. This can be remedied by projecting each $N_k$ onto its weight-zero piece under $N^0_{(k-1)}$. We can write this projection out as
\begin{equation}\label{lowering}
N_k = N_k^- + \sum_{\ell \geq 2} N_{k, -\ell} \, , \hspace{50pt} [N^0_{(k-1)}, \,  N_{k, -\ell} ]  = -\ell \op N_{k,- \ell}\, ,
\end{equation}
where $\ell$ specifies the weight under $N^0_{(k-1)}$, and the lowering operator $N_{k}^- = N_{k, 0}$ is fixed as the weight-zero piece. By projecting out the other components $N_{k,-\ell}$ we ensure that the resulting $\mathfrak{sl}(2)$-triples are commuting. It is now straightforward to complete the $\mathfrak{sl}(2)$-triples \eqref{commuting_triples} by solving 
$[N_k^0,N_k^+]= 2 N_k^+$ and $[N_k^+,N_k^-]=N^0_k$ for the raising operators $N^+_i$. For the purpose of computing $C_{\rm sl(2)}$ this will not be necessary.

Besides the $\mathfrak{sl}(2)$-triples we also need the boundary Hodge structure to construct the sl(2)-approximated Hodge star. The Hodge filtration $F^p_\infty$ defining this boundary Hodge structure can be obtained from any of the $\mathfrak{sl}(2)$-split filtrations $\tilde{F}^p_{(k)}$. By applying $N^-_{(k)} = N_1^- + \ldots +N_k^-$ on this filtration in a similar manner as the second equation in \eqref{recursion} we obtain
\begin{equation}\label{Fpinfty}
F^p_\infty = e^{i N^-_{(k)}} \tilde{F}^p_{(k)}\, .
\end{equation}
The pure Hodge structure is then obtained straightforwardly through \eqref{def_Hpq} and we can 
read off the operator $Q_\infty$ by 
\beq
 Q_{\infty} \omega = \frac{1}{2}\big(p- q\big) \omega\ , \qquad \omega \in H^{p,q}_\infty = F^p_\infty \cap \bar{F}^q_\infty\ .
\eeq 
Intuitively the appearance of this pure Hodge structure (rather than a mixed Hodge structure) follows from the fact that it corresponds to the Deligne splitting with a trivial nilpotent element $N_{(0)}^- = 0$, so it reduces to a pure Hodge structure as discussed in footnote~\ref{foot_check}. The boundary Hodge star operator can obtained through 
\begin{equation}\label{Cinfty}
C_\infty \op w_{p,q} = i^{p-q}\op w_{p,q}\, ,  \hspace{50pt} w_{p,q} \in H^{p,q}_\infty\,,
\end{equation}
or by recalling that $C_\infty  = e^{i \pi Q_\infty}$.

Finally, let us put the above building blocks together and construct the sl(2)-approximated Hodge star $C_{\rm sl(2)}$. Analogous to the one-modulus case \eqref{wd_306} we introduce a saxionic scaling operator $e(y)$ to capture the dependence on the saxions $y^k$. This allows us to define the sl(2)-approximated Hodge star as
\begin{equation}\label{boundarystar}
C_{\rm sl(2)} \equiv e^{+x^k N_k^-} e(y)^{-1}\, C_\infty \, e(y) e^{-x^k N_k^-}\, , \qquad e(y) = \exp\Big[ \frac{1}{2} \sum_k \log[y^k] \, N^0_k \Big]\, ,
\end{equation}
and the relation to the Hodge-star matrix \eqref{matrices_001} is again given by $\mathcal M_{\rm sl(2)} =\eta \op C_{\rm sl(2)}$. We then find that $C_{\rm sl(2)}$ gives a good approximation to the Hodge star operator $C$ in the strict asymptotic regime \eqref{strict-regime} with large $\lambda$, similar to the one-modulus case \eqref{Cinfty_limit}.


\subsection{An example computation of the sl(2)-approximation}
\label{sec_aht_example}
In this subsection we work out in detail the sl(2)-approximation for a two-moduli example. We consider the large complex-structure region for the mirror of the Calabi-Yau hypersurface inside $\mathbb{P}_{4}^{1,1,2,2,2}[8]$, 
which in this context was  studied originally in \cite{Candelas:1993dm, Hosono:1993qy, Hosono:1994ax}. 
A similar two-moduli model, $\mathbb{P}_4^{1,1,1,6,9}[18]$ of \cite{Hosono:1993qy, Candelas:1994hw}, has been worked out in 
\cite{Grimm:2018cpv} to which we refer for an additional example.


\subsubsection*{Periods and log-monodromy matrices at large complex structure}

Let us begin by recalling some generalities on periods near the large complex-structure point. 
The periods are encoded in a prepotential $\mathcal F$, which splits into a 
tree-level part,  one-loop corrections, and world-sheet instanton corrections when mapped via 
mirror symmetry to a quantity depending on curve volumes. 
At the large complex-structure point there are no essential exponential corrections and some of the leading polynomial corrections play no central role in the evaluation of the Hodge star.\footnote{More precisely, we note that 
there are generally linear and quadratic terms in $X^i$ appearing in \eqref{data_ex_prepot}. In contrast 
to the term proportional to $\chi$, their coefficients are real rational numbers. One can show that they can be absorbed 
by an additional $\mathfrak{sp}(6,\mathbb{Q})$-rotation of the three-form basis.} 
Thus, we can work at leading order with the prepotential
\begin{equation}
\label{data_ex_prepot}
\mathcal{F} = - \frac{1}{3!} \frac{\kappa_{ijk} X^i X^j X^k}{X^0} + \frac{1}{2}(X^0)^2 \frac{\chi \op\zeta(3)}{(2\pi\op i)^3} \, ,
\end{equation}
where $\kappa_{ijk}$ and $\chi$ are the triple-intersection numbers and Euler characteristic of the mirror 
Calabi-Yau threefold. Using the conventions shown in \eqref{kpot_100} and noting that $\mathcal F_I = \partial_I \mathcal F$, 
this prepotential results in the period vector
\begin{equation}
\Pi = \begin{pmatrix} 1 \\
t^i \\
\frac{1}{6} \kappa_{ijk} t^i t^j t^k +  \frac{i \chi \zeta(3)}{8\pi^3} \\
- \frac{1}{2} \kappa_{ijk} t^j t^k
\end{pmatrix}\, ,
\end{equation}
where we set $X^I = (1,t^i)$. 
Using the relation \eqref{monodromy_002} we can determine the monodromy matrix $T$, and 
via \eqref{def_log_mo} we find the log-monodromy matrices $N_a$ for $a=1,\ldots, h^{2,1} $. 
Together with  the leading term of the period vector (c.f.~\eqref{npot_001}), we have
\begin{equation}\label{logN_001}
N_a = \begin{pmatrix}
0 & 0 & 0 & 0 \\
\delta_{ai} & 0 & 0 & 0 \\
0 &  0 & 0 & -\delta_{aj} \\
0 & -\kappa_{aij}  & 0 & 0
\end{pmatrix}\, , \hspace{50pt}
 \mathbf{a}_0 = \left(1\, , \ 0 \, , \ \frac{i \chi \zeta(3)}{8\pi^3} \, , \ 0\right)^T.
\end{equation}
Note that for a large complex-structure limit the matrices $N_a$ always commute, 
and that they are elements of $\mathfrak{sp}(2h^{2,1}+2,\mathbb Z)$. 
Let us now specialize our discussion to  the Calabi-Yau threefold  $\mathbb{P}_{4}^{1,1,2,2,2}[8]$.
The relevant topological data can be found for instance in \cite{Candelas:1993dm}, which we recall as follows\footnote{
For a clearer presentation, we chose to interchange the complex-structure moduli $t^1 \leftrightarrow t^2$
as compared to \cite{Candelas:1993dm}.
}
\begin{equation}
\label{data_ex_001}
\kappa_{122}  = 4\, , \hspace{40pt} \kappa_{222} = 8 \, , \hspace{40pt} \chi = -168\, ,
\end{equation}
and all other intersection numbers vanishing. The log-monodromy matrices \eqref{logN_001} are then given by 
\begin{equation}
N_1 = \begin{pmatrix}
 0 & 0 & 0 & 0 & 0 & 0 \\
 1 & 0 & 0 & 0 & 0 & 0 \\
 0 & 0 & 0 & 0 & 0 & 0 \\
 0 & 0 & 0 & 0 & -1 & 0 \\
 0 & 0 & 0 & 0 & 0 & 0 \\
 0 & 0 & -4 & 0 & 0 & 0 
\end{pmatrix} , \hspace{40pt}
N_2 = \begin{pmatrix}
 0 & 0 & 0 & 0 & 0 & 0 \\
 0 & 0 & 0 & 0 & 0 & 0 \\
 1 & 0 & 0 & 0 & 0 & 0 \\
 0 & 0 & 0 & 0 & 0 & -1 \\
 0 & 0 & -4 & 0 & 0 & 0 \\
 0 & -4 & -8 & 0 & 0 & 0 
\end{pmatrix} .
\end{equation}
Having specified the relevant data about the periods near the large complex-structure point, we are now in the position to construct the sl(2)-approximation. In the following we work out this construction in explicit detail for the asymptotic regime  $y^1 \gg y^2 \gg 1$.


\subsubsection*{Step 1: sl(2)-splitting of $I^{p,q}_{(2)}$}

We begin our analysis with  both saxions $y^1$ and $y^2$ being large, which corresponds to 
considering the sum of log-monodromy matrices 
\eq{
\label{example_001}
N_{(2)}=N_1+N_2 \,.
}
As explained on page~\pageref{page_deligne}, in order to write down the associated Deligne splitting $I^{p,q}_{(2)}$ 
we need to determine the monodromy weight filtration $W_{\ell}(N_{(2)})$ and the limiting filtration $F^p_0$. 
We discuss these structures in turn:
\begin{itemize}

\item The monodromy weight filtration for the nilpotent operator \eqref{example_001} 
is  computed using  \eqref{Wfiltr}. 
Since $N_{(2)}$ is a finite-dimensional  and explicitly-known matrix this can easily be done using 
a computer-algebra program, and we find
 \begin{equation}\label{W(2)}
\arraycolsep2pt
\renewcommand{\arraystretch}{1.3}
 \begin{array}{lclcl}
 W_0(N_{(2)}) &=&W_1(N_{(2)}) &=& \mathrm{span}\bigl[  (0,0,0,1,0,0)\bigr]\, , \\
 W_2(N_{(2)}) &=& W_3(N_{(2)})&=&  \mathrm{span}\bigl[ (0,0,0,0,1,0)\,, (0,0,0,0,0,1) \bigr] \oplus
 W_0(N_{(2)}) \, ,  \\
 W_4(N_{(2)}) &=& W_5(N_{(2)})&=& \mathrm{span}\bigl[  (0,1,0,0,0,0)\,,  (0,0,1,0,0,0) \bigr] \oplus 
  W_2(N_{(2)})\, ,\\ 
 &&W_6(N_{(2)}) &=&  \mathrm{span}\bigl[(1,0,0,0,0,0) \bigr] \oplus 
 W_4(N_{(2)}) 
 \, .
 \end{array}
 \end{equation}
Note that this filtration corresponds  to the decomposition of the even homology into 
zero-, two-, four- and six-cycles on the mirror dual. To be precise, $W_{2p}(N_{(2)})$ is spanned by all 
even cycles of degree $2p$ or lower, which  turns out to be a general feature for the monodromy weight filtration of $N_{(h^{2,1}  )}$ at large complex structure.

\item For the limiting filtration \eqref{Fp0} we recall that the vector space $F^p$ is spanned by the first $3-p$ derivatives of the period vector. At large complex structure the derivative with respect to $t^i$ simply lowers a log-monodromy matrix $N_i$ in \eqref{npot_001}, since we can ignore derivatives of $\Gamma(z)$ near this boundary. Subsequently multiplying by 
$e^{-t^i N_i}$ from the left we find we are left with the vectors $\mathbf{a}_0$, $N_i \mathbf{a}_0, \ldots$ to span the spaces $F^p_0$. In other words, the vector space $F^p_0$ is obtained by taking the span of up to $3-p$ log-monodromy matrices $N_i$ acting on $\mathbf{a}_0$. We can represent this information about the limiting filtration $F^p_0$ succinctly in terms of a period matrix as
\begin{equation}\label{F(2)}
\Pi_{(2)} =  \begin{pmatrix} 
1 & 0 & 0 & 0 & 0 & 0 \\
 0 & 1 & 0 & 0 & 0 & 0 \\
 0 & 0 & 1 & 0 & 0 & 0 \\
 -\frac{21 i \zeta (3)}{\pi ^3} & 0 & 0 & 0 & 0 & 1 \\
 0 & 0 & 0 & 1 & 0 & 0 \\
 0 & 0 & 0 & 0 & 1 & 0 
 \end{pmatrix}.
\end{equation}
Here the first column corresponds to $F^3_0$, the first three columns to $F^2_0$, the first five columns to $F^1_0$, and all six columns $F^0_0 = H^3(Y_3)$.

\item Given the monodromy weight filtration \eqref{W(2)} and the limiting filtration \eqref{F(2)},
we can now compute the Deligne splitting via equation \eqref{Ipq}. Expressed in the 
diagrammatic form introduced in figure~\ref{fig_deligne},
with the associated vectors  understood as spanning the corresponding subspace, this 
yields
\begin{equation}\label{Ipq(2)}
I^{p,q}_{(2)}  = \begin{tikzpicture}[baseline={([yshift=-.5ex]current bounding box.center)},scale=1,cm={cos(45),sin(45),-sin(45),cos(45),(15,0)}]
  \draw[step = 1, gray, ultra thin] (0, 0) grid (3, 3);
  \draw[fill] (0, 0) circle[radius=0.04] node[right]{\small $(0,0,0,1, 0 , 0)$};
  \draw[fill] (1.05, 0.95) circle[radius=0.04] node[right]{\small $(0,0,0,0, 1 , 0)$};
  \draw[fill] (0.95, 1.05) circle[radius=0.04] node[left]{\small $(0,0,0,0, 0 , 1)$};
  \draw[fill] (1.95, 2.05) circle[radius=0.04] node[left]{\small $(0,1,0,0, 0 , 0)$};
  \draw[fill] (2.05, 1.95) circle[radius=0.04] node[right]{ \small $(0,0,1,0, 0 , 0)$};
  \draw[fill] (3, 3) circle[radius=0.04] node[right]{ \small $(1,0,0,-\frac{21 i \zeta(3)}{\pi^3}, 0 , 0)$};
\end{tikzpicture}
\end{equation}

\end{itemize}
However, note that that the Deligne splitting \eqref{Ipq(2)} is not $\mathbb{R}$-split. 
Indeed, under complex conjugation $I_{(2)}^{3,3}$ is shifted by a piece in $I_{(2)}^{0,0}$ as
\eq{\label{I33bar}
\bar{I}_{(2)}^{3,3} &= \mathrm{span} \bigl[ (1,0,0,+\tfrac{21 i \zeta(3)}{\pi^3}, 0 , 0)\bigr]\\
& = \mathrm{span} \bigl[ (1,0,0,-\tfrac{21 i \zeta(3)}{\pi^3}, 0 , 0)\bigr]\ \text{mod}\ \mathrm{span}\bigl[  \tfrac{42 i \zeta(3)}{\pi^3} (0,0,0,1, 0 , 0)\bigr]  = 
I_{(2)}^{3,3} \ \text{mod}\  I_{(2)}^{0,0}\, ,
}
and hence the complex-conjugation rule shown in  \eqref{Ipqbar} follows. 
We now want to perform a complex rotation of the period matrix $\Pi_{(2)}$ to make the Deligne splitting $\mathbb{R}$-split. This procedure is outlined between equations \eqref{Hcomplex} and \eqref{FpR}. 
\begin{itemize}

\item To begin with, we determine the grading operator $\cN^0_{(2)}$ as defined by \eqref{Hcomplex} for the Deligne splitting \eqref{Ipq(2)}. Using for instance a computer algebra progam, this  operator is computed as
\begin{equation}\label{H2complex}
\cN^0_{(2)}= \begin{pmatrix}
 3 & 0 & 0 & 0 & 0 & 0 \\
 0 & 1 & 0 & 0 & 0 & 0 \\
 0 & 0 & 1 & 0 & 0 & 0 \\
 -\frac{126 i \zeta (3)}{\pi ^3} & 0 & 0 & -3 & 0 & 0 \\
 0 & 0 & 0 & 0 & -1 & 0 \\
 0 & 0 & 0 & 0 & 0 & -1 
\end{pmatrix}\, ,
\end{equation}
where the imaginary component of $\cN^0_{(2)}$ corresponds  to the breaking of the $\mathbb{R}$-split property of the Deligne splitting $I^{p,q}_{(2)}$ in \eqref{I33bar}.

\item Next, we recall that the rotation is  implemented through a phase operator $\delta_2$, which  is fixed by the grading operator $\cN^0_{(2)}$ through \eqref{Hconjugate}, and $\delta_2$ can be computed explicitly by using \eqref{delta}. For the derived grading operator \eqref{H2complex} and Deligne splitting \eqref{Ipq(2)} we then  find the phase operator 
\begin{equation}
\delta_2 =  (\delta_2)_{-3,-3} = \begin{pmatrix}
 0 & 0 & 0 & 0 & 0 & 0 \\
 0 & 0 & 0 & 0 & 0 & 0 \\
 0 & 0 & 0 & 0 & 0 & 0 \\
- \frac{21 \zeta (3)}{\pi ^3} & 0 & 0 & 0 & 0 & 0 \\
 0 & 0 & 0 & 0 & 0 & 0 \\
 0 & 0 & 0 & 0 & 0 & 0 
\end{pmatrix}\, ,
\end{equation}
where we stressed that $\delta_2$ only has a $(-3,-3)$-component with respect to Deligne splitting $I^{p,q}_{(2)}$. By using \eqref{zeta} one therefore can already see that $\zeta_2=0$, so our rotation by $\delta$ will directly rotate us to the $\mathfrak{sl}(2)$-split. Following \eqref{recursion}, the period matrix of this limiting filtration $\tilde{F}^p_{(2)}$ is given by
\begin{equation}\label{sl2F2}
\tilde{\Pi}_{(2)} = e^{i \delta_2} \Pi_{(2)} =  \begin{pmatrix} 
1 & 0 & 0 & 0 & 0 & 0 \\
 0 & 1 & 0 & 0 & 0 & 0 \\
 0 & 0 & 1 & 0 & 0 & 0 \\
0 & 0 & 0 & 0 & 0 & 1 \\
 0 & 0 & 0 & 1 & 0 & 0 \\
 0 & 0 & 0 & 0 & 1 & 0 
 \end{pmatrix}.
\end{equation}

\item Combining this result  with the monodromy weight filtration \eqref{W(2)}, one straightforwardly shows that the $\mathfrak{sl}(2)$-splitting at the lowest hierarchy is spanned by
\begin{equation}\label{sl2Ipq(2)}
\tilde{I}_{(2)}^{p,q}  = \begin{tikzpicture}[baseline={([yshift=-.5ex]current bounding box.center)},scale=1,cm={cos(45),sin(45),-sin(45),cos(45),(15,0)}]
  \draw[step = 1, gray, ultra thin] (0, 0) grid (3, 3);

  \draw[fill] (0, 0) circle[radius=0.04] node[right]{\small $(0,0,0,1, 0 , 0)$};
  \draw[fill] (1.05, 0.95) circle[radius=0.04] node[right]{\small $(0,0,0,0, 1 , 0)$};
  \draw[fill] (0.95, 1.05) circle[radius=0.04] node[left]{\small $(0,0,0,0, 0 , 1)$};
  \draw[fill] (1.95, 2.05) circle[radius=0.04] node[left]{\small $(0,1,0,0, 0 , 0)$};
  \draw[fill] (2.05, 1.95) circle[radius=0.04] node[right]{ \small $(0,0,1,0, 0 , 0)$};
  \draw[fill] (3, 3) circle[radius=0.04] node[right]{ \small $(1,0,0,0, 0 , 0)$};
\end{tikzpicture}\, .
\end{equation}
Notice that the resulting $\mathfrak{sl}(2)$-split  $\tilde{I}_{(2)}^{p,p} $ can be interpreted precisely as the decomposition into $(p,p)$-forms on the mirror dual K\"ahler side, similar to our comment on the filtration $W(N_{(2)})$ below 
equation \eqref{W(2)}.

\end{itemize}


\subsubsection*{Step 2: sl(2)-splitting of $I^{p,q}_{(1)}$}

Following the algorithm outlined in figure~\ref{algorithm}, 
we next consider the hierarchy  set by $y^1$ with  $y^2 \ll y^1$. From a practical perspective this means we will be working with the Deligne splitting $I^{p,q}_{(1)}$ associated with the log-monodromy matrix $N_1$. 
We determine this splitting as follows:
\begin{itemize}

\item Similarly as above, we first compute the monodromy weight filtration \eqref{Wfiltr} associated with $N_1$. 
It takes the following form
 \eq{
 \label{W(1)}
 \arraycolsep2pt
\renewcommand{\arraystretch}{1.3}
 \begin{array}{@{}lclcl@{}}
 W_0(N_1)&=&W_1(N_1) &=&  0\, , \\
 W_2(N_1) &=& W_3(N_1) &=& \mathrm{span}\bigl[ (0,1,0,0,0,0)\,,  (0,0,0,0,0,1)\,, (0,0,0,1,0,0)\bigr]\, ,\\
 W_4(N_1) &=& W_5(N_1) &=& 
 \mathrm{span}\bigl[  (1,0,0,0,0,0)\,, (0,0,1,0,0,0)\, ,  (0,0,0,0,1,0) \bigr]
  \oplus W_2(N_1) \,,
  \\
 &&W_6(N_1) &=& W_4(N_1) \,.
 \\[-10pt]
\end{array}
}

\item Next, we determine the limiting filtration $F^p_{(1)}$ from the $\mathfrak{sl}(2)$-split filtration $\tilde{F}^p_{(2)}$ 
of the previous hierarchy through equation \eqref{recursion}. 
At the level of the period matrix this means we rotate \eqref{sl2F2}  as
\begin{equation}\label{F(1)}
\Pi_{(1)} = e^{i N_2} \tilde{\Pi}_{(2)}=
e^{i N_2} e^{i \delta_2} \Pi_{(2)}
= \begin{pmatrix}
 1 & 0 & 0 & 0 & 0 & 0 \\
 0 & 1 & 0 & 0 & 0 & 0 \\
 -i & 0 & 1 & 0 & 0 & 0 \\
 \frac{4 i}{3} & -2 & -4 & 0 & - i & 1 \\
 2 & 0 & 4 i & 1 & 0 & 0 \\
 4 & 4 i & 8 i & 0 & 1 & 0 \\
 \end{pmatrix}.
 \end{equation}

\item Using the definition of the Deligne splitting given in \eqref{Ipq} together with the monodromy weight filtration \eqref{W(1)} and the limiting filtration \eqref{F(1)}, we find that the Deligne splitting is spanned by
\begin{equation}\label{Ipq(1)}
I^{p,q}_{(1)} = 
\begin{tikzpicture}[baseline={([yshift=-.5ex]current bounding box.center)},scale=1,cm={cos(45),sin(45),-sin(45),cos(45),(15,0)}]
  \draw[step = 1, gray, ultra thin] (0, 0) grid (3, 3);
  \draw[fill] (0, 2) circle[radius=0.05] node[left]{\small $(0,1,0,-2,0,4i)$ };
  \draw[fill] (2, 0) circle[radius=0.05] node[right]{\small $(0,1,0,-2,0,-4i)$};
  \draw[fill] (1, 1) circle[radius=0.05] node[below]{\small $(0,1,0,2,0,0)$};
  \draw[fill] (2, 2) circle[radius=0.05] node[above]{\small $(1,-\frac{2i}{3},0,-\frac{4i}{3}, -2, -\frac{4}{3})$};
  \draw[fill] (1, 3) circle[radius=0.05] node[left]{\small $(1,0,-i,\frac{4i}{3},2,4)$ };
  \draw[fill] (3, 1) circle[radius=0.05] node[right]{\small  $(1,-\frac{4i}{3}, i, \frac{4i}{3}, 2, -\frac{4}{3})$};
\end{tikzpicture}\, .
\end{equation}

\end{itemize}
Note that none of the spaces in the upper part of the Deligne splitting \eqref{Ipq(1)} 
obey the $\mathbb{R}$-split conjugation rule. In  particular, under complex conjugation $I^{2,2}$ is related to itself up to a shift in $I^{1,1}$, while $I^{1,3}$ is related to $I^{3,1}$ up to a piece in $I^{2,0}$ and vice versa. To be precise, we find
\eq{
\bar{I}_{(1)}^{2,2} &= 
\mathrm{span}\bigl[(1,+\tfrac{2i}{3},0,+\tfrac{4i}{3}, -2, -\tfrac{4}{3}) \bigr] \\
&=
\mathrm{span}\bigl[ (1,-\tfrac{2i}{3},0,-\tfrac{4i}{3}, -2, -\tfrac{4}{3})\bigr]  \  \text{mod} \ 
\mathrm{span}\bigl[ \tfrac{4i}{3} (0,1,0,2,0,0) \bigr] = I_{(1)}^{2,2} \  \text{mod} \ I_{(1)}^{1,1}\, , 
\\
\bar{I}_{(1)}^{3,1} & =
\mathrm{span}\bigl[(1,+\tfrac{4i}{3}, -i, -\tfrac{4i}{3}, 2, -\tfrac{4}{3})\bigr]  \\
& = \mathrm{span}\bigl[ (1,0,-i,+\tfrac{4i}{3},2,4)\bigr] \  \text{mod} \ 
\mathrm{span}\bigl[ \tfrac{4i}{3} (0,1,0,-2,0,4i)\bigr] = I_{(1)}^{3,1} \  \text{mod} \ I_{(1)}^{2,0} \, ,
}
which is  in accordance with the complex conjugation rule \eqref{Ipqbar}.
We can now perform  the rotation to the $\mathfrak{sl}(2)$-split $\tilde{I}^{p,q}_{(1)}$:
\begin{itemize}

\item As before we begin by computing the grading operator defined in \eqref{Hcomplex} 
for the above Deligne splitting \eqref{Ipq(1)}. This grading operator is found to be
\begin{equation}
\cN^0_{(1)} = \begin{pmatrix}
 1 & 0 & 0 & 0 & 0 & 0 \\
 -\frac{4 i}{3} & -1 & -\frac{4}{3} & 0 & 0 & 0 \\
 0 & 0 & 1 & 0 & 0 & 0 \\
 0 & 0 & 0 & -1 & \frac{4 i}{3} & 0 \\
 0 & 0 & 0 & 0 & 1 & 0 \\
 0 & 0 & \frac{16 i}{3} & 0 & \frac{4}{3} & -1 \\
 \end{pmatrix}\, .
\end{equation}

\item Next, using \eqref{delta}, the phase operator is then computed from $\cN^0_{(1)}$ as
\begin{equation}
\delta_1 =(\delta_1 )_{-1,-1} \begin{pmatrix}
  0 & 0 & 0 & 0 & 0 & 0 \\
 -\frac{2}{3} & 0 & 0 & 0 & 0 & 0 \\
 0 & 0 & 0 & 0 & 0 & 0 \\
 0 & 0 & 0 & 0 & \frac{2}{3} & 0 \\
 0 & 0 & 0 & 0 & 0 & 0 \\
 0 & 0 & \frac{8}{3} & 0 & 0 & 0 \\
 \end{pmatrix}\, .
 \end{equation}
Note that similar to the Deligne splitting $I^{p,q}_{(2)}$ discussed above, we  find that $\zeta_1=0$ by using \eqref{zeta}, since $\delta$ only has a $(-1,-1)$-component. Therefore by rotating to the $\mathbb{R}$-split we again directly rotate to the $\mathfrak{sl}(2)$-split. At the level of the period matrix \eqref{F(1)} this means that $\tilde{F}^p_{(1)}$ can be represented as
\begin{equation}
\tilde{\Pi}_{(1)} = e^{i\delta_1} \Pi_{(1)} =  \begin{pmatrix}
 1 & 0 & 0 & 0 & 0 & 0 \\
 \frac{2 i}{3} & 1 & 0 & 0 & 0 & 0 \\
 -i & 0 & 1 & 0 & 0 & 0 \\
 0 & -2 & -\frac{4}{3} & -\frac{2 i}{3} & i & 1 \\
 2 & 0 & 4 i & 1 & 0 & 0 \\
 \frac{4}{3} & 4 i & \frac{16 i}{3} & 0 & 1 & 0 \\
 \end{pmatrix} \, .
\end{equation}

\item Finally, using \eqref{Ipq} for the filtration $\tilde{F}^p_{(1)}$ we obtain the $\mathfrak{sl}(2)$-splitting which 
is spanned by
\begin{equation}\label{sl2Ipq(1)}
\tilde{I}^{p,q}_{(1)} = 
\begin{tikzpicture}[baseline={([yshift=-.5ex]current bounding box.center)},scale=1,cm={cos(45),sin(45),-sin(45),cos(45),(15,0)}]
  \draw[step = 1, gray, ultra thin] (0, 0) grid (3, 3);
  \draw[fill] (2, 0) circle[radius=0.05] node[right]{\small $(0,1,0,-2,0,-4i)$};
  \draw[fill] (0, 2) circle[radius=0.05] node[left]{\small $(0,1,0,-2,0,4i)$};
  \draw[fill] (1, 1) circle[radius=0.05] node[below]{\small $(0,1,0,2,0,0)$};
  \draw[fill] (2, 2) circle[radius=0.05] node[above]{\small $(1,0,0,0, -2, -\frac{4}{3})$};
  \draw[fill] (3, 1) circle[radius=0.05] node[right]{\small $(1,-\frac{2i}{3},i,0,2,\frac{4}{3})$};
  \draw[fill] (1, 3) circle[radius=0.05] node[left]{\small $(1,\frac{2i}{3},-i,0,2,\frac{4}{3})$};
\end{tikzpicture}\, .
\end{equation}

\end{itemize}


\subsubsection*{Step 3: constructing the sl(2)-approximation}

We now construct the sl(2)-approximated Hodge star from the two $\mathfrak{sl}(2)$-splittings \eqref{sl2Ipq(1)} and \eqref{sl2Ipq(2)} determined above. We begin with the $\mathfrak{sl}(2)$-triples. The weight operators $N^0_{(i)}$ are fixed by the multiplication rule \eqref{def_weight} on the two $\mathfrak{sl}(2)$-splittings. For \eqref{sl2Ipq(1)} and \eqref{sl2Ipq(2)} these grading operators are respectively given by 
\begin{equation}
N^0_{(1)} = \begin{pmatrix} 
 1 & 0 & 0 & 0 & 0 & 0 \\
 0 & -1 & -\frac{4}{3} & 0 & 0 & 0 \\
 0 & 0 & 1 & 0 & 0 & 0 \\
 0 & 0 & 0 & -1 & 0 & 0 \\
 0 & 0 & 0 & 0 & 1 & 0 \\
 0 & 0 & 0 & 0 & \frac{4}{3} & -1 \\
 \end{pmatrix}\, , \qquad N^0_{(2)} = \begin{pmatrix}
  3 & 0 & 0 & 0 & 0 & 0 \\
 0 & 1 & 0 & 0 & 0 & 0 \\
 0 & 0 & 1 & 0 & 0 & 0 \\
 0 & 0 & 0 & -3 & 0 & 0 \\
 0 & 0 & 0 & 0 & -1 & 0 \\
 0 & 0 & 0 & 0 & 0 & -1 \\
  \end{pmatrix}\, .
\end{equation}
In order to construct the lowering operators $N_i^-$ we need to decompose the log-monodromy matrices $N_a$ 
with respect to the weight operators as described in \eqref{lowering}. For the first lowering operator we find simply that $N_1 = N_1^-$, since $N^0_{(0)}=0$. On the other hand, for $N_2$ we find that it decomposes with respect to $N^0_{(1)}$ as
\eq{
  N_2 = (N_2)_0 + (N_2)_{-2} \,,
}
where
\begin{equation}
(N_2)_0 = \begin{pmatrix}
 0 & 0 & 0 & 0 & 0 & 0 \\
 -\frac{2}{3} & 0 & 0 & 0 & 0 & 0 \\
 1 & 0 & 0 & 0 & 0 & 0 \\
 0 & 0 & 0 & 0 & \frac{2}{3} & -1 \\
 0 & 0 & -4 & 0 & 0 & 0 \\
 0 & -4 & -\frac{16}{3} & 0 & 0 & 0 \\
\end{pmatrix}\, , \qquad 
 \quad (N_2)_{-2} = \begin{pmatrix}
 0 & 0 & 0 & 0 & 0 & 0 \\
 \frac{2}{3} & 0 & 0 & 0 & 0 & 0 \\
 0 & 0 & 0 & 0 & 0 & 0 \\
 0 & 0 & 0 & 0 & -\frac{2}{3} & 0 \\
 0 & 0 & 0 & 0 & 0 & 0 \\
 0 & 0 & -\frac{8}{3} & 0 & 0 & 0 \\
 \end{pmatrix}\, .
\end{equation}
We then identify the lowering operator with the weight-zero piece as $N_2^- = (N_2)_0$.

Next we construct the boundary Hodge star $C_\infty$. The Hodge filtration $F^p_\infty$ associated with this boundary Hodge structure follows from \eqref{Fpinfty}, and we can write the period matrix associated to these vector spaces $F^p_\infty$ as
\begin{equation}
\tilde{\Pi}_{(0)} = e^{i N_1^-} \tilde{\Pi}_{(1)} = e^{i (N_1^-+N_2^-)} \tilde{\Pi}_{(2)} = \begin{pmatrix}
 1 & 0 & 0 & 0 & 0 & 0 \\
 \frac{i}{3} & 1 & 0 & 0 & 0 & 0 \\
 i & 0 & 1 & 0 & 0 & 0 \\
 -2 i & -2 & -\frac{16}{3} & -\frac{i}{3} & -i & 1 \\
 2 & 0 & -4 i & 1 & 0 & 0 \\
 \frac{16}{3} & -4 i & -\frac{28 i}{3} & 0 & 1 & 0 \\
 \end{pmatrix}\, .
\end{equation}
From this boundary Hodge filtration one finds that the boundary $(p,q)$-spaces $H^{p,q}_\infty = F^p_\infty \cap F^q_\infty$ are spanned by
\begin{equation}
\begin{aligned}
H^{3,0}_\infty &= \mathrm{span}\bigl[ ( 1  ,   -\tfrac{i}{3}  ,  -i  ,  2 i  ,  2  ,  \tfrac{16}{3} ) \bigr]\, , 
\\
H^{2,1}_\infty &= \mathrm{span}\bigl[( 1  ,   0  ,   -\tfrac{3 i}{8}  ,   -2 i  ,   -\tfrac{1}{2}  ,   -\tfrac{11}{6})\, ,\,  ( 0  ,   1  ,   -\tfrac{3}{8}  ,   0  ,   -\tfrac{3 i}{2}  ,   \tfrac{i}{2}  )\bigr]\, , \\
\end{aligned}
\end{equation}
and the other spaces follow by complex conjugation. The corresponding Hodge star operator is then determined by the multiplication rule  \eqref{Cinfty} as
\begin{equation}
C_\infty = \begin{pmatrix}
 0 & 0 & 0 & \frac{1}{2} & 0 & 0 \\
 0 & 0 & 0 & 0 & \frac{11}{18} & -\frac{1}{6} \\
 0 & 0 & 0 & 0 & -\frac{1}{6} & \frac{1}{4} \\
 -2 & 0 & 0 & 0 & 0 & 0 \\
 0 & -2 & -\frac{4}{3} & 0 & 0 & 0 \\
 0 & -\frac{4}{3} & -\frac{44}{9} & 0 & 0 & 0 \\
\end{pmatrix} .
\end{equation}
Having constructed the necessary building blocks, we are now ready to put together the sl(2)-approximated Hodge star operator according to \eqref{boundarystar}. Setting the axions to zero for simplicity, that is $x^i=0$, we find that 
\begin{equation}
\label{data_ex_csl2}
C_{\rm sl(2)} = \begin{pmatrix}
 0 & 0 & 0 & -\frac{1}{2 y_{1} y_{2}^2} & 0 & 0 \\
 0 & 0 & 0 & 0 & -\frac{y_{1}}{2 y_{2}^2}-\frac{1}{9 y_{1}} & \frac{1}{6 y_{1}} \\
 0 & 0 & 0 & 0 & \frac{1}{6 y_{1}} & -\frac{1}{4 y_{1}} \\
 2 y_{1} y_{2}^2 & 0 & 0 & 0 & 0 & 0 \\
 0 & \frac{2 y_{2}^2}{y_{1}} & \frac{4 y_{2}^2}{3 y_{1}} & 0 & 0 & 0 \\
 0 & \frac{4 y_{2}^2}{3 y_{1}} & \frac{8 y_{2}^2}{9 y_{1}}+4 y_{1} & 0 & 0 & 0 \\
\end{pmatrix},
\end{equation}
and the Hodge-star matrix $\mathcal M_{\rm sl(2)}$ is then again obtained as
$\mathcal M_{\rm sl(2)} = \eta\op C_{\rm sl(2)}$. 

It is instructive to compare the sl(2)-approximated Hodge star operator \eqref{data_ex_csl2} to the full Hodge star operator \eqref{hodgestar01} computed via the LCS prepotential \eqref{data_ex_prepot}. In particular, this allows us to see which terms are dropped in the sl(2)-approximation. We find that the full Hodge star operator is given by
\eq{\label{Cfull}
C &= \scalebox{0.8}{$\begin{pmatrix}
0 & 0 & \frac{6}{\cK} & 0 \\
0 & 0 & 0 & \frac{3}{2 \cK} G^{ij}\\
  - \frac{\cK}{6} & 0 & 0 & 0 \\
0 & -\frac{2\cK}{3} G_{ij}  & 0 & 0
\end{pmatrix}$} \\
&= \scalebox{0.8}{$\left(
\begin{array}{cccccc}
 0 & 0 & 0 & -\frac{3}{6 y_1 y_2^2+4 y_2^3} & 0 & 0 \\
 0 & 0 & 0 & 0 & -\frac{3 y_1^2+4 y_1 y_2+2 y_2^2}{6 y_1 y_2^2+4 y_2^3} &
   \frac{1}{6 y_1+4 y_2} \\
 0 & 0 & 0 & 0 & \frac{1}{6 y_1+4 y_2} & -\frac{3}{12 y_1+8 y_2} \\
 \frac{2}{3} y_2^2 (3 y_1+2 y_2) & 0 & 0 & 0 & 0 & 0 \\
 0 & \frac{6 y_2^2}{3 y_1+2 y_2} & \frac{4 y_2^2}{3 y_1+2 y_2} & 0 & 0 & 0 \\
 0 & \frac{4 y_2^2}{3 y_1+2 y_2} & \frac{4 \left(3 y_1^2+4 y_1 y_2+2
   y_2^2\right)}{3 y_1+2 y_2} & 0 & 0 & 0 \\
\end{array}
\right) $}\, , 
}
with $\cK = \cK_{ijk} y^i y^j y^k - \tfrac{3 \chi \zeta(3)}{16 \pi^3}   $ and K\"ahler metric
\begin{equation}
G_{ij} = \left(
\begin{array}{cc}
 \frac{144 y_2^4}{\left(8 y_2^2 (3 y_1+2 y_2)-3 \epsilon \right)^2} & \frac{96 y_2^4+36
   y_2 \epsilon }{\left(8 y_2^2 (3 y_1+2 y_2)-3 \epsilon \right)^2} \\
 \frac{96 y_2^4+36 y_2 \epsilon }{\left(8 y_2^2 (3 y_1+2 y_2)-3 \epsilon \right)^2} &
   \frac{96 y_2^2 \left(3 y_1^2+4 y_1 y_2+2 y_2^2\right)+36 \epsilon  (y_1+2
   y_2)}{\left(8 y_2^2 (3 y_1+2 y_2)-3 \epsilon \right)^2} \\
\end{array}
\right)\, ,
\end{equation} 
where in the second line of \eqref{Cfull} we dropped the $ \epsilon = \frac{ \chi\zeta(3)}{8 \pi^3}$ correction. Note by comparing \eqref{data_ex_csl2} and \eqref{Cfull}, we see that the sl(2)-approximation is more involved than simply dropping $\epsilon$, and in the limit $y_1 \gg y_2 \gg 1$ we exclude further polynomial corrections in $y_2/y_1$.


\section{Moduli stabilization in the asymptotic regime}\label{sec-mod-stab}
In this section we explain how the sl(2)-approximation can be used as a first step for finding flux vacua. We begin with an overview of the various degrees of approximation in section \ref{ssec:scheme}. Using this scheme we then stabilize moduli for three different examples in complex-structure moduli space in sections \ref{ssec:LCS2}, \ref{ssec:LCS3} and \ref{ssec:coniLCS}.

\subsection{An approximation scheme for finding flux vacua}\label{ssec:scheme}
In general, it is a rather difficult problem to stabilize moduli in flux compactifications, since the Hodge star operator in the extremization conditions \eqref{eom_003} depends in an intricate way on the underlying complex-structure moduli. In this work we propose to address this problem by approximating the moduli dependence of the scalar potential in three steps as
\tikzstyle{block} = [draw,, rectangle, 
    minimum height=3em, minimum width=2em, align=center]
\begin{equation}
\begin{tikzpicture}[baseline = -0.4ex]
\node[block, text width=4.2cm] (a) at (0,0) {(1) sl(2)-approximation};
\node[block, text width=3.5cm] (b) at (5,0) {(2) Nilpotent orbit approximation};
\node[block, text width=4.5cm] (c) at (10,0) {(3) Full series of \\
exponential corrections };

\draw [ thick, ->] (a) -- (b);
\draw [ thick, ->] (b) -- (c);
\end{tikzpicture}
\end{equation}
By using the sl(2)-approximation as a first step, we break the behavior of the scalar potential down into polynomial terms in the complex-structure moduli. This allows us to solve the relevant extremization conditions for the vevs of the moduli straightforwardly. Subsequently, we can include corrections to this polynomial behavior by switching to the nilpotent orbit approximation. Roughly speaking, this amounts to dropping all exponential corrections in the scalar potential,\footnote{We should stress that one has to be careful where these exponential corrections are dropped. While we can drop corrections at these orders in the scalar potential, essential exponential corrections have to be included in its defining K\"ahler potential and flux superpotential. These terms are required to be present for consistency of the period vector following the discussion of \eqref{Pi-exp2} (see also \cite{Bastian:2021eom, Bastian:2021hpc} for a more detailed explanation).} but including corrections to the simple polynomial terms of the sl(2)-approximation, yielding algebraic equations in the moduli. Taking the flux vacua found in the sl(2)-approximation as input, we then iterate a numerical program to extremize the scalar potential in the nilpotent orbit approximation. Finally, one can include exponentially small terms in the saxions in the scalar potential. In this work we aim to stabilize all moduli at the level of the nilpotent orbit, but in principle one could use these corrections to lift flat directions in the first two steps.

\subsubsection*{Sl(2)-approximation}
As a starting point, we study the extremization conditions for flux vacua in the sl(2)-ap\-prox\-i\-ma\-tion \eqref{strict-regime} for the saxions. In section \ref{sec_type_ii_boundary} we explained how the Hodge star in this regime can be approximated by the operator $C_{\rm sl(2)}$ given by \eqref{boundarystar}.  By writing the self-duality condition \eqref{eom_001} with $C_{\rm sl(2)}$ we obtain as extremization condition
\begin{equation}\label{self_dual}
C_{\rm sl(2)}(x,y) \ G_3
= i G_3 \, .
\end{equation}
This approximation drastically simplifies how the Hodge star operator depends on the moduli. For the axions the dependence enters only through factors of $e^{- x^k N_k^-} $. It is therefore useful to absorb these factors into $G_3$ by defining flux axion polynomials
\begin{equation}\label{axion_polynomials}
\rho_{\rm sl(2)}(x) \equiv  e^{- x^k N_k^-} G_3  \, .
\end{equation}
Note that $\rho_{\rm sl(2)}$ is invariant under monodromy transformations $x^i \to x^i +1$ when we shift the fluxes according to $G_3 \to e^{N_i^-} G_3$.

For the saxions we can make the dependence explicit by decomposing into eigenspaces of the weight operators $N_i^0$ of the $\mathfrak{sl}(2, \mathbb{R})$-triples. By applying this decomposition for $\rho_{\rm sl(2)}$ we obtain
\begin{equation}\label{rho_sl2}
\qquad \rho_{\rm sl(2)}(x^i)  = \sum_{\ell \in \cE}  \rho_\ell (x^i)\, , \qquad N_i^0 \rho_\ell(x^i) = \ell_i \rho_\ell(x^i)\, ,
\end{equation}
where $\ell = (\ell_1, \ldots , \ell_n)$, and $\cE$ denotes the index set that specifies which values the $\ell_i$ can take.  Generally one has bounds $-d \leq \ell_i \leq d$, with $d$ the complex dimension of the underlying Calabi-Yau manifold, which here is $d=3$. This decomposition \eqref{rho_sl2} allows us to write out the self-duality condition \eqref{self_dual} in terms of the sl(2)-approximated Hodge star \eqref{boundarystar} componentwise as
\begin{equation}\label{self_dual_sl2}
y_1^{\ell_1} \cdots y_n^{\ell_n} \, C_\infty \, \rho_\ell (x^i) =     \rho_{-\ell} (x^i) \, , \qquad \ell \in \cE \, ,
\end{equation}
where $C_\infty$ is the boundary Hodge star operator independent of the limiting coordinates.\footnote{The operator \eqref{Cinfty} is also constructed as a product of the sl(2)-approximation. It can be understood as a Hodge star operator attached to the boundary for the Hodge structure \eqref{Fpinfty} arising in the asymptotic limit.} With \eqref{self_dual_sl2} we now have obtained a simple set of polynomial equations that one can solve straightforwardly.

\subsubsection*{Nilpotent orbit approximation}
For the next step in our approximation scheme we proceed to the nilpotent orbit. Recall that this approximation amounts to dropping some exponential terms in the $(3,0)$-form periods. In order to identify the essential corrections which remain one has to use \eqref{nilp-form} and \eqref{Fp0}. In the end, this yields a set of vector spaces $H^{p,q}_{\rm nil} $ that can be spanned by a basis valued polynomially in the coordinates $x^i, y^i$. Generally speaking, the corresponding Hodge star operator $C_{\rm nil}$ then depends on these coordinates through algebraic functions, in contrast to more general transcendental functions when exponential corrections are included. On the other hand, in comparison to the sl(2)-approximation one finds that $C_{\rm nil}$ contains series of corrections to $C_{\rm sl(2)}$ when expanding in $y_1/y_2, \ldots, y_{n-1}/y_n, 1/y_n \ll 1$. For now, let us simply state the self-duality condition \eqref{eom_001} at the level of the nilpotent orbit as
\begin{equation}\label{self_dual_nil}
C_{\rm nil} \, G_3 = i G_3\, .
\end{equation}
Next, we write out the dependence of $C_{\rm nil}$ on the coordinates in a similar manner as we did for the sl(2)-approximation. The dependence on the axions again factorizes as
\begin{equation}
C_{\rm nil} (x^i, y^i)= e^{x^k N_k} \, \hat{C}_{\rm nil} (y^i)\,  e^{-x^k N_k} \, , \qquad \hat{C}_{\rm nil}  (y^i)= C_{\rm nil}(0, y^i) \, ,
\end{equation}
where here the log-monodromy matrices $N_i$ appear instead of the lowering operators $N_i^-$ of the $\mathfrak{sl}(2,\mathbb{R})$-triples in \eqref{boundarystar}. In analogy to \eqref{axion_polynomials} we then absorb the axion dependence into flux axion polynomials as
\begin{equation}
\rho_{\rm nil} (x^i) = e^{-x^k N_k} G_3\, ,
\end{equation}
which are invariant under monodromy transformations $x^i \to x^i + 1$ if we redefine the fluxes according to $G_3 \to e^{N_i}G_3$. The dependence of $\hat{C}_{\rm nil}$ on the saxions $y^i$ is considerably more complicated than that of the sl(2)-approximated Hodge star operator $C_{\rm sl(2)}$, so we do not write this out further.\footnote{A systematic approach to incorporate these corrections in $y_1/y_2, \ldots, y_{n-1}/y_n, 1/y_n $ to the Hodge star operator was put forward in the holographic perspective of \cite{Grimm:2020cda, Grimm:2021ikg}.} The self-duality condition \eqref{self_dual_nil} can then be rewritten as
\begin{equation}\label{self_dual_nil_expanded}
\hat{C}_{\rm nil}(y^i)\,  \rho_{\rm nil}(x^i) = i \rho_{\rm nil}(x^i)\, .
\end{equation}
This extremization condition can be expanded in any basis of choice for the fluxes by writing out $\rho_{\rm nil}$ componentwise, yielding a system of algebraic equations in the moduli. Our approach in this work is to take solutions of \eqref{self_dual_sl2} as input, and then slowly vary the moduli vevs to find a solution to \eqref{self_dual_nil_expanded}.

\subsubsection*{Exponential corrections}
As a final remark, let us comment on the importance of exponential corrections in the saxions. Following the discussion of \eqref{Pi-exp2} based on \cite{Bastian:2021eom, Bastian:2021hpc}, we can divide the set of exponential corrections into terms that are essential in the period vector and those that are not. The former are included in the nilpotent orbit approximation, while the latter represent subleading corrections.
\begin{itemize}
\item \textit{Essential exponential corrections.} These corrections have to be included in the K\"ahler potential and superpotential when deriving the scalar potential from \eqref{pot_04} in the nilpotent orbit approximation. They give contributions to the polynomial part of the scalar potential, which arise due to non-trivial cancelations with the inverse K\"ahler metric. One is thus allowed to drop exponential terms in the saxions only once the scalar potential has been computed. In particular, we want to point out that these corrections therefore have to be included in the F-term  conditions $D_I W=0$, while in the self-duality condition \eqref{eom_001} exponential terms can be ignored. 
The fact that the self-duality condition turns moduli stabilization manifestly into an algebraic problem is the reason we prefer to work in this paper with those extremization conditions over the vanishing F-terms constraints.

\item \textit{Non-essential exponential corrections.} These terms in the periods can only produce exponential corrections in the scalar potential.  They are relevant when the nilpotent orbit approximation does not yet suffice to stabilize all moduli, in which case one can use them to lift the remaining flat directions by means of e.g.~racetrack potentials. In this work we stabilize all moduli via \eqref{self_dual_nil_expanded} at the level of the nilpotent orbit, so these corrections can be ignored as long as the saxions take sufficiently large values.

\end{itemize}

\subsection{Large complex-structure limit for $h^{2,1} = 2$}\label{ssec:LCS2}

Let us consider the example discussed in section~\ref{sec_aht_example}. 
We recall the triple intersection numbers of the mirror threefold  from 
\eqref{data_ex_001} as 
\begin{equation}
\kappa_{122}  = 4\, , \hspace{40pt} \kappa_{222} = 8\, ,
\end{equation}
which specify the  K\"ahler metric in the complex-structure sector 
via the prepotential \eqref{data_ex_prepot}. The sl(2)-approximated 
Hodge-star operator in the regime
 $y^1 \gg y^2 \gg 1$
has been shown in \eqref{data_ex_csl2},
where we note that the prepotential contains more detailed information 
about the moduli-space geometry 
than the sl(2)-approximation. In order to 
compare these two approaches, we define the relative difference 
\eq{
\label{diff_01}
  \Delta = \frac{|| \phi^{\rm nil} - \phi^{\rm sl(2)}||}{|| \phi^{\rm nil}||} \,,
}
where $\phi=(y^i,x^i,s,c)$ and where for simplicity we used the Euclidean norm.
To illustrate this point, let us give an explicit example:
\eq{
  \renewcommand{\arraystretch}{1.2}
  \arraycolsep10pt
  \begin{array}{||
  l@{\hspace{2pt}}c@{\hspace{2pt}}l
  ||
  l@{\hspace{2pt}}c@{\hspace{2pt}}l
  ||
  l@{\hspace{2pt}}c@{\hspace{2pt}}l
  ||}
  \hline\hline
  \multicolumn{3}{||c||}{\mbox{fluxes}}
  &
  \multicolumn{3}{c||}{\mbox{sl(2)-approximation}}  
  &
  \multicolumn{3}{c||}{\mbox{large complex structure}}  
  \\ \hline
  h^I &=& (0,0,1) 
  & y^i &=& (9.72,2.79) 
  & y^i &=& (7.85,2.79)   
  \\
  h_I &=& (-162,0,0) 
  & x^i &=& (0.00,0.00) 
  & x^i &=& (0.00,0.00)   
  \\
  f^I &=& (1,0,0) 
  &  s &=& 0.93
  &  s &=& 0.93  
  \\
  f_I &=& (0,1,37) 
  & c &=& 0.00
  & c &=& 0.00  
  \\
  \hline
  N_{\rm flux}&=&-199 & \multicolumn{6}{c||}{\Delta=0.22}
  \\
  \hline\hline
  \end{array}
}
The choice of $H_3$ and $F_3$ fluxes together with their tadpole contribution (c.f.~ equation \eqref{tad_01})
is shown in the first column, in the second column we show the location of the minimum 
in the sl(2)-approximation, and in the third column the location of the minimum 
determined using the full prepotential at large complex structure is shown.
These two loci agree reasonably-well even for the small hierarchy of three, and their relative 
difference $\Delta$ is $22\%$.

Next, we want to investigate how well the sl(2)-approximation to the Hodge-star operator 
agrees with the large complex-structure result depending on the hierarchy of the 
saxions. We implement the hierarchy through a parameter $\lambda$ as follows
\eq{\label{scaling_two_moduli}
  y^1=\lambda^2\,, \hspace{40pt}
  y^2=\lambda\,, 
}  
and for larger  $\lambda$ we expect a better agreement between the two approaches. 
We have considered three different families of fluxes characterized by 
different initial choices for $h^I$, $f^I$ and the axions $x^i$. 
The dependence of the relative difference $\Delta$ on the hierarchy parameter $\lambda$ 
is shown  in figure~\ref{fig_lcs2_01}, and
in figure~\ref{fig_lcs2_02} we shown the dependence of the (absolute value) of the tadpole
contribution $N_{\rm flux}$ on $\lambda$. 
We  make the following observations:
\begin{itemize}

\item As it is expected, for a large hierarchy $\lambda$ the sl(2)-approximation 
agrees better with the large complex-structure result. Somewhat unexpected 
is however how quickly a reasonable agreement is achieved, for instance, for a 
hierarchy of $\lambda=6$ the two approaches agree up to a difference of $15\%$. 

\item Furthermore, it is also expected that when approaching the boundary
in moduli space the tadpole contribution increases, however, the rate 
with which the tadpole increases is higher than naively expected.
For the family corresponding to the green curve in figures~\ref{plot_lcs_all}
the tadpole dependence can be fitted as
\eq{
  N_{\rm flux} = 8.27\, \lambda^{3.99}+34.69 \,,
}
which is in good agreement with the data for $\lambda\geq 2$. Thus, for 
these examples, when approaching the large complex-structure limit the tadpole
contribution increases rapidly. The scaling in $\lambda$ of the tadpole can also be understood from the weights of the fluxes under the $\mathfrak{sl}(2)$-triples. The heaviest charge is $(1,0,0,0,0,0)$ with weights $(\ell_1,\ell_2)=(1,2)$ under $N^0_1,N^0_2$. Using \eqref{boundarystar} for the asymptotic behavior of the Hodge star and plugging in \eqref{scaling_two_moduli} for the saxions we find that
\begin{equation}
N_{\rm flux} \sim (y^1)^{\ell_1} (y^2)^{\ell_2} = \lambda^4\, .
\end{equation}
\end{itemize}
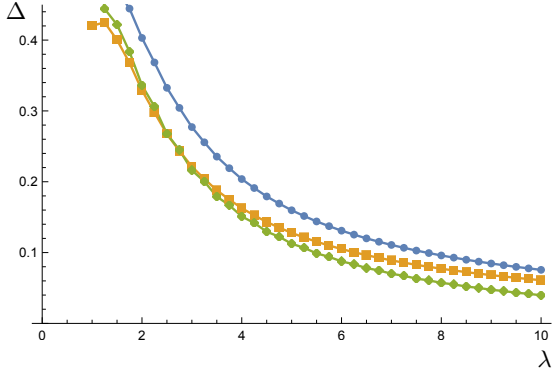
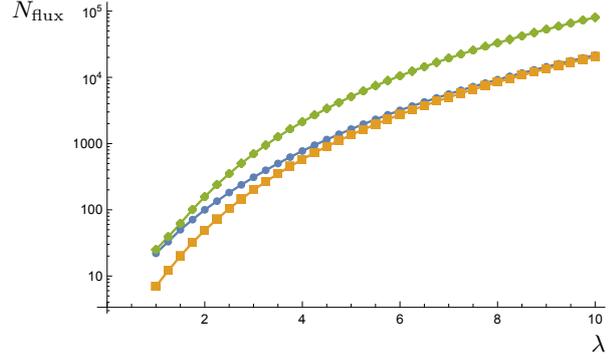
\begin{figure}
\centering
\vspace*{15pt}
\begin{subfigure}{0.45\textwidth}
\centering
\includegraphics[width=200pt]{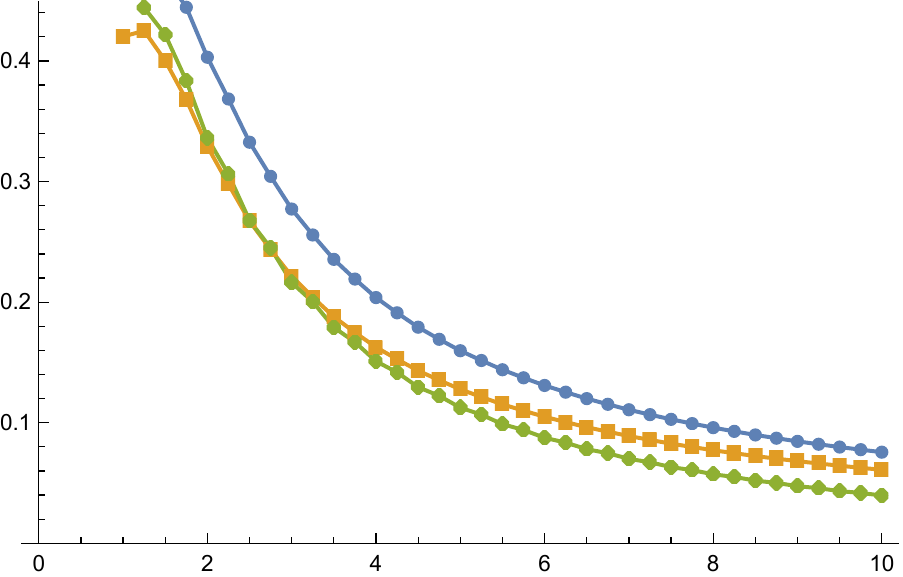}%
\begin{picture}(0,0)
\put(-6,-9){\footnotesize$\lambda$}
\put(-205,122){\footnotesize$\Delta$}
\end{picture}
\caption{Dependence of $\Delta$ on $\lambda$.\label{fig_lcs2_01}}
\end{subfigure}
\hspace{20pt}
\begin{subfigure}{0.45\textwidth}
\centering
\includegraphics[width=200pt]{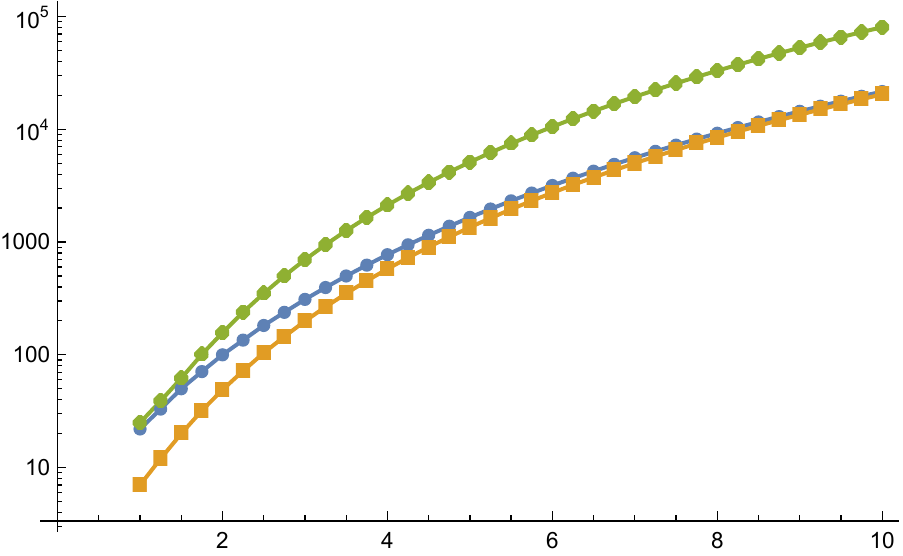}%
\begin{picture}(0,0)
\put(-6,-9){\footnotesize$\lambda$}
\put(-223,117){\footnotesize$N_{\rm flux}$}
\end{picture}
\caption{Log-dependence of $N_{\rm flux}$ on $\lambda$.\label{fig_lcs2_02}}
\end{subfigure}
\caption{Large complex-structure limit for $h^{2,1}=2$: 
dependence of the relative difference $\Delta$ and of the tadpole contribution 
$N_{\rm flux}$ on the hierarchy parameter $\lambda$. The plots show three different families, 
which all show a similar behaviour.\label{plot_lcs2_all}}
\end{figure}


\subsection{Large complex-structure limit for $h^{2,1} = 3$}\label{ssec:LCS3}

As a second  example we consider the large complex-structure limit in the case $h^{2,1}=3$, for which the 
prepotential $\mathcal F$ 
appearing in \eqref{pm} is cubic
\eq{
  \label{prepot}
  \mathcal F = -\frac{1}{3!} \frac{\kappa_{ijk} X^i X^j X^k}{X^0} \,, 
}
with $ t^i= x^i+i\op y^i=\frac{X^i}{X^0}$.
For our specific model we chose  the following non-trivial triple intersection numbers
\eq{
  \kappa_{123} =3 \,, \qquad
  \kappa_{133}=4 \,,\qquad
  \kappa_{233}=4\,,\qquad
  \kappa_{333}=5\,,
}
and the the corresponding K\"ahler metric in the complex-structure sector is well-defined inside the K\"ahler 
cone characterized by $\mbox{Im}\op t^i = y^i>0$.

As our sl(2)-approximation for this example we consider the regime $y^1 \gg y^2 \gg y^3 \gg 1$. Using the algorithm laid out in \ref{sec_tech_details}, we construct the sl(2)-approximated Hodge star \eqref{boundarystar}. Let us give the relevant building blocks here. The $\mathfrak{sl}(2)$-triples are given by
\begin{equation}
\begin{aligned}
N_1^- &=\scalebox{0.75}{$  \left(
\begin{array}{cccccccc}
 0 & 0 & 0 & 0 & 0 & 0 & 0 & 0 \\
 1 & 0 & 0 & 0 & 0 & 0 & 0 & 0 \\
 0 & 0 & 0 & 0 & 0 & 0 & 0 & 0 \\
 0 & 0 & 0 & 0 & 0 & 0 & 0 & 0 \\
 0 & 0 & 0 & 0 & 0 & -1 & 0 & 0 \\
 0 & 0 & 0 & 0 & 0 & 0 & 0 & 0 \\
 0 & 0 & 0 & -3 & 0 & 0 & 0 & 0 \\
 0 & 0 & -3 & -4 & 0 & 0 & 0 & 0 \\
\end{array}
\right)$}\, ,  & N^0_1 &= \scalebox{0.75}{$ \left(
\begin{array}{cccccccc}
 1 & 0 & 0 & 0 & 0 & 0 & 0 & 0 \\
 0 & -1 & 0 & -\frac{4}{3} & 0 & 0 & 0 & 0 \\
 0 & 0 & 1 & 0 & 0 & 0 & 0 & 0 \\
 0 & 0 & 0 & 1 & 0 & 0 & 0 & 0 \\
 0 & 0 & 0 & 0 & -1 & 0 & 0 & 0 \\
 0 & 0 & 0 & 0 & 0 & 1 & 0 & 0 \\
 0 & 0 & 0 & 0 & 0 & 0 & -1 & 0 \\
 0 & 0 & 0 & 0 & 0 & \frac{4}{3} & 0 & -1 \\
\end{array}
\right)$} \,,\\
N_2^- &= \scalebox{0.75}{$  \left(
\begin{array}{cccccccc}
 0 & 0 & 0 & 0 & 0 & 0 & 0 & 0 \\
 0 & 0 & 0 & 0 & 0 & 0 & 0 & 0 \\
 1 & 0 & 0 & 0 & 0 & 0 & 0 & 0 \\
 0 & 0 & 0 & 0 & 0 & 0 & 0 & 0 \\
 0 & 0 & 0 & 0 & 0 & 0 & -1 & 0 \\
 0 & 0 & 0 & -3 & 0 & 0 & 0 & 0 \\
 0 & 0 & 0 & 0 & 0 & 0 & 0 & 0 \\
 0 & -3 & 0 & -4 & 0 & 0 & 0 & 0 \\
\end{array}
\right)$} \,, &N^0_2 &=\scalebox{0.75}{$ \left(
\begin{array}{cccccccc}
 1 & 0 & 0 & 0 & 0 & 0 & 0 & 0 \\
 0 & 1 & 0 & 0 & 0 & 0 & 0 & 0 \\
 0 & 0 & -1 & -\frac{4}{3} & 0 & 0 & 0 & 0 \\
 0 & 0 & 0 & 1 & 0 & 0 & 0 & 0 \\
 0 & 0 & 0 & 0 & -1 & 0 & 0 & 0 \\
 0 & 0 & 0 & 0 & 0 & -1 & 0 & 0 \\
 0 & 0 & 0 & 0 & 0 & 0 & 1 & 0 \\
 0 & 0 & 0 & 0 & 0 & 0 & \frac{4}{3} & -1 \\
\end{array}
\right)$} \,,\\
N_3^- &= \scalebox{0.75}{$ \left(
\begin{array}{cccccccc}
 0 & 0 & 0 & 0 & 0 & 0 & 0 & 0 \\
 -\frac{2}{3} & 0 & 0 & 0 & 0 & 0 & 0 & 0 \\
 -\frac{2}{3} & 0 & 0 & 0 & 0 & 0 & 0 & 0 \\
 1 & 0 & 0 & 0 & 0 & 0 & 0 & 0 \\
 0 & 0 & 0 & 0 & 0 & \frac{2}{3} & \frac{2}{3} & -1 \\
 0 & 0 & -3 & -2 & 0 & 0 & 0 & 0 \\
 0 & -3 & 0 & -2 & 0 & 0 & 0 & 0 \\
 0 & -2 & -2 & -\frac{8}{3} & 0 & 0 & 0 & 0 \\
\end{array}
\right)$}\, , \hspace{30pt}&N^0_3 &= \scalebox{0.75}{$ \left(
\begin{array}{cccccccc}
 1 & 0 & 0 & 0 & 0 & 0 & 0 & 0 \\
 0 & 1 & 0 & \frac{4}{3} & 0 & 0 & 0 & 0 \\
 0 & 0 & 1 & \frac{4}{3} & 0 & 0 & 0 & 0 \\
 0 & 0 & 0 & -1 & 0 & 0 & 0 & 0 \\
 0 & 0 & 0 & 0 & -1 & 0 & 0 & 0 \\
 0 & 0 & 0 & 0 & 0 & -1 & 0 & 0 \\
 0 & 0 & 0 & 0 & 0 & 0 & -1 & 0 \\
 0 & 0 & 0 & 0 & 0 & -\frac{4}{3} & -\frac{4}{3} & 1 \\
\end{array}
\right)$} \,, \\
\end{aligned}
\end{equation}
and the boundary Hodge star by
\begin{equation}
C_\infty = \scalebox{0.85}{$  \left(
\begin{array}{cccccccc}
 0 & 0 & 0 & 0 & \frac{1}{3} & 0 & 0 & 0 \\
 0 & 0 & 0 & 0 & 0 & \frac{13}{27} & \frac{4}{27} & -\frac{2}{9} \\
 0 & 0 & 0 & 0 & 0 & \frac{4}{27} & \frac{13}{27} & -\frac{2}{9} \\
 0 & 0 & 0 & 0 & 0 & -\frac{2}{9} & -\frac{2}{9} & \frac{1}{3} \\
 -3 & 0 & 0 & 0 & 0 & 0 & 0 & 0 \\
 0 & -3 & 0 & -2 & 0 & 0 & 0 & 0 \\
 0 & 0 & -3 & -2 & 0 & 0 & 0 & 0 \\
 0 & -2 & -2 & -\frac{17}{3} & 0 & 0 & 0 & 0 \\
\end{array}
\right) $} \, .
\end{equation}
Let us now again compare the sl(2)-approximation with the full prepotential at large complex structure, 
and display the following explicit example:
\eq{
  \renewcommand{\arraystretch}{1.2}
  \arraycolsep10pt
  \begin{array}{||
  l@{\hspace{2pt}}c@{\hspace{2pt}}l
  ||
  l@{\hspace{2pt}}c@{\hspace{2pt}}l
  ||
  l@{\hspace{2pt}}c@{\hspace{2pt}}l
  ||}
  \hline\hline
  \multicolumn{3}{||c||}{\mbox{fluxes}}
  &
  \multicolumn{3}{c||}{\mbox{sl(2)-approximation}}  
  &
  \multicolumn{3}{c||}{\mbox{large complex structure}}  
  \\ \hline
  h^I &=& (0,0,0,1) 
  & y^i &=& (8.01,4.01,2.00) 
  & y^i &=& (6.39,2.69,2.18)   
  \\
  h_I &=& (-192,0,0,0) 
  & x^i &=& (0.00,0.00,0.00) 
  & x^i &=& (0.00,0.00,0.00)   
  \\
  f^I &=& (1,0,0,0) 
  &  s &=& 1.00
  &  s &=& 1.08  
  \\
  f_I &=& (0,2,8,55) 
  & c &=& 0.00
  & c &=& 0.00  
  \\
  \hline
  N&=&-247 & \multicolumn{6}{c||}{\Delta=0.13}
  \\
  \hline\hline
  \end{array}
}
Similarly as above, the choice of $H_3$ and $F_3$ fluxes 
is shown in the first column, in the second column we show the minimum 
in the sl(2)-approximation, and the third column contains the location of the minimum 
using the full prepotential.
These  loci agree reasonably-well even for the small hierarchy of two, and their relative 
difference $\Delta$ is $13\%$.

Next, we investigate how well the sl(2)-approximation to the Hodge-star operator 
agrees with the large complex-structure result depending on the hierarchy of the 
saxions. We implement the hierarchy through as follows
\eq{\label{scaling_three_moduli}
  y^1=\lambda^3\,, \hspace{40pt}
  y^2=\lambda^2\,, \hspace{40pt}
  y^3=\lambda\,, 
}  
and for larger  $\lambda$ we expect a better agreement between the two approaches. 
We have again considered three different families of fluxes characterized by 
different initial choices for $h^I$, $f^I$ and the axions $x^i$. 
The dependence of the relative difference $\Delta$ on the hierarchy parameter $\lambda$ 
is shown  in figure~\ref{fig_lcs_01}, and
in figure~\ref{fig_lcs_02} we shown the dependence of the (absolute value) of the tadpole
contribution $N_{\rm flux}$ on $\lambda$. 
We  make the following observations:
\begin{itemize}

\item For a large hierarchy $\lambda$ the sl(2)-approximation 
agrees well with the large complex-structure result. For a 
hierarchy of $\lambda=4$ the two approaches agree up to a difference of $15\%$. 

\item When approaching the boundary
in moduli space the tadpole contribution increases,
and for the family corresponding to the green curve in figures~\ref{plot_lcs_all}
the tadpole dependence can be fitted as
\eq{
  N_{\rm flux} = 3.21\, \lambda^{5.97}+351.16 \,,
}
which is in good agreement with the data for $\lambda\geq 3$. Again we can understand this scaling from the weights of the fluxes under the $\mathfrak{sl}(2)$-triples. The heaviest charge $(1,0,0,0,0,0)$ has weights $\ell_1=\ell_2=\ell_3=1$ under the grading operators $H_i$. Using \eqref{boundarystar} for the asymptotic behavior of the Hodge star and plugging in \eqref{scaling_three_moduli} for the saxions we find that
\begin{equation}
N_{\rm flux} \sim (y^1)^{\ell_1} (y^2)^{\ell_2} (y^3)^{\ell_3} = \lambda^6\, .
\end{equation}

\end{itemize}
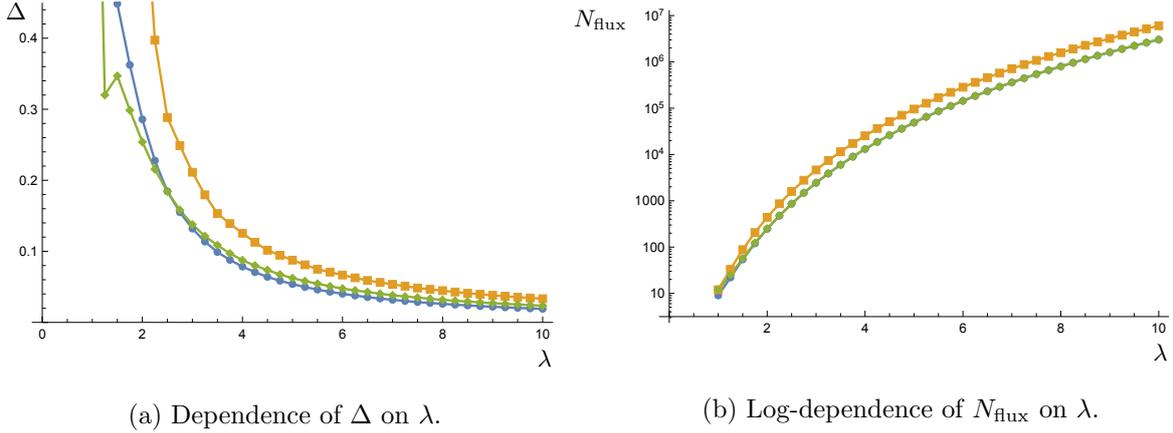
\begin{figure}
\centering
\vspace*{15pt}
\begin{subfigure}{0.45\textwidth}
\centering
\includegraphics[width=200pt]{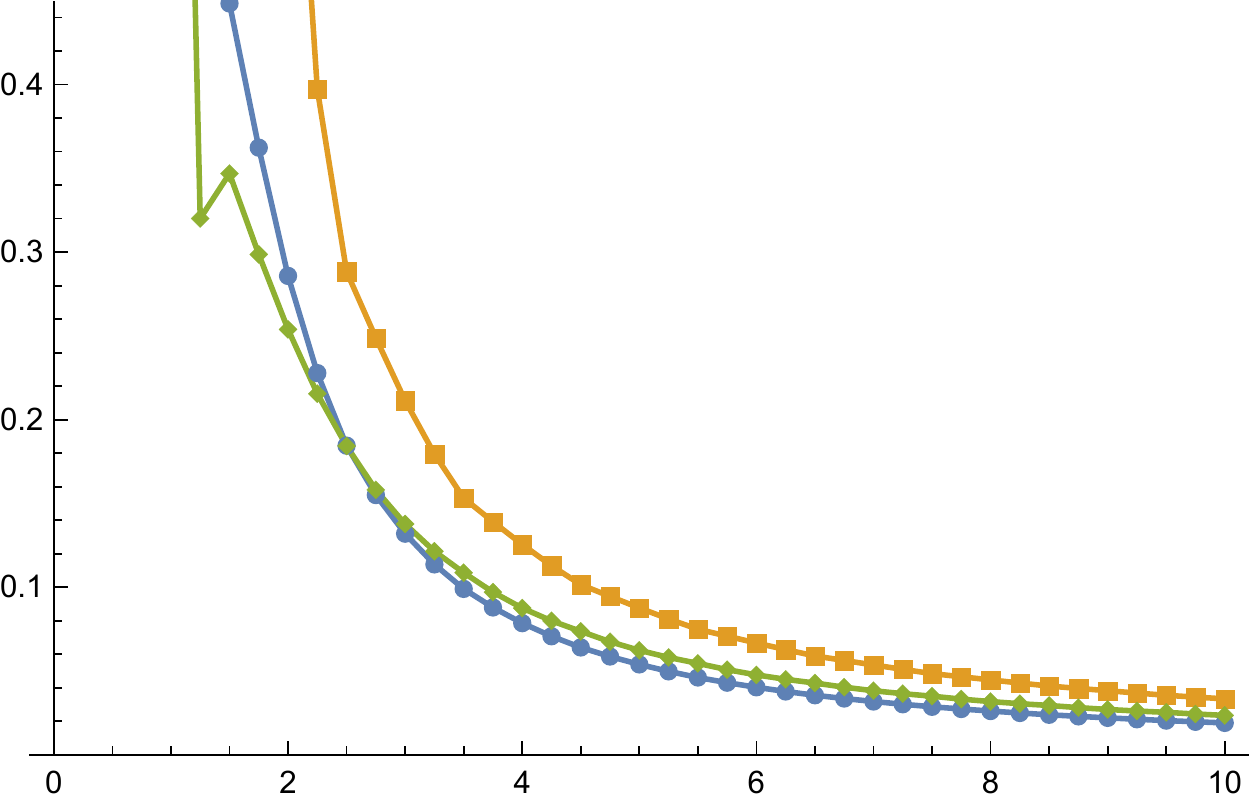}%
\begin{picture}(0,0)
\put(-6,-9){\footnotesize$\lambda$}
\put(-205,122){\footnotesize$\Delta$}
\end{picture}
\caption{Dependence of $\Delta$ on $\lambda$.\label{fig_lcs_01}}
\end{subfigure}
\hspace{20pt}
\begin{subfigure}{0.45\textwidth}
\centering
\includegraphics[width=200pt]{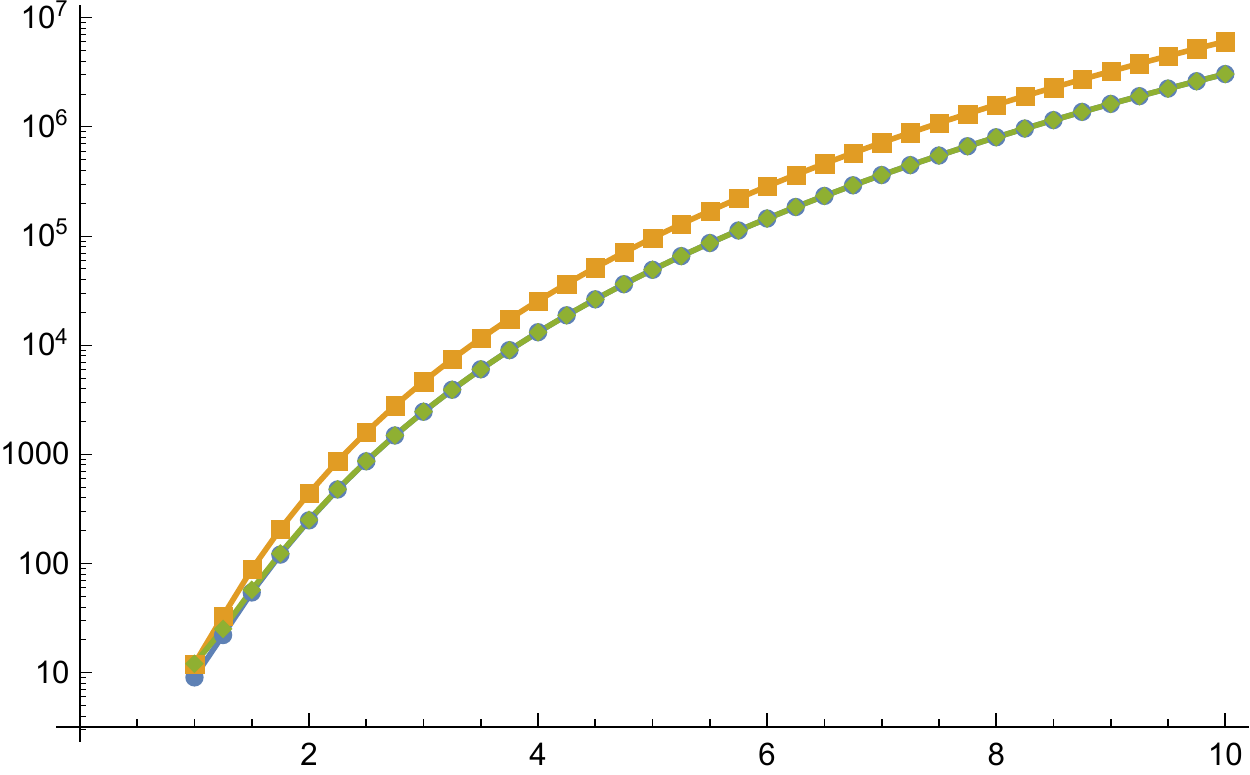}%
\begin{picture}(0,0)
\put(-6,-9){\footnotesize$\lambda$}
\put(-223,117){\footnotesize$N_{\rm flux}$}
\end{picture}
\caption{Log-dependence of $N_{\rm flux}$ on $\lambda$.\label{fig_lcs_02}}
\end{subfigure}
\caption{Large complex-structure limit: dependence of the relative difference $\Delta$ and of the tadpole contribution 
$N_{\rm flux}$ on the hierarchy parameter $\lambda$. The plots show three different families, 
which all show a similar behaviour. Note that in \ref{fig_lcs_02} the green and blue 
line are almost overlapping.\label{plot_lcs_all}}
\end{figure}


\subsection{Conifold--large complex-structure limit for $h^{2,1} = 3$}\label{ssec:coniLCS}

As a third example we consider a combined conifold and large complex-structure limit. 
We choose  $h^{2,1}=3$, and send one saxion to a conifold locus in moduli 
space and the remaining two saxions to the large complex-structure point. 
This example has been considered before in \cite{Blumenhagen:2020ire}, but here we neglect the 
instanton contributions to the prepotential. In particular, for our purposes it
is sufficient to consider the following  prepotential  
\eq{
\label{prepot_002}
  \mathcal F =  -\frac{1}{3!} \frac{\kappa_{ijk} X^i X^j X^k}{X^0}  + \frac{1}{2}\op a_{ij} \op X^i X^j + b_i\op X^i X^0 
  + c\op \bigl(X^0\bigr)^2 - \frac{1}{2\pi \op i} \bigl(X^3\bigr)^2 \log\left[ \frac{X^3}{X^0} \right] ,
}
where $i,j,k,=1,\ldots,3$. The non-trivial triple intersection numbers $\kappa_{ijk}$ and 
 constants $a_{ij}$ and $b_i$ are given by
\eq{
  \arraycolsep2pt
  \begin{array}{lcl}
  \kappa_{111} &=& 8\,, \\
  \kappa_{112} &=& 2\,, \\
  \kappa_{113} &=& 4\,, \\
  \kappa_{123} &=& 1\,, \\
  \kappa_{133} &=& 2\,, 
  \end{array}
  \hspace{40pt}
  a_{33} = \textstyle{ \frac{1}{2} + \frac{3-2\log[2\pi]}{2\pi\op i}\,,}
  \hspace{40pt}
  \begin{array}{lcl}
  b_{1} &=& \frac{23}{6}\,, \\
  b_{2} &=& 1\,, \\
  b_{3} &=& \frac{23}{12}\,.
  \end{array}
}
The constant $c$ can be set to zero for the limit we are interested in for simplicity. After computing 
the periods $\partial_I \mathcal F$ and the matrix $\partial_I\partial_J\mathcal F$, we 
set $t^i = X^i/X^0$ and $X^0=1$, and we 
perform a further field redefinition of the form 
\eq{
  t^1 \to \tilde t^1  \,,\qquad
  t^2 \to \tilde t^2+ 4\op \tilde t^3  \,,\qquad
  t^3 \to e^{2\pi \op i\op \tilde t^3} \,,
}
where the tildes will be omitted in the following. 
The domain of our coordinates is then specified by $\mbox{Im}\op t^i = y^i>0$. 

Let us now consider the sl(2)-approximation in the regime $y^3 \gg y^2 \gg y^1 \gg 1$. Using the periods following from the prepotential \eqref{prepot_002} and the algorithm of section \ref{sec_tech_details} we construct the sl(2)-approximated Hodge star \eqref{boundarystar}. The relevant building blocks are the $\mathfrak{sl}(2)$-triples
\begin{equation}
\begin{aligned}
N_1^- &= \scalebox{0.75}{$ \left(
\begin{array}{cccccccc}
 0 & 0 & 0 & 0 & 0 & 0 & 0 & 0 \\
 1 & 0 & 0 & 0 & 0 & 0 & 0 & 0 \\
 -\frac{4}{3} & 0 & 0 & 0 & 0 & 0 & 0 & 0 \\
 0 & 0 & 0 & 0 & 0 & 0 & 0 & 0 \\
 5 & 0 & 0 & 0 & 0 & -1 & \frac{4}{3} & 0 \\
 0 & -\frac{16}{3} & -2 & -\frac{8}{3} & 0 & 0 & 0 & 0 \\
 0 & -2 & 0 & -1 & 0 & 0 & 0 & 0 \\
 0 & -\frac{8}{3} & -1 & -\frac{4}{3} & 0 & 0 & 0 & 0 \\
\end{array}
\right)$} \,,  \hspace{30pt}&N^0_1 &=\scalebox{0.75}{$ \left(
\begin{array}{cccccccc}
 2 & 0 & 0 & 0 & 0 & 0 & 0 & 0 \\
 0 & 0 & 0 & 0 & 0 & 0 & 0 & 0 \\
 0 & \frac{8}{3} & 2 & \frac{4}{3} & 0 & 0 & 0 & 0 \\
 0 & 0 & 0 & 0 & 0 & 0 & 0 & 0 \\
 0 & \frac{31}{3} & 4 & \frac{31}{6} & -2 & 0 & 0 & 0 \\
 \frac{31}{3} & 0 & 0 & 0 & 0 & 0 & -\frac{8}{3} & 0 \\
 4 & 0 & 0 & 0 & 0 & 0 & -2 & 0 \\
 \frac{31}{6} & 0 & 0 & 0 & 0 & 0 & -\frac{4}{3} & 0 \\
\end{array}
\right)$}  \,, \\
N_2^- &= \scalebox{0.75}{$\left(
\begin{array}{cccccccc}
 0 & 0 & 0 & 0 & 0 & 0 & 0 & 0 \\
 0 & 0 & 0 & 0 & 0 & 0 & 0 & 0 \\
 0 & 0 & 0 & 0 & 0 & 0 & 0 & 0 \\
 0 & 0 & 0 & 0 & 0 & 0 & 0 & 0 \\
 0 & 0 & 0 & 0 & 0 & 0 & 0 & 0 \\
 0 & 0 & 0 & 0 & 0 & 0 & 0 & 0 \\
 0 & 0 & 0 & 0 & 0 & 0 & 0 & 0 \\
 0 & 0 & 0 & \frac{1}{2} & 0 & 0 & 0 & 0 \\
\end{array}
\right)$}  \,,  &N^0_2 &=\scalebox{0.75}{$\left(
\begin{array}{cccccccc}
 0 & 0 & 0 & 0 & 0 & 0 & 0 & 0 \\
 0 & 0 & 0 & -\frac{1}{2} & 0 & 0 & 0 & 0 \\
 0 & 0 & 0 & 0 & 0 & 0 & 0 & 0 \\
 0 & 0 & 0 & 1 & 0 & 0 & 0 & 0 \\
 0 & 0 & 0 & 0 & 0 & 0 & 0 & 0 \\
 0 & 0 & 0 & 0 & 0 & 0 & 0 & 0 \\
 0 & 0 & 0 & 0 & 0 & 0 & 0 & 0 \\
 0 & 0 & 0 & 1 & 0 & \frac{1}{2} & 0 & -1 \\
\end{array}
\right)$}\, , \\
N_3^- &= \scalebox{0.75}{$\left(
\begin{array}{cccccccc}
 0 & 0 & 0 & 0 & 0 & 0 & 0 & 0 \\
 0 & 0 & 0 & 0 & 0 & 0 & 0 & 0 \\
 4 & 0 & 0 & 0 & 0 & 0 & 0 & 0 \\
 0 & 0 & 0 & 0 & 0 & 0 & 0 & 0 \\
 8 & 0 & 0 & 0 & 0 & 0 & -4 & 0 \\
 0 & -8 & 0 & -4 & 0 & 0 & 0 & 0 \\
 0 & 0 & 0 & 0 & 0 & 0 & 0 & 0 \\
 0 & -4 & 0 & -2 & 0 & 0 & 0 & 0 \\
\end{array}
\right)$}  \,, &N^0_3 &= \scalebox{0.75}{$ \left(
\begin{array}{cccccccc}
 1 & 0 & 0 & 0 & 0 & 0 & 0 & 0 \\
 0 & 1 & 0 & \frac{1}{2} & 0 & 0 & 0 & 0 \\
 0 & -\frac{8}{3} & -1 & -\frac{4}{3} & 0 & 0 & 0 & 0 \\
 0 & 0 & 0 & 0 & 0 & 0 & 0 & 0 \\
 0 & 5 & 0 & \frac{5}{2} & -1 & 0 & 0 & 0 \\
 5 & 0 & 0 & 0 & 0 & -1 & \frac{8}{3} & 0 \\
 0 & 0 & 0 & 0 & 0 & 0 & 1 & 0 \\
 \frac{5}{2} & 0 & 0 & 0 & 0 & -\frac{1}{2} & \frac{4}{3} & 0 \\
\end{array}
\right)$}  \,, 
\end{aligned}
\end{equation}
and the boundary Hodge star
\begin{equation}
C_\infty =\scalebox{0.85}{$  \left(
\begin{array}{cccccccc}
 0 & -\frac{23}{24} & -\frac{1}{4} & -\frac{23}{48} & \frac{1}{4} & 0 & 0 & 0 \\
 -\frac{5}{16} & 0 & 0 & \frac{1}{2} & 0 & \frac{5}{8} & -\frac{1}{6} & -1 \\
 -\frac{43}{12} & 0 & 0 & 0 & 0 & -\frac{1}{6} & \frac{38}{9} & 0 \\
 0 & 0 & 0 & -1 & 0 & -1 & 0 & 2 \\
 -\frac{281}{32} & 0 & 0 & 0 & 0 & \frac{5}{16} & \frac{43}{12} & 0 \\
 0 & -\frac{1745}{144} & -\frac{31}{24} & -\frac{1745}{288} & \frac{23}{24} & 0 & 0 & 0 \\
 0 & -\frac{31}{24} & -\frac{1}{2} & -\frac{31}{48} & \frac{1}{4} & 0 & 0 & 0 \\
 0 & -\frac{1745}{288} & -\frac{31}{48} & -\frac{2321}{576} & \frac{23}{48} & -\frac{1}{2} & 0 & 1 \\
\end{array}
\right)$}\, .
\end{equation}
To compare moduli stabilization within the sl(2)-approximation and 
within the conifold-large complex structure limit (coni-LCS), we consider
first the following example
\eq{
  \renewcommand{\arraystretch}{1.2}
  \arraycolsep10pt
  \begin{array}{||
  l@{\hspace{2pt}}c@{\hspace{2pt}}l
  ||
  l@{\hspace{2pt}}c@{\hspace{2pt}}l
  ||
  l@{\hspace{2pt}}c@{\hspace{2pt}}l
  ||}
  \hline\hline
  \multicolumn{3}{||c||}{\mbox{fluxes}}
  &
  \multicolumn{3}{c||}{\mbox{sl(2)-approximation}}  
  &
  \multicolumn{3}{c||}{\mbox{coni-LCS}}  
  \\ \hline
  h^I &=& (1,0,0,1) 
  & y^i &=& (1.88,3.78,8.20) 
  & y^i &=& (1.88,2.61,6.92)   
  \\
  h_I &=& (3,-60,1,-30) 
  & x^i &=& (-0.14,-1.27,0.99) 
  & x^i &=& (-0.14,-1.27,1.35)   
  \\
  f^I &=& (0,1,0,0) 
  &  s &=& 1.05
  &  s &=& 1.05  
  \\
  f_I &=& (132,32,0,18) 
  & c &=& 0.07
  & c &=& 0.07  
  \\
  \hline
  N&=&-210 & \multicolumn{6}{c||}{\Delta=0.22}
  \\
  \hline\hline
  \end{array}
}
The hierarchy in the sl(2)-approximation has been chosen as a factor of two, 
the relative difference to the moduli stabilized via the coni-LCS prepotential 
\eqref{prepot_002} for this example is $22\%$.

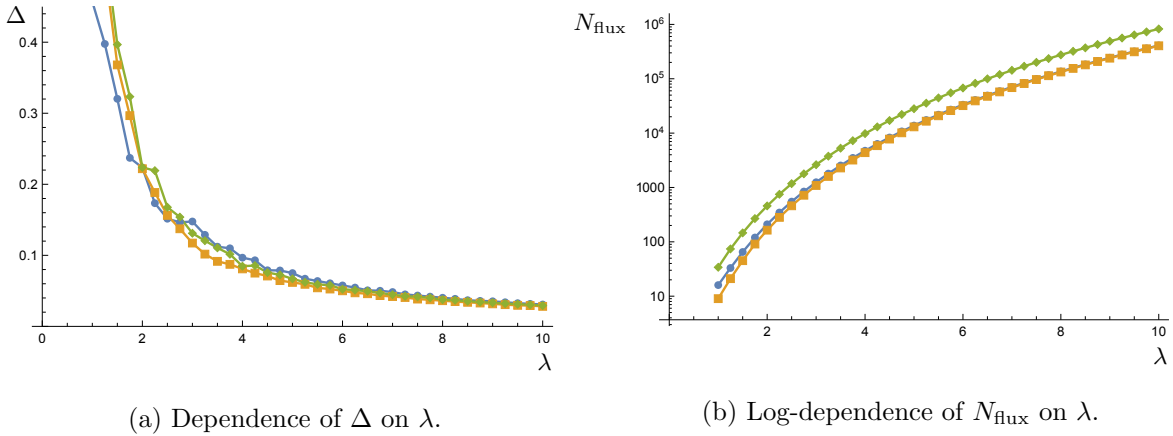
\begin{figure}
\centering
\vspace*{15pt}
\begin{subfigure}{0.45\textwidth}
\centering
\includegraphics[width=200pt]{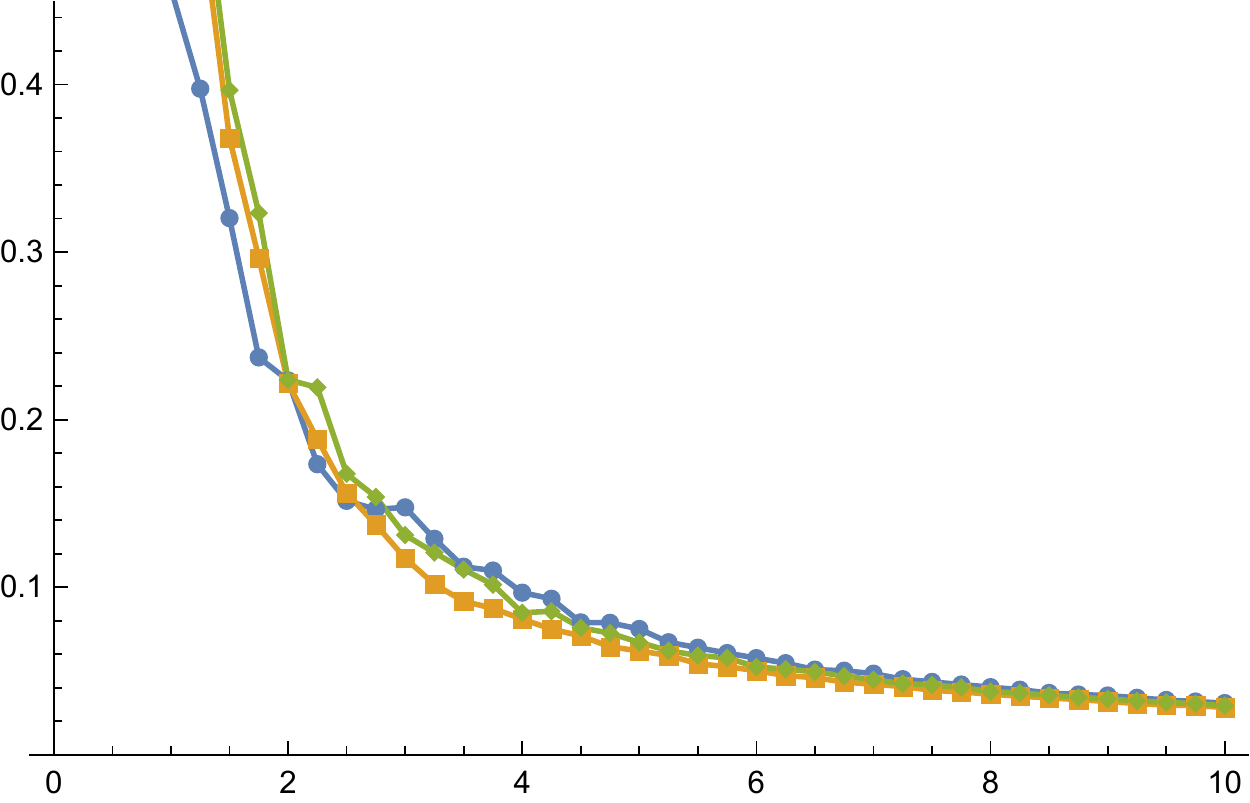}%
\begin{picture}(0,0)
\put(-6,-9){\footnotesize$\lambda$}
\put(-205,122){\footnotesize$\Delta$}
\end{picture}
\caption{Dependence of $\Delta$ on $\lambda$.\label{fig_clcs_01}}
\end{subfigure}
\hspace{20pt}
\begin{subfigure}{0.45\textwidth}
\centering
\includegraphics[width=200pt]{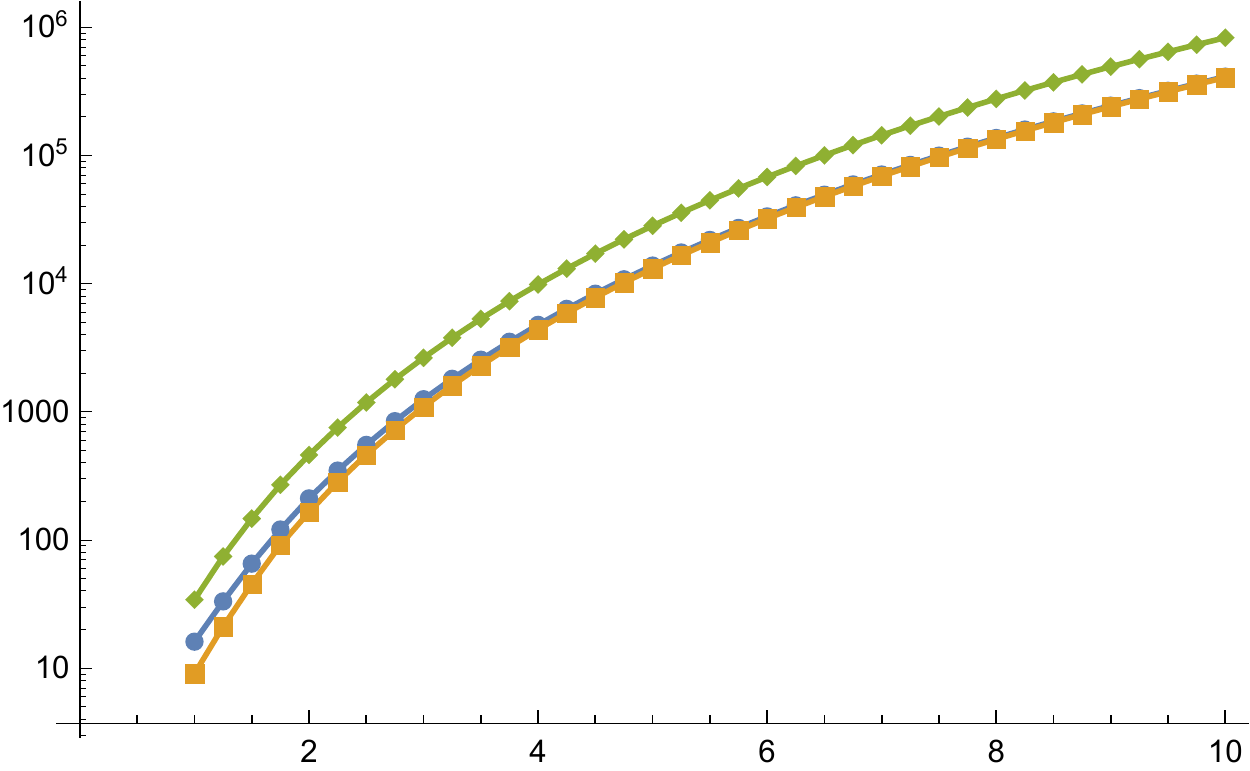}%
\begin{picture}(0,0)
\put(-6,-9){\footnotesize$\lambda$}
\put(-223,117){\footnotesize$N_{\rm flux}$}
\end{picture}
\caption{Log-dependence of $N_{\rm flux}$ on $\lambda$.\label{fig_clcs_02}}
\end{subfigure}
\caption{Conifold--large complex-structure limit: dependence of the relative difference $\Delta$ and of the tadpole contribution 
$N_{\rm flux}$ on the hierarchy parameter $\lambda$. The plots show three different families, 
which all show a similar behaviour. \label{plot_clcs_all}}
\end{figure}

Next, we want to study how well the sl(2)-approximation of the Hodge-star operator
captures moduli stabilization via the coni-LCS prepotential. We follow a strategy similar 
to the previous example and implement 
a hierarchy for the saxions through a parameter $\lambda$ as
\eq{\label{scaling_coniLCS}
  y^1=\lambda\,, \hspace{40pt}
  y^2=\lambda^2\,, \hspace{40pt}
  y^3=\lambda^3\,. 
}  
As before, we expect that  for larger  $\lambda$ the agreement between the two approaches
improves. 
We have again considered three different families of fluxes characterized by 
different initial choices for $h^I$, $f^I$ and the axions $x^i$. 
The dependence of the relative difference $\Delta$ on the hierarchy parameter $\lambda$ 
is shown  in figure~\ref{fig_clcs_01}, and
in figure~\ref{fig_clcs_02} we shown the dependence of the (absolute value) of the tadpole
contribution $N_{\rm flux}$ on $\lambda$. 
Our observations agree with the previous example, in particular:
\begin{itemize}

\item we see that even for a small hierarchy of $\lambda=4$ the relative difference between 
the stabilized moduli is only around $10\%$.

\item The tadpole contribution \eqref{tad_01} grows rapidly with the hierarchy parameter 
$\lambda$. For instance, for the orange curve in figure~\ref{fig_clcs_02} we obtain a fit 
\eq{
  N_{\rm flux} = 4.29\, \lambda^{4.97}+96.71 \,,
}
which is in good agreement with the data for $\lambda\geq3$. Therefore, also for
this example the tadpole contribution increases rapidly when approaching the boundary
in moduli space. Again this scaling is understood from the growth of the Hodge star for the heaviest charge $(1, 0, 0, 0, 0, 23/6, 1, 23/12)$. Using \eqref{boundarystar} we find that
\begin{equation}
N_{\rm flux} \sim (y^1)^{\ell_1} (y^2)^{\ell_2} (y^3)^{\ell_3} = \lambda^5\, ,
\end{equation}
where we used that its weights under the $N^0_i$ are $(\ell_1,\ell_2,\ell_3)=(2,0,1)$, and plugged in the scaling \eqref{scaling_coniLCS} of the saxions in $\lambda$.

\end{itemize}

\section{Moduli stabilization in F-theory}\label{F-theory}

We now want to apply the techniques discussed in the above sections to F-theory compactifications on elliptically fibered Calabi-Yau fourfolds. We begin with a brief review of the scalar potential induced by four-form flux, and in section \ref{ssec:lcs} we specialize our discussion to the large complex-structure regime. This  then serves as the starting point for an F-theory example 
which we study in section \ref{ssec:linear} and which has been discussed before in \cite{Marchesano:2021gyv}. For reviews on the subject of F-theory compactifications and its flux vacua we refer the reader to \cite{Denef:2008wq, Weigand:2018rez}.


\subsubsection*{Supergravity description}

We begin by considering  M-theory compactifications on Calabi-Yau fourfolds $Y_4$ with $G_4$-flux turned on. This gives rise to an effective $\mathcal{N}=2$ supergravity theory in three dimensions, where the flux induces a scalar potential for the complex-structure and K\"ahler structure moduli \cite{Haack:2001jz}. One can then lift this setting to a four-dimensional $\mathcal{N}=1$ supergravity theory by requiring $Y_4$ to be elliptically fibered and shrinking the volume of the torus fiber \cite{Denef:2008wq, Weigand:2018rez, Grimm:2010ks}. The scalar potential obtained in this way reads
\begin{equation}\label{eq:potentialFtheory}
V = \frac{1}{\cV_4^3} \Big( \int_{Y_4} G_4 \wedge \star G_4 - \int_{Y_4} G_4 \wedge G_4 \Big)\, , 
\end{equation}
where $\cV_4$ denotes the volume of the Calabi-Yau fourfold $Y_4$ and $\star$ is the corresponding Hodge-star operator. The scalar potential \eqref{eq:potentialFtheory} depends both on the complex-structure and K\"ahler moduli through the $\star$ in the first term, and the overall volume factor $\cV_4$ gives an additional dependence on the K\"ahler moduli. We also note that the flux $G_4$ is constrained by the tadpole cancellation condition as \cite{Sethi:1996es}
\begin{equation}
\frac{1}{2} \int_{Y_4} G_4 \wedge G_4 = \frac{\chi(Y_4)}{24}\, .
\end{equation}
Let us now focus on the complex-structure sector of this theory and mostly ignore the K\"ahler moduli in the following. This requires us to assume that $G_4$ is an element of the primitive cohomology $H^4_{\rm p}(Y, \mathbb{Z})$ \cite{Haack:2001jz}, which can be expressed as the condition $J \wedge G_4 = 0$ with $J$ being the K\"ahler two-form of $Y_4$. The K\"ahler and superpotential giving rise to the scalar potential \eqref{eq:potentialFtheory} can then be written as \cite{Gukov:1999ya, Haack:2001jz}
\begin{equation}\label{fourfold_potentials}
K = - \log \int_{Y_4} \Omega \wedge \bar{\Omega}\, , \hspace{50pt} W  =  \int_{Y_4} \Omega \wedge G_4\, ,
\end{equation}
where $\Omega$ is the (up to rescaling) unique $(4,0)$-form on $Y_4$. Minima of the scalar potential \eqref{eq:potentialFtheory} are  found by either solving for vanishing F-terms or imposing a self-duality condition on $G_4$ \cite{Haack:2001jz}, and these constraints read 
\begin{equation}\label{fourfold_extremization}
D_I W = \partial_I W + K_I W = 0 \, ,  \hspace{50pt} \star G_4 = G_4\, ,
\end{equation}
respectively.
One easily checks that this leads to a vanishing potential \eqref{eq:potentialFtheory} at the minimum, giving rise to a Minkowski  vacuum.


\subsection{Large complex-structure regime}\label{ssec:lcs}

To make our discussion more explicit, we now  specialize to a particular region in complex-structure moduli space. 
More concretely, we consider the large complex-structure regime  of a Calabi-Yau fourfold $Y_4$.


\subsubsection*{Moduli space geometry}

Using homological mirror symmetry the superpotential can be expressed  in terms of the central charges of B-branes wrapping even-degree cycles of the mirror-fourfold $X_4$, and we refer to the references \cite{Grimm:2009ef,CaboBizet:2014ovf, Gerhardus:2016iot, Cota:2017aal} for a more detailed discussion. Following the conventions of \cite{Marchesano:2021gyv}, we expand the periods of the holomorphic four-form $\Omega$ around the large complex-structure point as
\begin{equation}\label{fourfold_periods}
\Pi = \begin{pmatrix}
1 \\
-t^I \\
\frac{1}{2} \eta_{\mu \nu} \zeta^\nu_{IJ} t^I t^J \\
-\frac{1}{6} \cK_{IJKL} t^J t^K t^L +i K_I^{(3)}\\
\frac{1}{24} \cK_{IJKL} t^I t^J t^K t^L -i K_I^{(3)} t^I
\end{pmatrix},
\end{equation}
where $\cK_{IJKL}$ are the quadruple intersection numbers of $X_4$ and the coefficients $K_I^{(3)}$ arise from integrating the third Chern class. In formulas this reads
\begin{equation}\label{eq:quadruple}
\cK_{IJKL} = D_I \cdot D_J \cdot D_K \cdot D_L\, , \hspace{50pt} K_I^{(3)} = \frac{\zeta(3)}{8\pi^3}\  \int_{X_4} c_3(X_4) \wedge J_I \, ,
\end{equation}
where  $D_I$ denotes a basis of divisor classes for $X_4$ that generate its K\"ahler cone\footnote{We assume the K\"ahler cone to be simplicial, i.e.~the number of generators is equal to $h^{1,1}(X_4)$.} and $J_I \in H^2(X_4, \mathbb{Z})$ denote their Poincar\'e dual two-forms. Under mirror symmetry the $h^{1,1}(X_4)$ K\"ahler moduli $t^i$ of $X_4$ are identified with the $h^{3,1}(Y_4)$ complex-structure moduli of $Y_4$, and the coefficients $K_I^{(3)}$ can be interpreted as perturbative corrections to the periods, similar to the correction involving the Euler characteristic  in the prepotential \eqref{data_ex_prepot} for the threefold case. Furthermore, we introduced a tensor $\zeta^\mu_{IJ}$ to expand all intersections of divisor classes $D_I \cdot D_J$ into a basis of four-cycles $H_\mu$ as
\begin{equation}\label{zeta_tensor}
D_I \cdot D_J = \zeta^\mu_{IJ} H_\mu \, .
\end{equation}
The intersection of two four-cycles $H_\mu \cdot H_\nu$ is  denoted by
\begin{equation}
\eta_{\mu \nu} =  H_\mu \cdot H_\nu\,,
\end{equation} 
and comparing with \eqref{eq:quadruple} leads to the following relation for the intersection numbers 
\begin{equation}
\cK_{IJKL} = \zeta^\mu_{IJ} \op \eta_{\mu\nu} \op \zeta^\nu_{KL} \, .
\end{equation}
The  superpotential  \eqref{fourfold_potentials} shown above can be expressed as $W = G_4 \op\Sigma\op \Pi$, where $\Sigma$
is the matrix coupling the periods to the flux quanta which is  given by
\begin{equation}
\Sigma = \begin{pmatrix}
0 & 0 & 0 & 0 & 1 \\
0 & 0 & 0 & -\delta^I_J & 0 \\
0 & 0 & \eta^{\mu \nu} & 0 & 0 \\
0 & -\delta_I^J & 0 & 0 & 0 \\
1 & 0 & 0 & 0 & 0 \\
\end{pmatrix}.
\end{equation}
Note that we used $I,\mu$ to label the rows and $J,\nu$ for columns. The $G_4$-flux written in these conventions takes the form 
\begin{equation}
G_4 = \left( \, m\,,  \hspace{4pt} m^I \,,  \hspace{4pt} \hat{m}_\mu \,,  \hspace{4pt}  e_I \,, \hspace{4pt}  e\, \right),
\end{equation}
where the flux-quanta are (half-)interger quantized ,
and the contribution of the fluxes to the tadpole cancellation condition is  given by 
\eq{
  N_{\rm flux} = \frac{1}{2} \int_{Y_4} G_4 \wedge G_4 = \frac{1}{2} \op G_4 \Sigma \op G_4\,.
}  
Let us stress that the period vector \eqref{fourfold_periods} is not expanded in an integral basis, which means that special care has to be taken with the quantized fluxes coupling to these periods in the superpotential \eqref{fourfold_potentials}. This quantization was worked out in \cite{Gerhardus:2016iot}, and has been reformulated as a shift of the flux quanta in \cite{Marchesano:2021gyv}. For our purposes having integer-valued fluxes will not be important, so we simply refer to the listed references for more details. 
Finally, it is instructive to decompose the periods in \eqref{fourfold_periods} according to the nilpotent orbit form similar to
the threefold case discussed above. In particular, we can read off the log-monodromy matrices $N_A$ and leading term 
$a_0$ of the periods as
\begin{equation}\label{fourfold_logs}
\Pi = e^{t^I N_I} a_0 \, , \qquad N_A = \begin{pmatrix}
0 & 0& 0 & 0 & 0 \\
-\delta_{AI} & 0 & 0 & 0 & 0 \\
0 & -\eta_{\mu \rho} \zeta^\rho_{AJ} & 0 & 0 & 0 \\
0 & 0 & -\zeta^\nu_{AI} & 0 & 0 \\
0 & 0 & 0 & -\delta_{AJ} & 0 
\end{pmatrix}, \qquad a_0  = \begin{pmatrix}
1 \\
0 \\
0 \\
K^{(3)}_I \\
0
\end{pmatrix}.
\end{equation}


\subsubsection*{Periods for an elliptically fibered mirror}

Let us now specialize to an elliptically fibered mirror fourfold $X_4$, which corresponds to the example considered in the following section. For fourfolds with this fibration structure the topological data of $X_4$ is determined by the base $B_3$, and one can explicitly construct a basis for the four-cycles $H_\mu$ as discussed in \cite{Cota:2017aal}. In the following we work in the conventions of \cite{Marchesano:2021gyv}. 
We start by constructing a basis for the two- and six-cycles of $X_4$ from the base $B_3$. For the generators of the Mori cone of $B_3$ corresponding to two-cycles we write $C'^a$, while the dual four-cycles that generate the K\"ahler cone are denoted by divisors $D'_a$. The index runs as $a = 1, \ldots , h^{1,1}(B_3)$ with $h^{1,1}(B_3) = h^{1,1}(X_4)-1$. Denoting the projection of the fibration by $\pi$ and the divisor class of the section by $E$, we can then generate the Mori cone of $X_4$ by
\begin{equation}
C^0, \qquad C^a = E \cdot \pi^{-1} C'^a \, ,
\end{equation}
where $C^0$ corresponds to the class of the fiber. We generate the K\"ahler cone of $X_4$ by the dual basis of divisor classes
\begin{equation}\label{divisor_basis}
D_0 = E + \pi^* c_1(B_3)\, , \qquad D_a = \pi^* D'_a\, ,
\end{equation}
where $c_1(B_3) = c_1^a D_a'$ denotes the first Chern class of the base $B_3$. We can recover the intersection numbers of $B_3$ from those of $X_4$ as
\begin{equation}
\cK_{abc} = D'_a \cdot D'_b \cdot D'_c = D_0 \cdot D_a \cdot D_b \cdot D_c = \cK_{0abc}\, .
\end{equation}
Having constructed a basis for the two- and six-cycles, let us next consider the four-cycles $H_{\mu}$. We can generate these four-cycles via the divisors $D'_a$ and curves $C'^a$ of $B_3$ as
\begin{equation}\label{fourfold_fourcycles}
H_a = D_0 \cdot \pi^{-1}(D'_a) = D_0 \cdot D_a \, , \qquad H_{\hat{a}} = \pi^{-1} (C'^a)\, , \qquad a,\hat{a} = 1, \ldots , h^{1,1}(B_3) \,,
\end{equation}
where we split $\mu = (a, \hat{a})$. As a final task we construct the tensors $\zeta^\mu_{ij}$ and $\eta_{\mu\nu}$ appearing in the periods \eqref{fourfold_periods}. The tensor $\zeta^\mu_{IJ}$ relating intersections of two-cycles $D_I \cdot D_J$ to the four-cycle basis $H_\mu$ by \eqref{zeta_tensor} is found to be 
\begin{equation}
\zeta_{0b}^a =\delta_{ab}\, , \qquad \zeta^{\hat{a}}_{bc} = \cK_{abc} \, , \qquad \zeta^a_{00} = c_1^a\, ,
\end{equation}
and all other components either vanish or are fixed by symmetry. The intersections $\eta_{\mu \nu}$ between the four-cycles $H_\mu$ are given by
\begin{equation}
 \eta_{ab} = \cK_{abc}c_1^c \, , \qquad \eta_{a\hat{b}} = \delta_{ab} \, , \qquad \eta_{\hat{a} \hat{b}} = 0\, .
 \end{equation}


\subsection{The linear scenario -- construction} \label{ssec:linear}

With the expressions introduced above, we can now discuss a more concrete setting.
We consider the triple fibration $T^2 \to \mathbb{P}^1 \to \mathbb{P}^1 \to \mathbb{P}^1 $ as the mirror fourfold, and 
for  the toric construction  we refer to \cite{Mayr:1996sh}. In \cite{Marchesano:2021gyv} this geometry was used to realize a particular moduli stabilization scheme called the linear scenario. Here we discuss this setup from the perspective of the sl(2)-approximation,
and we comment on the scaling of the corresponding tadpole-contribution of the fluxes.


\subsubsection*{Moduli space geometry}

The relevant topological data of the above fourfold $X_4$ can be summarized by the following intersection numbers 
\begin{equation}\label{fourfold_volume}
\begin{aligned}
\cK_{ABCD} y^A y^B y^D y^D &= y_L( 32 y_0^3 + 24 y_0 y_1 y_2 +12 y_0 y_2^2 + 24 y_0^2 y_1 + 26 y_0^2 y_2 ) \\
& \ \ \ + 64 y_0^2 y_1 y_2 + 24 y_0 y_1^2 y_2  + 8 y_0 y_2^3
+ 24 y_0 y_1 y_2^2 + 64 y_0^3 y_1 \\
& \ \ \  + 24 y_0^2 y_1^2 + 36 y_0^2 y_2^2 + 72 y_0^3 y_2 + 52 y_0^4 \\
& = 4\op y_L\op \cK_L + f(y^\alpha)\, ,
\end{aligned}
\end{equation}
and by the integrated third Chern class
\begin{equation}
K_A^{(3)}  y^A =-  \frac{\zeta(3)}{8\pi^3} \Big(3136 y^0  + 480 y^L +  960 y^1 + 1080 y^2 \Big)\, ,
\end{equation}
and the base $B_3$ of the elliptic fibration has first Chern class
\begin{equation}
c_1(B_3) =  D_1 + 2D_2\, .
\end{equation}
The divisor $D_0$ corresponds to the zero section of the elliptic fibration as given in  the basis \eqref{divisor_basis}, while $D_L$ corresponds to the class of the Calabi-Yau threefold fiber over $\mathbb{P}^1$. In the last line of \eqref{fourfold_volume} we singled out the terms dependent on the saxion $y^L$  by writing 
\begin{equation}\label{fourfold_functions}
 \cK_L = \sum_{\alpha,\beta,\gamma} \cK_{Labc} y^\alpha y^\beta y^\gamma\, , \qquad f(y^a) = \sum_{\alpha,\beta,\gamma,\delta} \cK_{\alpha\beta\gamma\delta}   y^\alpha y^\beta y^\gamma y^\delta \, ,
\end{equation}
where $f(y^a)$ contains the remaining terms only depending on $y^\alpha=y^0,y^1,y^2$. Let us also note that we sort our complex-structure moduli in the order $(t^0, t^L, t^1 , t^2)$ for the construction of the log-monodromy matrices as described by \eqref{fourfold_logs}, and split the indices as $A=(L, \alpha)$ with $\alpha=0,1,2$. Furthermore, we use the four-cycle basis as described by \eqref{fourfold_fourcycles}, where we let the indices run as $a, \hat{a} = L, 1, 2$.\footnote{Note that this four-cycle basis differs from the one used in \cite{Marchesano:2021gyv}, where another basis was taken using the fibration structure of the threefold fiber rather than the elliptic fiber.} 


\subsubsection*{Saxion hierarchies and the sl(2)-approximation}

Let us now introduce particular scalings for the saxions $y_A$. For the linear scenario of \cite{Marchesano:2021gyv} there is a hierarchy $y_L \gg y_\alpha $ which we realize by
\begin{equation}\label{scaling_linear_scenario}
y_L = \lambda^3\, , \hspace{50pt} y_\alpha = \lambda \, \hat{y}_\alpha\, .
\end{equation}
In order to understand this regime from the perspective of the sl(2)-approximation we need to consider a further hierarchy $y_L \gg y_1 \gg y_2 \gg y_0$, which we implement via an additional scaling parameter $\rho$ as
\begin{equation}\label{scaling_linear_sl2}
y_L = \lambda^3 \rho \, , \hspace{50pt} 
y_0 = \lambda  \, , \hspace{50pt}
y_1 = \lambda \rho^2 \, , \hspace{50pt}
y_2 = \lambda \rho\, .
\end{equation}
Setting for instance $\rho = \lambda$ we find scalings $(y_L, y_0,y_1,y_2) = (\lambda^4 , \lambda, \lambda^3, \lambda^2)$ as needed for the hierarchy of the sl(2)-approximation, while for $\rho = 1$ we reduce to a scaling of the form of \eqref{scaling_linear_scenario}. Typically we will keep both the scaling in $\rho$ and $\lambda$ explicit rather than making a choice for $\rho$. We can then use the scaling in $\lambda$ found in the sl(2)-approximation to make statements about the linear scenario. In order to keep the discussion here concise we included the data specifying the sl(2)-approximated Hodge star in appendix \ref{app:linearscenario}. It is, however, instructive to give the eigenspaces of the weight operators $N^{0}_i$, which have been summarized in table \ref{linear_scenario_eigenspaces}.

\begin{table}[t]
\centering
\renewcommand*{\arraystretch}{1.8}
\begin{tabular}{| c | c |}
\hline weights  & charges \\ \hline  \hline
$(-1,-3)$ & $(0, 0, 0, 0, 0, 0, 0, 0, 0, 0, 0, 0, 0, 0, 0, 1) $ \\ \hline
$(-1,-1)$ & \begin{minipage}{0.43\textwidth}\centering\vspace*{0.2cm}
$(0, 0, 0, 0, 0, 0, 0, 0, 0, 0, 0, 1, 0, 0, 0, 0)$\\
$ (0, 0, 0, 0, 0, 0, 0, 0, 0, 0, 0, 0, 0, 1, 0, 0)$ \\
$(0, 0, 0, 0, 0, 0, 0, 0, 0, 0, 0, 0, 0, 0, 1, 0 ) $ \vspace*{0.2cm}
\end{minipage} \\ \hline
$(-1,1)$ & \begin{minipage}{0.43\textwidth}\centering\vspace*{0.2cm}
$(0, 0, 0, 0, 0, 0, 1, 0, 0, 0, 0, 0, 0, 0, 0, 0)$ \\
$(0, 0, 0, 0, 0, 0, 0, 1, 0, 0, 0, 0, 0, 0, 0, 0)$ \\
$(0, 0, 0, 0, 0, 0, 0, 0, 1, 0, 0, 0, 0, 0, 0, 0)$ \vspace*{0.2cm}
\end{minipage} \\ \hline
$(-1,3)$ & $( 0, 0, 1, 0, 0, 0, 0, 0, 0, 0, 0, 0, 0, 0, 0, 0)$ \\ \hline
$(1,-3)$ & $(0, 0, 0, 0, 0, 0, 0, 0, 0, 0, 0, 1, 1, 1, \tfrac{1}{2}, 0)$ \\ \hline
$(1,-1)$ & \begin{minipage}{0.43\textwidth}\centering\vspace*{0.2cm}
$(0, 0, 0, 0, 0, 1, 0, 0, -\tfrac{1}{2}, 1, -1, 0, 0, 0, 0, 0)$\\
$(0, 0, 0, 0, 0, 0, 1, 0, 1, -\tfrac{3}{2}, 1, 0, 0, 0, 0, 0)$\\
$(0, 0, 0, 0, 0, 0, 0, 1, -1, 1, 0, 0, 0, 0, 0, 0)$ \vspace*{0.2cm}
\end{minipage} \\ \hline
$(1,1)$ & \begin{minipage}{0.43\textwidth}\centering\vspace*{0.2cm}
$(0, 1, 0, 0, -2, 0, 0, 0, 0, 0, 0, 0, 0, 0, 0, 0)$\\
$(0, 0, 1, 0, -2, 0, 0, 0, 0, 0, 0, 0, 0, 0, 0, 0)$ \\
$(0, 0, 0, 1, -2, 0, 0, 0, 0, 0, 0, 0, 0, 0, 0, 0)$\vspace*{0.2cm}
\end{minipage} \\ \hline
 $(1,3)$ & $(1, 0, 0, 0, 0, 0, 0, 0, 0, 0, 0, 0, 0, 0, 0, 0)$ \\ \hline
\end{tabular}
\caption{\label{linear_scenario_eigenspaces} Eigenspaces of the weight operators given in \eqref{linear_operators_weights}. In order to connect to the scaling of the linear scenario \eqref{scaling_linear_scenario} we grouped the weights $\ell_\alpha$ under $N^0_\alpha$ together as $(\ell_L, \ell_0+\ell_1+\ell_2)$. }
\end{table}


\subsubsection*{Parametric scaling of the tadpole}

Regarding this model, the aim in  \cite{Marchesano:2021gyv} was to make a flux-choice which keeps the tadpole contribution 
$N_{\rm flux}$ finite
asymptotically. The parametric growth of the tadpole due to a flux is specified by its weights under the $\mathfrak{sl}(2, \mathbb{R})$-triples and we recall that for a charge in the eigenspace  $q \in V_{\ell_1 \ell_2 \ell_3 \ell_4}$ its Hodge norm \eqref{boundarystar} grows as
\begin{equation}
\| q \|^2 \sim y_L^{\ell_L}\op y_0^{\ell_0} \op y_1^{\ell_1}\op y_2^{\ell_2} \, .
\end{equation}
Using the scalings of the saxions shown in \eqref{scaling_linear_sl2}, we find for large $\lambda$ that $3\ell_L+\ell_0+\ell_1+\ell_2 \leq 0$ must hold for our flux quanta. From \eqref{linear_operators_weights} we can compute that $\ell_A = \pm 1$ for all weights, so the fluxes with weights $\ell_L=1$ and $\ell_0+\ell_1+\ell_2 =-1,1,3$ should be turned off. Inspecting table \ref{linear_scenario_eigenspaces} we then find that this corresponds to the flux choice 
\begin{equation}\label{linear_fluxes}
m=0\,,  \qquad m_i = (0,m_L,0,0), \qquad \hat{m}^\mu = (0, \hat{m}^1, \hat{m}^2, \hat{m}^{\hat{L}},0,0)\, , 
\end{equation}
and all other fluxes unconstrained. This matches precisely with the fluxes that were turned off in \cite{Marchesano:2021gyv},  motivated from the sl(2)-approximation.
(Note that we chose a different four-cycle basis as compared to \cite{Marchesano:2021gyv} so the choice of $\hat{m}^\mu $ 
takes a slightly different form.)


\subsubsection*{Flat directions}

Given the above choice of fluxes, we can now investigate the stabilization of the saxions within various approximation schemes.  
Via the self-duality condition of the $G_4$-flux, the axions are fixed as
as \cite{Marchesano:2021gyv}
\begin{equation}\label{linear_axions}
\begin{aligned}
x^\alpha &= - \frac{\hat{m}^\alpha}{m^L} \, ,  \qquad &x^L &=  - \frac{e}{e_L} - \frac{1}{3 e_L (m^L)^2 } \Big( \cK_{Labc} \hat{m}^a \hat{m}^b \hat{m}^c - 3 e_a \hat{m}^a m^L \Big)\, .
\end{aligned}
\end{equation}
However, for simplicity we  set these axions to zero in the following, that is $x^L=x^0=x^1=x^2=0$, 
which means we impose $e=\hat{m}^1=\hat{m}^2= \hat{m}^{\hat{L}}=0$ on the flux quanta in addition to \eqref{linear_fluxes}.
\begin{itemize}

\item \textit{sl(2)-approximation}: For the specified choice of fluxes \eqref{linear_fluxes} with vanishing axions we find $(1, -3)$ and $(-1, 3)$ as allowed weights for $(\ell_L, \ell_0+\ell_1+\ell_2)$. 
Using the sl(2)-approximated Hodge star \eqref{boundarystar} in the self-duality condition \eqref{fourfold_extremization}, this 
leads to
\begin{equation}
e_L = - \frac{y_0 \op y_1 \op y_2}{2\op y_L} m_L \, , \qquad e_\alpha = 2 e_L\, .
\end{equation}
Notice that the saxions only appear as the combination $y_0 y_1 y_2 / y_L$, so there are three saxionic directions left unconstrained. In particular, when plugging in the scaling of the linear scenario \eqref{scaling_linear_scenario} we see that $\lambda$ drops out completely. We can stabilize these seemingly flat directions by including corrections to the sl(2)-approximation.

\item \textit{Nilpotent orbit approximation}: We next include the full set of polynomial terms in the periods, in  particular, we investigate the difference between including the corrections $K^{(3)}_i$ or not. The relevant extremization conditions for the saxions now read \cite{Marchesano:2021gyv}
\begin{equation}\label{linear_solve}
\begin{aligned}
e_L &= -\frac{\cK}{6} g_{LL} m^L = \Big( -\frac{ \cK_L}{t_L}+ \frac{ f }{24 t_L^2 }  -\frac{ 
K_L^{(3)} \zeta(3)}{64 \pi^3 t_L}\Big) m^L+ \cO\Big(\frac{1}{\lambda^4}\Big) \, , \\
e_\alpha &= \frac{m^L}{6}  \cK_L \partial_\alpha \Big( \frac{f}{4 \cK_L} \Big) -  \frac{ 9 K^{(3)}_L  \cK_{L\alpha} m^L }{8  \cK_L }+ \cO\Big(\frac{1}{\lambda^2}\Big) \, ,
\end{aligned}
\end{equation}
where $\cK_{L \alpha} = \sum_{\beta, \gamma} \cK_{L \alpha \beta \gamma} y^\beta y^\gamma$. We also specified the order in $\lambda$ at which corrections to these equations enter under the scaling \eqref{scaling_linear_scenario}. Note that only $K^{(3)}_L$ appears here, so it dominates over the other corrections $K^{(3)}_\alpha$ in the expansion in $\lambda$. 
Since it is rather difficult to solve \eqref{linear_solve} explicitly for the saxions, we will instead consider some specific flux quanta $e_L, e_\alpha$ to exemplify that the saxions can indeed be stabilized. 
As flux quanta we take
\begin{equation}
m_L =  623\, , \qquad e_A = (-4698, -3072, -2760, -1566 )\, ,
\end{equation}
for which the saxions \eqref{scaling_linear_scenario} are  stabilized at
\begin{equation}
\lambda = 6 \, , \qquad \hat{y}^\alpha = 1\, .
\end{equation}

\item \textit{Nilpotent orbit approximation without $K^{(3)}_L$}: We also compute the eigenvalues of the mass matrix $K^{ab} \partial_a \partial_b V$ while formally setting $K^{(3)}_L = 0$ and  $K^{(3)}_\alpha = 0$. This yields the canonically-normalized masses
\begin{equation}\label{masses_without}
m^2  = \bigl (
3.4 \cdot 10^{18}, \ 1.7 \cdot 10^{13}, \ 1.7 \cdot 10^{13}, \ 1.7 \cdot 10^{13}, \ 1.6 \cdot 10^{13}, \ 3.8 \cdot 10^{11}, \ 2.7 \cdot 10^{11}, \ 0) \, ,
\end{equation}
and hence for this setting there is one flat direction.

\item \textit{Nilpotent orbit approximation with $K^{(3)}_L$}: We then include the correction $K^{(3)}_L $ but still set $K^{(3)}_\alpha = 0$ and find that the flat direction now acquires a mass
\begin{equation}\label{masses_with}
m^2 = \bigl(3.4 \cdot 10^{18}, \ 1.7 \cdot 10^{13}, \ 1.7 \cdot 10^{13}, \ 1.7 \cdot 10^{13}, \ 1.6 \cdot 10^{13}, \ 3.6 \cdot 10^{11}, \ 2.5 \cdot 10^{11}, \ 2.3 \cdot 10^7 \bigr) ,
\end{equation}
while the other masses are only affected slightly. The now non-zero mass is significantly smaller as compared to the other moduli, since its mass scale is set by the correction $K_L^{(3)}$ rather than the leading polynomial terms.

\end{itemize}


\subsubsection*{Type IIB2 case}

In order to understand the above observations better, let us reduce the F-theory setting to Type IIB string theory. This brings us to the IIB2 moduli stabilization scheme of \cite{Marchesano:2021gyv}, which was originally considered in \cite{Palti:2008mg}. In order to  match 
the two configurations we set $f=0$ and $K^{(3)}_\alpha = 0$, and identify $K_{L\alpha \beta \gamma}$ with the intersection numbers of the Calabi-Yau threefold and $K^{(3)}_L$ as the Euler characteristic correction. Furthermore, the dilaton $t^L$ is interpreted as the axio-dilaton $\tau$. We will keep the notation of the F-theory setting, and simply drop the terms that are absent in the Type IIB setup, for which the self-duality condition \eqref{linear_solve}  reduces to
\begin{equation}\label{IIB_solve}
\begin{aligned}
e_L & = \Big( -\frac{1}{6} \frac{\cK_L}{y_L}  -\frac{ 
K_L^{(3)} \zeta(3)}{64 \pi^3 y_L}\Big) m^L\, , \qquad &e_\alpha &=  -  \frac{ 9 K^{(3)}_L  \cK_{L\alpha} m^L }{8  \cK_L }\, .
\end{aligned}
\end{equation}
When we drop the correction $K_L^{(3)}$ and plug in the saxion scaling \eqref{scaling_linear_scenario} we see that $\lambda$ cancels out of these equations. However, subsequently including the correction fixes $\lambda$ as
\begin{equation}\label{lambda_vev}
\lambda^3 = \frac{64 \pi ^3}{K_L^{(3)} \zeta(3)} \Big( - \frac{1}{6} \cK_{L\alpha \beta \gamma}\hat{y}^\alpha \hat{y}^\beta \hat{y}^\gamma - \frac{e_L}{m^L}  \Big)\, .
\end{equation}
Thus in the IIB2 scenario the modulus $\lambda$ parametrizes precisely the flat direction we encountered before, which is stabilized by including perturbative corrections. In the general F-theory setup this symmetry was more difficult to spot in \eqref{scaling_linear_scenario} due to the presence of the function $f(\hat{y}^\alpha)$ which alters this parametrization. Nevertheless, by studying the effect of $K_L^{(3)}$ on the masses in \eqref{masses_without} and \eqref{masses_with} we were able to observe this feature.


\subsection{The linear scenario -- discussion} 

We now want to discuss the linear scenario in view of relations to swampland conjectures, 
to the finiteness theorem and to the tadpole conjecture.


\subsubsection*{Relations to swampland conjectures and finiteness theorem}
The above hierarchy in the masses of the linear scenario is quite interesting in light of some swampland conjectures. Say we place a cutoff scale between the moduli masses stabilized in \eqref{masses_without} and the field direction stabilized by the $K^{(3)}_L$ effect in \eqref{masses_with}. From the mass hierarchy it then follows that we can integrate out all axion fields (and three saxion fields), while one runaway direction remains. In the IIB2 case this direction is parametrized via the saxionic modulus $\lambda$ by \eqref{scaling_linear_scenario}, while in the more general F-theory setup this parametrization is slightly more involved. Either way, by restricting to this valley of the scalar potential we obtain a pseudomoduli space containing a single saxion field and no axions. In turn, we can send $\lambda \to \infty$, resulting in an infinite distance limit. Generically this path does not lift to a geodesic in the original higher-dimensional moduli space, so this provides us with an interesting class of examples for the Convex Hull Distance Conjecture of \cite{Calderon-Infante:2020dhm}. On the other hand, this infinite distance limit is rather intriguing from the perspective of the Distant Axionic String Conjecture \cite{Lanza:2020qmt, Lanza:2021qsu}. It predicts the emergence of asymptotically tensionless strings coupled to axion fields, but in this effective field theory all axions have already been integrated out due to the cutoff. It would be interesting to return to this puzzle in the near future, and see if a more detailed study of axion strings in the background of such scalar potentials elucidates this matter.

Let us also comment on the linear scenario in the context of the general finiteness theorems for flux vacua satisfying 
the self-duality condition \cite{BakkerGrimmSchnellTsimerman,Grimm:2020cda}. A first observation is 
that the tadpole $N_{\rm flux} = - e_L m^L$ seems to be independent of the flux $e_\alpha$. However, it is not hard to 
see that $e_\alpha$ cannot be chosen freely and 
there are only finitely many choices allowed in this setting. The second equation in \eqref{IIB_solve} implies that if we 
want to increase $e_\alpha$, we either have to increase $m^L$ or decrease the moduli. In the former case, we immediately 
see that the tadpole grows, while in the latter case we reach a point where our use of the asymptotic results are no longer applicable. 
Furthermore, we check that the possible vevs for the saxions are bounded from above, which is another necessary condition for the finiteness of solutions. 
Inserting the scaling \eqref{scaling_linear_scenario} into the first equation of \eqref{IIB_solve} we see that the $\hat{y}^\alpha$ are bounded from above. Namely, otherwise $e_L$ grows as we increase the volume $\cK_{L\alpha \beta \gamma} \hat{y}^\alpha \hat{y}^\beta \hat{y}^\gamma $, resulting in a diverging tadpole $N_{\rm flux} = - e_L m^L$. Using similar reasoning we find that $\lambda$ is bounded through the condition \eqref{lambda_vev}. 

Finally, let us also have a closer look at the flux vacuum loci themselves. In particular, consider the case $K_L^{(3)}=0$. 
For such geometries the modulus $\lambda$ is unfixed due to the absence of corrections in the nilpotent orbit approximation. 
This means the resulting flux vacua are not point-like, but rather infinite lines stretching to the boundary of moduli space.\footnote{It could be the case that the 
inclusion of exponential corrections stabilizes this flat direction. There are, however, examples where such corrections are absent \cite{BakkerGrimmSchnellTsimerman}. 
In that regard it is interesting to point out the work of \cite{Palti:2020qlc}, where the absence of instantons in special cases was related to higher-supersymmetric settings.} 
It is interesting to point out that vacuum loci of this type do not need to be algebraic. This implies that if one aims to describe the structure of all flux vacua one has to 
leave the world of algebraic geometry. Remarkably, a delicate and powerful extension of algebraic geometry that can be use to describe flux vacua is provided by using 
tame topology and o-minimality \cite{BakkerGrimmSchnellTsimerman}. The resulting tame form of geometry manifest the notion of finiteness and removes many pathologies that 
are allowed in geometric settings based on ordinary topology.


\subsubsection*{Tadpole conjecture I -- without axions}

We now focus on the tadpole contribution of the fluxes and its scaling with the number of 
moduli \cite{Bena:2020xrh,Bena:2021wyr}. For simplicity we restrict our attention to the IIB2 case, 
and our discussion follows closely the line of arguments first presented in \cite{Plauschinn:2021hkp}.
In particular, we will be using statistical data for Calabi-Yau threefolds obtained  in \cite{Demirtas:2018akl}
to show that the linear scenario is, under certain genericity assumptions, compatible with the tadpole conjecture.
Our main argument can then be summarized  as follows:
\begin{itemize}

\item Since we restrict our discussion to the large complex-structure limit and ignore instanton corrections, 
we need to ensure that the latter are sufficiently suppressed. To implement this constraint 
on the mirror-dual side, we require that all holomorphic curves $\cC$ have a volume greater than a 
constant $c$. Similarly we require that all divisors $D$ and the mirror threefold $\tilde Y_3$ itself to have volumes greater than $c$,
and we use these conditions to define the stretched K\"ahler 
cone \cite{Demirtas:2018akl}, i.e.~we consider those $J$ that satisfy 
\beq \label{Kahler-cone}
  \int_C J > c\ , \qquad  \frac{1}{2}\int_D J^2 > c\ , \qquad  \frac{1}{6}\int_{\tilde Y_3} J^3 > c\ . 
\eeq
Note that (after applying mirror-symmetry) the asymptotic regime introduced at the beginning of section \ref{sec_tech_details} lies in the stretched K\"ahler cone with $c=1$. 

\item In \cite{Demirtas:2018akl} the authors analyzed the Kreuzer-Skarke list \cite{Kreuzer:2000xy} 
and determined properties associated with the stretched K\"ahler cone. Via mirror symmetry, these results on the K\"ahler-moduli 
side then translate to the large complex-structure limit we are interested in.

\item  We note that the K\"ahler form $J$ can be expanded in any basis of the second cohomology.  
For example, we can use the generators $\omega_\alpha$ of a simplicial subcone 
as discussed in more detail in \cite{Grimm:2019bey} and 
write $J=y^\alpha\omega_\alpha$. Setting $c=1$ in \eqref{Kahler-cone} then implies $y^\alpha>1$ and we obtain the 
above parametrization of the asymptotic regime. However, in \cite{Demirtas:2018akl} the 
analysis was carried out in a basis naturally arising in the explicit construction of Calabi-Yau threefold examples. Denoting this
basis by $[D_\alpha]$ we expand $J= v^\alpha [D_\alpha]$. We note that the $v^\alpha$ now obey certain non-trivial inequalities 
for $J$ to be an element of the stretched K\"ahler cone \eqref{Kahler-cone}. Clearly, also the intersection numbers $\kappa_{\alpha \beta \gamma} = D_\alpha \cdot D_\beta \cdot D_\gamma$ have to be 
evaluated in this different basis and the statistical statements of \cite{Demirtas:2018akl} concerning their behaviour is generically true in this `special' basis.

\item Of interest to us are the number of non-zero entries in the triple intersection numbers $\kappa_{\alpha\beta\gamma}$ 
and a measure for how stretched the K\"ahler cone is at large $h^{1,1}$. To associate a number to the latter, we introduce 
the distance $|v| = \sqrt{\sum_a(v^a)^2}$ from the origin of the K\"ahler cone to a K\"ahler form $J= v^a [D_a]$, and 
denote by $d_{\rm min}$ the minimal distance between the origin of the K\"ahler cone and the tip of the stretched K\"ahler cone. 
In  \cite{Demirtas:2018akl}  these have been determined for the Kreuzer-Skarke database and their dependence 
on $h^{1,1}$ have been determined. Under mirror symmetry this dependence translates into\footnote{
When using the basis $\omega_{\alpha}$ to expand the K\"ahler form then the dependence of the quantities in \eqref{rel_9379274}
on $h^{2,1}$ will be different. Since this information is not readily available, we use the basis employed in 
\cite{Demirtas:2018akl}.}
\eq{
  \label{rel_9379274}
   \#(\kappa_{\alpha\beta\gamma}\neq 0) \gtrsim 6.5\op h^{2,1} + 25\,,
   \hspace{40pt}
    d_{\rm min} \simeq 10^{-1.4} \op (h^{2,1})^{2.5}  \,.
}
Furthermore,  for large $h^{2,1}$ the size of the entries of $\kappa_{\alpha\beta\gamma}$ is of order $\mathcal O(10)$ 
which can be inferred from figure 2 of \cite{Demirtas:2018akl}.  For generic situations and in the large $h^{2,1}$ limit 
we then obtain the following rough scaling behavior (see \cite{Plauschinn:2021hkp} for details on the derivation) 
\eq{
  \label{fits_001}
  \cK_L \sim  (h^{2,1})^{-1/2} \op | v |^3 \,,
  \hspace{50pt}
  \cK_{L\alpha} \sim (h^{2,1})^{-1} \op | v  |^2
  \hspace{50pt}
  \cK_L^{(3)} \sim h^{2,1} \,.
}
We included here the quantity $K_L^{(3)}$ which contains the Euler characteristic.
Note that this scaling behaviour matches roughly the statistical analysis presented in \cite{Demirtas:2018akl} for the minimal 
total volume and divisor volumes in the stretched K\"ahler cone when replacing $|v|$ with $d_{\rm min}$. 
  
\end{itemize}
Let us now use \eqref{fits_001} and determine the moduli-dependent expression 
in the second relation in equation \eqref{IIB_solve}. 
For generic situations and for large $h^{2,1}$ we can make the following estimate 
\eq{
  \label{rel_8394}
   \frac{ 9K^{(3)}_L  \cK_{L\alpha}}{8\cK_L } \sim (h^{2,1})^{1/2} \op | v |^{-1} \lesssim (h^{2,1})^{-2} \,,
}
where in the second step we  applied the bound 
 $| v | \geq d_{\rm min}$ and where we ignored numerical factors. 
This scaling can now be used in the self-duality condition  \eqref{IIB_solve}, where the second condition 
together with \eqref{rel_8394} translates to 
\eq{
e_\alpha \lesssim (h^{2,1})^{-2} \op m^L\,.
}
However, 
since the fluxes $e_{\alpha}$ and $m^L$ are integer quantized
it follows that for non-zero $e_{\alpha}$ and $m^L$ the flux $m^L$ has to scale at least as $(h^{2,1})^2$.
For the tadpole contribution this implies 
\eq{
  N_{\rm flux} = - e_L m^L  \sim (h^{2,1})^2 \,,
}
which is in agreement with the tadpole conjecture.

While this result is non-trivial, let us emphasize that we have used fitted statistical data to make this estimate. 
Some of our steps, most notably \eqref{rel_8394}, are true only approximately and are made under the assumption that 
conspicuous cancellations are absent. 
Comparing \eqref{fits_001} with statistical results on the divisor volumes obtained in \cite{Demirtas:2018akl} shows 
agreement, but ideally we would like to have available results on the minimal value that $\cK_{L}/\cK_{L\alpha} $ can take in the stretched K\"ahler cone directly. 
Moreover, it is conceivable that certain families of compactification spaces exist that show a slightly different scaling behaviour. 
Hence, we do not claim that the IIB2 moduli stabilization scheme of \cite{Marchesano:2021gyv} cannot produce examples 
that violate the tadpole conjecture -- however, under the assumptions stated above the 
scheme of \cite{Marchesano:2021gyv} generically satisfies the tadpole conjecture
(in the case of vanishing axions). 
Note that a similar observation has  been made independently in \cite{Lust:2021xds}.


\subsubsection*{Tadpole conjecture II -- including axions}

We finally want to discuss the case of non-vanishing axions.
Having non-zero fluxes $\hat{m}^\mu$ introduces additional terms, and the modified self-duality equation in F-theory reads
\begin{equation}
m^L e_\alpha -\frac{1}{2} \cK_{L \alpha \beta \gamma} \hat{m}^\beta \hat{m}^\gamma = (m^L)^2 \bigg( \frac{1}{6}  \cK_L \partial_\alpha \Big( \frac{f}{4 \cK_L} \Big)  -  \frac{ 9 K^{(3)}_L  \cK_{L\alpha} }{8  \cK_L } \bigg)\, .
\end{equation} 
For the type IIB limit we set  $f=0$, and together with the first relation in \eqref{IIB_solve} the self-duality condition 
can be expressed as
\eq{
\label{rel_0987}
e_L  = \Big( -\frac{1}{6} \frac{\cK_L}{y_L}  -\frac{ 
K_L^{(3)} \zeta(3)}{64 \pi^3 y_L}\Big) m^L\, , 
\qquad
m^L e_\alpha -\frac{1}{2} \cK_{L \alpha \beta \gamma} \hat{m}^\beta \hat{m}^\gamma = -  \frac{ 9 K^{(3)}_L  \cK_{L\alpha} (m^L)^2 }{8  \cK_L }\,.
}
We now want to repeat our reasoning from above, for which we have to distinguish two cases:
\begin{itemize}

\item We first consider the case in which  the combination 
$m^L e_\alpha -\frac{1}{2} \cK_{L \alpha \beta \gamma} \hat{m}^\beta \hat{m}^\gamma $
appearing in the second relation of \eqref{rel_0987} is generic, in particular, for large $m^L$ this combination 
is a (half-)integer of order $m^L$ (or larger). Using then the scaling \eqref{rel_8394}, we conclude 
again that 
\eq{
  m^L \sim (h^{2,1})^2 
  \hspace{40pt}\Longrightarrow\hspace{40pt}
  N_{\rm flux} \sim (h^{2,1})^2 \,,
}
as in our discussion for vanishing axions. Hence, also in this situation the tadpole conjecture is satisfied.

\item As a second case we consider a situation where the combination
$m^L e_\alpha -\frac{1}{2} \cK_{L \alpha \beta \gamma} \hat{m}^\beta \hat{m}^\gamma $
is fine-tuned to a non-vanishing $\mathcal O(1)$ half-integer. Here we only need to require 
a linear scaling of $m^L$ with $h^{2,1}$, which leads to 
\eq{
  m^L \sim h^{2,1}
  \hspace{40pt}\Longrightarrow\hspace{40pt}
  N_{\rm flux} \sim h^{2,1} \,.
}
The scaling behaviour of the tadpole conjecture is therefore satisfied also in this non-generic situation.

\end{itemize}


\section{Conclusions}
In this work we have presented a novel strategy for systematically stabilizing complex-structure moduli  in Type IIB and F-theory flux compactifications. For generic regions in field space the underlying scalar potential is known to be a complicated transcendental function of the complex-structure moduli. However, a remarkably universal structure emerges 
when we move towards an asymptotic regime, i.e.~when we approach the boundaries of the moduli space. Motivated by key results from asymptotic Hodge theory, we laid out an approximation scheme to study flux vacua in such settings. This procedure approximates the dependence on the moduli in three steps, which are given by: (1) the sl(2)-approximation, (2) the nilpotent orbit approximation, and (3) the full series of exponential corrections. The first two steps both drop exponential corrections to the scalar potential, therefore yielding a system of algebraic extremization conditions for the moduli. In fact, in the sl(2)-approximation the self-duality condition reduces even further to a simple set of polynomial equations. By iterating through these degrees of approximation and improving the moduli vevs at each step, we turn moduli stabilization into an organized procedure where the complicated dependence on the moduli has been broken down systematically. 

It is important to stress that our approach can be realized in every 
asymptotic regime and generally we find that both approximations (1) and (2) are non-trivial. In particular, determining the nilpotent orbit 
approximation (2) might be familiar for the large complex structure regime, where exponential 
corrections nicely decouple. However, this decoupling does not happen on most other boundaries and dropping exponential corrections becomes a non-trivial 
step that has to be performed when determining the nilpotent orbit. The derivation of the sl(2)-approximation~(1) is non-trivial in essentially all 
asymptotic regions including the large complex structure regime.      
For this reason, in section~\ref{sec_type_ii_boundary} we gave an extensive introduction that explains pedagogically how to construct this sl(2)-approximation step-by-step. Furthermore, in order to exemplify this story, we worked out an explicit two-moduli example in great detail. The goal of this approximation is to derive a representation of sl(2)$^n$ and the boundary Hodge decomposition 
that arise for a $n$-parameter limit in complex-structure moduli space.
In order to do that one has to impose a hierarchy among the complex-structure saxions, and if one restricts attention to this strict asymptotic regime 
corrections proportional to ratios between the saxions can be dropped. The sl(2)$^n$-representation and the boundary Hodge decomposition can then be 
computed algorithmically by iterating through the saxion hierarchies. The resulting data allowed us to introduce an sl(2)-approximated version of the Hodge star 
denoted by $C_{\rm sl(2)}$. It encodes in a simple fashion how 
the Hodge norm behaves in the strict asymptotic regime. Furthermore, the properties of $C_{\rm sl(2)}$ 
allow us to write down polynomial conditions that need to be satisfied for flux vacua in this approximation.

The sl(2)-approximation turns the original very complicated problem of solving the flux vacuum conditions into a tractable 
polynomial task. We explicitly carried out its construction for a number of examples, and showed that the vacua found with this method  resemble the vacua found in the nilpotent orbit approximation rather well --- even if one imposes only a moderate hierarchy of the saxions. 
Raising the hierarchy to an $\mathcal O(10)$-factor we found that the location of vacua in approximation (1) and (2) differ only by few percent. This was true 
in a two-moduli and two three-moduli examples in which we derived the sl(2)-approximation, i.e.~both in the large complex-structure and the conifold-LCS regime. 
Combining these observations with the fact that the approximation (2) and (3) differ only by exponentially suppressed corrections \cite{Schmid}, we thus realize that our proposed 
stabilization scheme gives a favourable numerical approach for finding vacua. It would be very interesting to explore if this gives a novel way to 
perform moduli stabilizations in settings with many moduli and compare the efficiency with direct searches for vacua. We expect 
that the step-wise approach reduces the computational complexity (see \cite{Denef:2006ad, Halverson:2018cio} for an in-depth discussion of the arising challenges). 
Furthermore, it would be exciting to combine our approach with recent efforts to use machine learning algorithms in studying the 
string landscape \cite{Abel:2014xta, Ruehle:2017mzq, Halverson:2019tkf, Betzler:2019kon, Cole:2019enn, AbdusSalam:2020ywo,CaboBizet:2020cse, Bena:2021wyr, Krippendorf:2021uxu}.

In this work we have also argued that our strategy can be directly extended to F-theory flux compactifications on Calabi-Yau fourfolds. We have 
given the relevant information needed to generally determine the sl(2)$^n$-representation and boundary Hodge decomposition which 
includes the relations in appendix \ref{fourfold_rel}. 
To prepare for giving an explicit example, we then specialized to the large complex-structure regime where the defining information 
of the scalar potential can be given in terms of the topological data of the mirror fourfold. The example that we studied in detail 
was the Calabi-Yau fourfold geometry considered in  \cite{Marchesano:2021gyv} and we focused on the realization of the so-called linear moduli stabilization scenario. 
In order to analyze this geometry from the perspective of asymptotic Hodge theory, we computed the nilpotent orbit and then determined the 
sl(2)-approximation.
We then required the tadpole contribution of the fluxes to remain parametrically finite in the strict asymptotic regime, leading to constraints on the flux quanta matching with \cite{Marchesano:2021gyv}. Subsequently we investigated the stabilization of moduli in the various degrees of approximation at our disposal. In the sl(2)-approximation only one combination of saxions was stabilized (while it already fixes all axions), and by systematically including subleading corrections we could check how the remaining flat directions were lifted. In particular, we reaffirmed the importance of 
the coefficient $K_L^{(3)}$ that is included when going form the sl(2)-approximation to the nilpotent orbit. This correction reduces to the Euler characteristic term in the type IIB case. Restricting 
to cases in which it vanishes we found sets of flux vacua that are real lines stretching to the boundary.
 
Based on this analysis of the linear scenario we also made some additional observations related to recent swampland conjectures and finiteness results. 
Firstly, we noted that describing all vacuum loci in a systematic fashion requires one to go beyond the usual algebro-geometric tools, and instead prompts one to use techniques furnished by tame topology and o-minimality \cite{BakkerGrimmSchnellTsimerman}. We argued that, as it has to be the case, the linear scenario satisfies the finiteness theorem of \cite{BakkerGrimmSchnellTsimerman}. 
Secondly, we pointed out that flux vacua being real lines stretching to the boundary provide us with interesting examples that decouple axions from their counterparts, allowing for infinite distance limits that could serve as non-trivial testing grounds for the axionic string conjecture of \cite{Lanza:2020qmt, Lanza:2021qsu}, the convex hull distance 
conjecture \cite{Calderon-Infante:2020dhm}, and the claims on supersymmetric protection \cite{Palti:2020qlc}. Thirdly, we investigated the validity of the tadpole conjecture \cite{Bena:2020xrh} for the linear scenario. By using statistical arguments following \cite{Plauschinn:2021hkp} we were able to confirm that, under certain genericity assumptions, these models are compatible with the tadpole conjecture. These results suggest that the linear scenario provides no inherent structure to generically 
give counter-examples to the tadpole conjecture --- if one aims to find such a counter-example, one would have to perform a 
thorough search through the available Calabi-Yau databases. It remains an open question whether an explicit 
counter-example exists.

Let us close by commenting on some further directions that exploit the general nature of our approach. 
To begin with, we can delve deeper into step (1) of our approximation scheme, the sl(2)-approximation. By using the classification of boundaries one can turn moduli stabilization into an abstract problem: the extremization conditions are a simple set of polynomial equations, and the allowed exponents are fixed by the possible boundary types. We can then determine at which boundaries the 
sl(2)-approximation stabilizes all moduli. 
For two-moduli fourfolds this has already been worked out in \cite{Grimm:2019ixq}, but a more general investigation is still missing. 
For step (2) of our scheme, the nilpotent orbit approximation, there is also an abstract approach left unexplored. It has recently been shown that based on \cite{CKS} 
the corrections completing the sl(2)-approximation into the nilpotent orbit approximation can also be incorporated systematically 
by using the holographic perspective put forward in \cite{Grimm:2020cda, Grimm:2021ikg}. It would be interesting to study moduli stabilization using the same strategy. 
Finally, we note that our in our search for flux vacua with approximation steps (1) and (2) we have only included polynomial corrections and essential exponential corrections. 
Finding exact vacua including the whole series of exponential corrections remains a challenging open task. Luckily, in the asymptotic regime, we can judge how 
close we are to the full answer due to the existence of the outlined approximation scheme.


\vskip1em
\subsubsection*{Acknowledgments}
It is a pleasure to thank Brice Bastian, Mariana Gra\~na, Alvaro Herr\'aez, Stefano Lanza, Severin L\"ust, Fernando Marchesano, Jeroen Monnee, Eran Palti, David Prieto, Christian Schnell, Irene Valenzuela and Max Wiesner for useful discussions and correspondence. TG and DH
are partly supported by the Dutch Research Council (NWO) via a Start-Up grant and a VICI grant, while EP is
supported by a Heisenberg grant of the \textit{Deutsche Forschungsgemeinschaft} (DFG,
German Research Foundation) with project-number 430285316.


\newpage
\appendix
\section{Sl(2)-approximation for linear scenario example}\label{app:linearscenario}
In this section we summarize the relevant building blocks for the sl(2)-approximated Hodge star \eqref{boundarystar}. This consists of the weight and lowering operators $H_i, N^-_i$ of the $\mathfrak{sl}(2)$-triples, and the boundary Hodge star $C_\infty$.
The weight operators are given by
\begingroup
\def\myFigureScale{0.62}
\allowdisplaybreaks
\eq{
\label{linear_operators_weights}
\scalebox{\myFigureScale}{$N^0_0$} &\scalebox{\myFigureScale}{$= \left(
\begin{array}{cccccccccccccccc}
 1 & 0 & 0 & 0 & 0 & 0 & 0 & 0 & 0 & 0 & 0 & 0 & 0 & 0 & 0 & 0 \\
 0 & -1 & 0 & 0 & 0 & 0 & 0 & 0 & 0 & 0 & 0 & 0 & 0 & 0 & 0 & 0 \\
 0 & 0 & 1 & 0 & 0 & 0 & 0 & 0 & 0 & 0 & 0 & 0 & 0 & 0 & 0 & 0 \\
 0 & 1 & 0 & 1 & 0 & 0 & 0 & 0 & 0 & 0 & 0 & 0 & 0 & 0 & 0 & 0 \\
 0 & 2 & 0 & 0 & 1 & 0 & 0 & 0 & 0 & 0 & 0 & 0 & 0 & 0 & 0 & 0 \\
 0 & 0 & 0 & 0 & 0 & 1 & 0 & 0 & 0 & -2 & -3 & 0 & 0 & 0 & 0 & 0 \\
 0 & 0 & 0 & 0 & 0 & 0 & 1 & 0 & -2 & -4 & -6 & 0 & 0 & 0 & 0 & 0 \\
 0 & 0 & 0 & 0 & 0 & 0 & 0 & 1 & -3 & -6 & -6 & 0 & 0 & 0 & 0 & 0 \\
 0 & 0 & 0 & 0 & 0 & 0 & 0 & 0 & -1 & 0 & 0 & 0 & 0 & 0 & 0 & 0 \\
 0 & 0 & 0 & 0 & 0 & 0 & 0 & 0 & 0 & -1 & 0 & 0 & 0 & 0 & 0 & 0 \\
 0 & 0 & 0 & 0 & 0 & 0 & 0 & 0 & 0 & 0 & -1 & 0 & 0 & 0 & 0 & 0 \\
 0 & 0 & 0 & 0 & 0 & 0 & 0 & 0 & 0 & 0 & 0 & 1 & 0 & -1 & -2 & 0 \\
 0 & 0 & 0 & 0 & 0 & 0 & 0 & 0 & 0 & 0 & 0 & 0 & -1 & 0 & 0 & 0 \\
 0 & 0 & 0 & 0 & 0 & 0 & 0 & 0 & 0 & 0 & 0 & 0 & 0 & -1 & 0 & 0 \\
 0 & 0 & 0 & 0 & 0 & 0 & 0 & 0 & 0 & 0 & 0 & 0 & 0 & 0 & -1 & 0 \\
 0 & 0 & 0 & 0 & 0 & 0 & 0 & 0 & 0 & 0 & 0 & 0 & 0 & 0 & 0 & -1 \\
\end{array}
\right) , $}
\\
\scalebox{\myFigureScale}{$N^0_L$} &\scalebox{\myFigureScale}{$= \left(
\begin{array}{cccccccccccccccc}
 1 & 0 & 0 & 0 & 0 & 0 & 0 & 0 & 0 & 0 & 0 & 0 & 0 & 0 & 0 & 0 \\
 0 & 1 & 0 & 0 & 0 & 0 & 0 & 0 & 0 & 0 & 0 & 0 & 0 & 0 & 0 & 0 \\
 0 & -2 & -1 & -2 & -1 & 0 & 0 & 0 & 0 & 0 & 0 & 0 & 0 & 0 & 0 & 0 \\
 0 & 0 & 0 & 1 & 0 & 0 & 0 & 0 & 0 & 0 & 0 & 0 & 0 & 0 & 0 & 0 \\
 0 & 0 & 0 & 0 & 1 & 0 & 0 & 0 & 0 & 0 & 0 & 0 & 0 & 0 & 0 & 0 \\
 0 & 0 & 0 & 0 & 0 & 1 & 0 & 0 & 0 & 0 & 0 & 0 & 0 & 0 & 0 & 0 \\
 0 & 0 & 0 & 0 & 0 & 2 & -1 & 0 & 0 & 0 & 2 & 0 & 0 & 0 & 0 & 0 \\
 0 & 0 & 0 & 0 & 0 & 1 & 0 & -1 & 0 & 2 & 3 & 0 & 0 & 0 & 0 & 0 \\
 0 & 0 & 0 & 0 & 0 & 0 & 0 & 0 & -1 & -2 & -1 & 0 & 0 & 0 & 0 & 0 \\
 0 & 0 & 0 & 0 & 0 & 0 & 0 & 0 & 0 & 1 & 0 & 0 & 0 & 0 & 0 & 0 \\
 0 & 0 & 0 & 0 & 0 & 0 & 0 & 0 & 0 & 0 & 1 & 0 & 0 & 0 & 0 & 0 \\
 0 & 0 & 0 & 0 & 0 & 0 & 0 & 0 & 0 & 0 & 0 & -1 & 2 & 0 & 0 & 0 \\
 0 & 0 & 0 & 0 & 0 & 0 & 0 & 0 & 0 & 0 & 0 & 0 & 1 & 0 & 0 & 0 \\
 0 & 0 & 0 & 0 & 0 & 0 & 0 & 0 & 0 & 0 & 0 & 0 & 2 & -1 & 0 & 0 \\
 0 & 0 & 0 & 0 & 0 & 0 & 0 & 0 & 0 & 0 & 0 & 0 & 1 & 0 & -1 & 0 \\
 0 & 0 & 0 & 0 & 0 & 0 & 0 & 0 & 0 & 0 & 0 & 0 & 0 & 0 & 0 & -1 \\
\end{array}
\right), $}\\
\scalebox{\myFigureScale}{$N^0_1$} &\scalebox{\myFigureScale}{$= \left(
\begin{array}{cccccccccccccccc}
 1 & 0 & 0 & 0 & 0 & 0 & 0 & 0 & 0 & 0 & 0 & 0 & 0 & 0 & 0 & 0 \\
 0 & 1 & 0 & 0 & 0 & 0 & 0 & 0 & 0 & 0 & 0 & 0 & 0 & 0 & 0 & 0 \\
 0 & 2 & 1 & 2 & 1 & 0 & 0 & 0 & 0 & 0 & 0 & 0 & 0 & 0 & 0 & 0 \\
 0 & -2 & 0 & -1 & -1 & 0 & 0 & 0 & 0 & 0 & 0 & 0 & 0 & 0 & 0 & 0 \\
 0 & 0 & 0 & 0 & 1 & 0 & 0 & 0 & 0 & 0 & 0 & 0 & 0 & 0 & 0 & 0 \\
 0 & 0 & 0 & 0 & 0 & -1 & 0 & 0 & 0 & 0 & 2 & 0 & 0 & 0 & 0 & 0 \\
 0 & 0 & 0 & 0 & 0 & -2 & 1 & 0 & 0 & 0 & 2 & 0 & 0 & 0 & 0 & 0 \\
 0 & 0 & 0 & 0 & 0 & -1 & 1 & -1 & 2 & 2 & 3 & 0 & 0 & 0 & 0 & 0 \\
 0 & 0 & 0 & 0 & 0 & 0 & 0 & 0 & 1 & 2 & 1 & 0 & 0 & 0 & 0 & 0 \\
 0 & 0 & 0 & 0 & 0 & 0 & 0 & 0 & 0 & -1 & -1 & 0 & 0 & 0 & 0 & 0 \\
 0 & 0 & 0 & 0 & 0 & 0 & 0 & 0 & 0 & 0 & 1 & 0 & 0 & 0 & 0 & 0 \\
 0 & 0 & 0 & 0 & 0 & 0 & 0 & 0 & 0 & 0 & 0 & -1 & -2 & 2 & 0 & 0 \\
 0 & 0 & 0 & 0 & 0 & 0 & 0 & 0 & 0 & 0 & 0 & 0 & -1 & 0 & 0 & 0 \\
 0 & 0 & 0 & 0 & 0 & 0 & 0 & 0 & 0 & 0 & 0 & 0 & -2 & 1 & 0 & 0 \\
 0 & 0 & 0 & 0 & 0 & 0 & 0 & 0 & 0 & 0 & 0 & 0 & -1 & 1 & -1 & 0 \\
 0 & 0 & 0 & 0 & 0 & 0 & 0 & 0 & 0 & 0 & 0 & 0 & 0 & 0 & 0 & -1 \\
\end{array}
\right),$}\\
\scalebox{\myFigureScale}{$N^0_2$} &\scalebox{\myFigureScale}{$= \left(
\begin{array}{cccccccccccccccc}
 1 & 0 & 0 & 0 & 0 & 0 & 0 & 0 & 0 & 0 & 0 & 0 & 0 & 0 & 0 & 0 \\
 0 & 1 & 0 & 0 & 0 & 0 & 0 & 0 & 0 & 0 & 0 & 0 & 0 & 0 & 0 & 0 \\
 0 & 0 & 1 & 0 & 0 & 0 & 0 & 0 & 0 & 0 & 0 & 0 & 0 & 0 & 0 & 0 \\
 0 & 1 & 0 & 1 & 1 & 0 & 0 & 0 & 0 & 0 & 0 & 0 & 0 & 0 & 0 & 0 \\
 0 & -2 & 0 & 0 & -1 & 0 & 0 & 0 & 0 & 0 & 0 & 0 & 0 & 0 & 0 & 0 \\
 0 & 0 & 0 & 0 & 0 & -1 & 0 & 0 & 0 & 2 & 1 & 0 & 0 & 0 & 0 & 0 \\
 0 & 0 & 0 & 0 & 0 & 0 & -1 & 0 & 2 & 4 & 2 & 0 & 0 & 0 & 0 & 0 \\
 0 & 0 & 0 & 0 & 0 & 0 & -1 & 1 & 1 & 2 & 0 & 0 & 0 & 0 & 0 & 0 \\
 0 & 0 & 0 & 0 & 0 & 0 & 0 & 0 & 1 & 0 & 0 & 0 & 0 & 0 & 0 & 0 \\
 0 & 0 & 0 & 0 & 0 & 0 & 0 & 0 & 0 & 1 & 1 & 0 & 0 & 0 & 0 & 0 \\
 0 & 0 & 0 & 0 & 0 & 0 & 0 & 0 & 0 & 0 & -1 & 0 & 0 & 0 & 0 & 0 \\
 0 & 0 & 0 & 0 & 0 & 0 & 0 & 0 & 0 & 0 & 0 & -1 & 0 & -1 & 2 & 0 \\
 0 & 0 & 0 & 0 & 0 & 0 & 0 & 0 & 0 & 0 & 0 & 0 & -1 & 0 & 0 & 0 \\
 0 & 0 & 0 & 0 & 0 & 0 & 0 & 0 & 0 & 0 & 0 & 0 & 0 & -1 & 0 & 0 \\
 0 & 0 & 0 & 0 & 0 & 0 & 0 & 0 & 0 & 0 & 0 & 0 & 0 & -1 & 1 & 0 \\
 0 & 0 & 0 & 0 & 0 & 0 & 0 & 0 & 0 & 0 & 0 & 0 & 0 & 0 & 0 & -1 \\
\end{array}
\right).$}
}
The lowering operators are given by
\eq{
\scalebox{\myFigureScale}{$N_0^-$} &\scalebox{\myFigureScale}{$= \left(
\begin{array}{cccccccccccccccc}
 0 & 0 & 0 & 0 & 0 & 0 & 0 & 0 & 0 & 0 & 0 & 0 & 0 & 0 & 0 & 0 \\
 -1 & 0 & 0 & 0 & 0 & 0 & 0 & 0 & 0 & 0 & 0 & 0 & 0 & 0 & 0 & 0 \\
 0 & 0 & 0 & 0 & 0 & 0 & 0 & 0 & 0 & 0 & 0 & 0 & 0 & 0 & 0 & 0 \\
 \frac{1}{2} & 0 & 0 & 0 & 0 & 0 & 0 & 0 & 0 & 0 & 0 & 0 & 0 & 0 & 0 & 0 \\
 1 & 0 & 0 & 0 & 0 & 0 & 0 & 0 & 0 & 0 & 0 & 0 & 0 & 0 & 0 & 0 \\
 0 & -2 & 0 & -1 & -\frac{3}{2} & 0 & 0 & 0 & 0 & 0 & 0 & 0 & 0 & 0 & 0 & 0 \\
 0 & -4 & -1 & -2 & -3 & 0 & 0 & 0 & 0 & 0 & 0 & 0 & 0 & 0 & 0 & 0 \\
 0 & -\frac{9}{2} & -\frac{3}{2} & -3 & -3 & 0 & 0 & 0 & 0 & 0 & 0 & 0 & 0 & 0 & 0 & 0 \\
 0 & 0 & -1 & 0 & 0 & 0 & 0 & 0 & 0 & 0 & 0 & 0 & 0 & 0 & 0 & 0 \\
 0 & -\frac{1}{2} & 0 & -1 & 0 & 0 & 0 & 0 & 0 & 0 & 0 & 0 & 0 & 0 & 0 & 0 \\
 0 & -1 & 0 & 0 & -1 & 0 & 0 & 0 & 0 & 0 & 0 & 0 & 0 & 0 & 0 & 0 \\
 0 & 0 & 0 & 0 & 0 & 0 & -\frac{1}{2} & -1 & 2 & 4 & \frac{9}{2} & 0 & 0 & 0 & 0 & 0 \\
 0 & 0 & 0 & 0 & 0 & -1 & 0 & 0 & 0 & 1 & \frac{3}{2} & 0 & 0 & 0 & 0 & 0 \\
 0 & 0 & 0 & 0 & 0 & 0 & -1 & 0 & 1 & 2 & 3 & 0 & 0 & 0 & 0 & 0 \\
 0 & 0 & 0 & 0 & 0 & 0 & 0 & -1 & \frac{3}{2} & 3 & 3 & 0 & 0 & 0 & 0 & 0 \\
 0 & 0 & 0 & 0 & 0 & 0 & 0 & 0 & 0 & 0 & 0 & -1 & 0 & \frac{1}{2} & 1 & 0 \\
\end{array}
\right) ,$}\\
\scalebox{\myFigureScale}{$N_L^- $} &\scalebox{\myFigureScale}{$= \left(
\begin{array}{cccccccccccccccc}
 0 & 0 & 0 & 0 & 0 & 0 & 0 & 0 & 0 & 0 & 0 & 0 & 0 & 0 & 0 & 0 \\
 0 & 0 & 0 & 0 & 0 & 0 & 0 & 0 & 0 & 0 & 0 & 0 & 0 & 0 & 0 & 0 \\
 -1 & 0 & 0 & 0 & 0 & 0 & 0 & 0 & 0 & 0 & 0 & 0 & 0 & 0 & 0 & 0 \\
 0 & 0 & 0 & 0 & 0 & 0 & 0 & 0 & 0 & 0 & 0 & 0 & 0 & 0 & 0 & 0 \\
 0 & 0 & 0 & 0 & 0 & 0 & 0 & 0 & 0 & 0 & 0 & 0 & 0 & 0 & 0 & 0 \\
 0 & 0 & 0 & 0 & 0 & 0 & 0 & 0 & 0 & 0 & 0 & 0 & 0 & 0 & 0 & 0 \\
 0 & -2 & 0 & 0 & -1 & 0 & 0 & 0 & 0 & 0 & 0 & 0 & 0 & 0 & 0 & 0 \\
 0 & -3 & 0 & -1 & -1 & 0 & 0 & 0 & 0 & 0 & 0 & 0 & 0 & 0 & 0 & 0 \\
 0 & -1 & 0 & 0 & 0 & 0 & 0 & 0 & 0 & 0 & 0 & 0 & 0 & 0 & 0 & 0 \\
 0 & 0 & 0 & 0 & 0 & 0 & 0 & 0 & 0 & 0 & 0 & 0 & 0 & 0 & 0 & 0 \\
 0 & 0 & 0 & 0 & 0 & 0 & 0 & 0 & 0 & 0 & 0 & 0 & 0 & 0 & 0 & 0 \\
 0 & 0 & 0 & 0 & 0 & -1 & 0 & 0 & 0 & 0 & 0 & 0 & 0 & 0 & 0 & 0 \\
 0 & 0 & 0 & 0 & 0 & 0 & 0 & 0 & 0 & 0 & 0 & 0 & 0 & 0 & 0 & 0 \\
 0 & 0 & 0 & 0 & 0 & 0 & 0 & 0 & 0 & 0 & -1 & 0 & 0 & 0 & 0 & 0 \\
 0 & 0 & 0 & 0 & 0 & 0 & 0 & 0 & 0 & -1 & -1 & 0 & 0 & 0 & 0 & 0 \\
 0 & 0 & 0 & 0 & 0 & 0 & 0 & 0 & 0 & 0 & 0 & 0 & -1 & 0 & 0 & 0 \\
\end{array}
\right),$} \\
\scalebox{\myFigureScale}{$N_1^- $} &\scalebox{\myFigureScale}{$= \left(
\begin{array}{cccccccccccccccc}
 0 & 0 & 0 & 0 & 0 & 0 & 0 & 0 & 0 & 0 & 0 & 0 & 0 & 0 & 0 & 0 \\
 0 & 0 & 0 & 0 & 0 & 0 & 0 & 0 & 0 & 0 & 0 & 0 & 0 & 0 & 0 & 0 \\
 1 & 0 & 0 & 0 & 0 & 0 & 0 & 0 & 0 & 0 & 0 & 0 & 0 & 0 & 0 & 0 \\
 -1 & 0 & 0 & 0 & 0 & 0 & 0 & 0 & 0 & 0 & 0 & 0 & 0 & 0 & 0 & 0 \\
 0 & 0 & 0 & 0 & 0 & 0 & 0 & 0 & 0 & 0 & 0 & 0 & 0 & 0 & 0 & 0 \\
 0 & -2 & 0 & 0 & -1 & 0 & 0 & 0 & 0 & 0 & 0 & 0 & 0 & 0 & 0 & 0 \\
 0 & -2 & 0 & 0 & -1 & 0 & 0 & 0 & 0 & 0 & 0 & 0 & 0 & 0 & 0 & 0 \\
 0 & -3 & -1 & -1 & -1 & 0 & 0 & 0 & 0 & 0 & 0 & 0 & 0 & 0 & 0 & 0 \\
 0 & 1 & 0 & 0 & 0 & 0 & 0 & 0 & 0 & 0 & 0 & 0 & 0 & 0 & 0 & 0 \\
 0 & -1 & 0 & 0 & 0 & 0 & 0 & 0 & 0 & 0 & 0 & 0 & 0 & 0 & 0 & 0 \\
 0 & 0 & 0 & 0 & 0 & 0 & 0 & 0 & 0 & 0 & 0 & 0 & 0 & 0 & 0 & 0 \\
 0 & 0 & 0 & 0 & 0 & 1 & -1 & 0 & 0 & 0 & 0 & 0 & 0 & 0 & 0 & 0 \\
 0 & 0 & 0 & 0 & 0 & 0 & 0 & 0 & 0 & 0 & -1 & 0 & 0 & 0 & 0 & 0 \\
 0 & 0 & 0 & 0 & 0 & 0 & 0 & 0 & 0 & 0 & -1 & 0 & 0 & 0 & 0 & 0 \\
 0 & 0 & 0 & 0 & 0 & 0 & 0 & 0 & -1 & -1 & -1 & 0 & 0 & 0 & 0 & 0 \\
 0 & 0 & 0 & 0 & 0 & 0 & 0 & 0 & 0 & 0 & 0 & 0 & 1 & -1 & 0 & 0 \\
\end{array}
\right),$} \\
\scalebox{\myFigureScale}{$N_2^- $} &\scalebox{\myFigureScale}{$=\left(
\begin{array}{cccccccccccccccc}
 0 & 0 & 0 & 0 & 0 & 0 & 0 & 0 & 0 & 0 & 0 & 0 & 0 & 0 & 0 & 0 \\
 0 & 0 & 0 & 0 & 0 & 0 & 0 & 0 & 0 & 0 & 0 & 0 & 0 & 0 & 0 & 0 \\
 0 & 0 & 0 & 0 & 0 & 0 & 0 & 0 & 0 & 0 & 0 & 0 & 0 & 0 & 0 & 0 \\
 \frac{1}{2} & 0 & 0 & 0 & 0 & 0 & 0 & 0 & 0 & 0 & 0 & 0 & 0 & 0 & 0 & 0 \\
 -1 & 0 & 0 & 0 & 0 & 0 & 0 & 0 & 0 & 0 & 0 & 0 & 0 & 0 & 0 & 0 \\
 0 & -2 & 0 & -1 & -\frac{1}{2} & 0 & 0 & 0 & 0 & 0 & 0 & 0 & 0 & 0 & 0 & 0 \\
 0 & -4 & -1 & -2 & -1 & 0 & 0 & 0 & 0 & 0 & 0 & 0 & 0 & 0 & 0 & 0 \\
 0 & -\frac{5}{2} & -\frac{1}{2} & -1 & -\frac{1}{2} & 0 & 0 & 0 & 0 & 0 & 0 & 0 & 0 & 0 & 0 & 0 \\
 0 & 0 & 0 & 0 & 0 & 0 & 0 & 0 & 0 & 0 & 0 & 0 & 0 & 0 & 0 & 0 \\
 0 & \frac{1}{2} & 0 & 0 & 0 & 0 & 0 & 0 & 0 & 0 & 0 & 0 & 0 & 0 & 0 & 0 \\
 0 & -1 & 0 & 0 & 0 & 0 & 0 & 0 & 0 & 0 & 0 & 0 & 0 & 0 & 0 & 0 \\
 0 & 0 & 0 & 0 & 0 & 0 & \frac{1}{2} & -1 & 0 & 0 & \frac{1}{2} & 0 & 0 & 0 & 0 & 0 \\
 0 & 0 & 0 & 0 & 0 & 0 & 0 & 0 & 0 & -1 & -\frac{1}{2} & 0 & 0 & 0 & 0 & 0 \\
 0 & 0 & 0 & 0 & 0 & 0 & 0 & 0 & -1 & -2 & -1 & 0 & 0 & 0 & 0 & 0 \\
 0 & 0 & 0 & 0 & 0 & 0 & 0 & 0 & -\frac{1}{2} & -1 & -\frac{1}{2} & 0 & 0 & 0 & 0 & 0 \\
 0 & 0 & 0 & 0 & 0 & 0 & 0 & 0 & 0 & 0 & 0 & 0 & 0 & \frac{1}{2} & -1 & 0 \\
\end{array}
\right).$}
}
Finally, the Hodge star operator of the boundary Hodge structure is given by
\eq{
\scalebox{\myFigureScale}{$
C_\infty = \left(
\begin{array}{cccccccccccccccc}
 1 & 0 & 0 & 0 & 0 & 0 & 0 & 0 & 0 & 0 & 0 & 0 & 0 & 0 & 0 & 0 \\
 0 & 1 & 0 & 0 & 0 & 0 & 0 & 0 & 0 & 0 & 0 & 0 & 0 & 0 & 0 & 0 \\
 0 & 0 & 1 & 0 & 0 & 0 & 0 & 0 & 0 & 0 & 0 & 0 & 0 & 0 & 0 & 0 \\
 0 & 1 & 0 & 1 & 1 & 0 & 0 & 0 & 0 & 0 & 0 & 0 & 0 & 0 & 0 & 0 \\
 0 & -2 & 0 & 0 & -1 & 0 & 0 & 0 & 0 & 0 & 0 & 0 & 0 & 0 & 0 & 0 \\
 0 & 0 & 0 & 0 & 0 & -1 & 0 & 0 & 0 & 2 & 1 & 0 & 0 & 0 & 0 & 0 \\
 0 & 0 & 0 & 0 & 0 & 0 & -1 & 0 & 2 & 4 & 2 & 0 & 0 & 0 & 0 & 0 \\
 0 & 0 & 0 & 0 & 0 & 0 & -1 & 1 & 1 & 2 & 0 & 0 & 0 & 0 & 0 & 0 \\
 0 & 0 & 0 & 0 & 0 & 0 & 0 & 0 & 1 & 0 & 0 & 0 & 0 & 0 & 0 & 0 \\
 0 & 0 & 0 & 0 & 0 & 0 & 0 & 0 & 0 & 1 & 1 & 0 & 0 & 0 & 0 & 0 \\
 0 & 0 & 0 & 0 & 0 & 0 & 0 & 0 & 0 & 0 & -1 & 0 & 0 & 0 & 0 & 0 \\
 0 & 0 & 0 & 0 & 0 & 0 & 0 & 0 & 0 & 0 & 0 & -1 & 0 & -1 & 2 & 0 \\
 0 & 0 & 0 & 0 & 0 & 0 & 0 & 0 & 0 & 0 & 0 & 0 & -1 & 0 & 0 & 0 \\
 0 & 0 & 0 & 0 & 0 & 0 & 0 & 0 & 0 & 0 & 0 & 0 & 0 & -1 & 0 & 0 \\
 0 & 0 & 0 & 0 & 0 & 0 & 0 & 0 & 0 & 0 & 0 & 0 & 0 & -1 & 1 & 0 \\
 0 & 0 & 0 & 0 & 0 & 0 & 0 & 0 & 0 & 0 & 0 & 0 & 0 & 0 & 0 & -1 \\
\end{array}
\right).$}
}
\endgroup


\section{Operator relations for Calabi-Yau fourfolds}\label{fourfold_rel}
In the construction of the sl(2)-approximation two operators $\eta$ and $\zeta$ appear, which can be expressed componentwise in terms of $\delta$ with respect to the decomposition \eqref{delta_decomp}. In this appendix we summarize these relations for Calabi-Yau fourfolds. For the purposes of this work we only need $\zeta$, but the expressions for $\eta$ are included  for completeness.

Before we state these relations, let us briefly recall how $\eta$ and $\zeta$ are  fixed in terms of $\delta$. The main relation between $\eta$, $\zeta$ and $\delta$ is given by \cite{CKS}
\begin{equation}\label{delta-zeta-eta}
 e^{i \delta} = e^{\zeta} \big(1 + \sum_{k \geq 1} P_k(C_2, \ldots, C_{k+1}) \big) \, ,
\end{equation}
where the polynomials $P_k(C_2, \ldots, C_{k+1}) $ are defined recursively as
\begin{equation}
P_0 =1 \, , \qquad P_k = -\frac{1}{k} \sum_{j=1}^k P_{k-j} C_{j+1}\, ,
\end{equation}
and coefficients $C_k$ are given in terms of $\eta$ as
\begin{equation}
C_{k+1}(\eta) = i \sum_{p,q} b^{k-1}_{p-1, q-1}  \eta_{-p,-q}\, , \qquad (1-x)^p (1+x)^q = \sum_{p,q,k} b^{k}_{p, q} x^k\, .
\end{equation}
Solving \eqref{delta-zeta-eta} componentwise for $\zeta_{-p,-q}$ and $\eta_{-p,-q}$ in terms of $\delta_{-r,-s}$ is still a rather non-trivial task. 
In the mathematics literature \cite{Kato} this was solved for the components of $\zeta$ for threefolds.\footnote{Furthermore, general expressions for the components $\zeta_{-p,-q}$ and $\eta_{-p,-q}$ in terms of $\delta_{-p,-q}$ were derived, modulo commutators of $\delta_{-r,-s}$ ($r\leq p$ and $s\leq q$) left undetermined.} These results were extended in \cite{Grimm:2021ikg}, where the components of $\eta$ were given in the threefold case. Following the strategy of these works \cite{Kato, Grimm:2021ikg}, we proceed and derive the componentwise expressions for $\eta$ and $\zeta$ for fourfolds. We refer to these articles for a more detailed explanation on how to work out these operator relations. We can express $\zeta$ componentwise in terms of $\delta$ as
\begin{align}\label{fourfold_zeta}
\zeta_{-1,-1}&= \zeta_{-2,-2}= 0\, , \quad \zeta_{-1,-2}= -\frac{i}{2} \delta_{-1,-2}\, , \quad \zeta_{-1,-3} = -\frac{3i}{4} \delta_{-1,-3}\, ,  \quad \zeta_{-1,-4} = -\frac{7i}{8} \delta_{-1,-4}\, ,\nn  \\
\zeta_{-2,-3} &=-\frac{3i}{8} \delta_{-2,-3}+ \frac{1}{8} [\delta_{-1,-2}, \delta_{-1,-1} ]\, , \qquad \zeta_{-2,-4} = -\frac{5i}{8} \delta_{-2,-4} + \frac{1}{4} [ \delta_{-1,-3}, \delta_{-1,-1} ]\, , \nn
\\
 \zeta_{-3,-3} &= \frac{1}{8}[\delta_{-2,-2}, \delta_{-1,-1}]\, ,  \\
\zeta_{-3,-4}  &= -\frac{5i}{16} \delta_{-3,-4} +\frac{3}{16} [\delta_{-2,-3} , \delta_{-1,-1} ] +\frac{3}{16} [\delta_{-1,-3}, \delta_{-2,-1} ] +\frac{i}{48} [\delta_{-1,-1}, [\delta_{-1,-1}, \delta_{-1,-2}]]\, ,  \nn \\
\zeta_{-4,-4} &= \frac{3}{16} [ \delta_{-3,-3}, \delta_{-1,-1} ] +\frac{3}{32} [\delta_{-3,-2} , \delta_{-1,-2}]+\frac{3}{32} [\delta_{-2,-3} , \delta_{-2,-1}]+\frac{i}{32} [\delta_{-1,-1}, [\delta_{-2,-1}, \delta_{-1,-2} ]] \, , \nn
\end{align}
while $\eta$ is expressed componentwise as
\begin{align}\label{fourfold_eta}
\eta_{-1,-1} &= - \delta_{-1,-1}\, , \qquad \eta_{-1,-2} = - \delta_{-1,-2}\, , \qquad \eta_{-1,-3} = - \frac{3}{4}\delta_{-1,-3}\, , \qquad \eta_{-1,-4} = - \frac{1}{2}\delta_{-1,-4}\, , \nn \\
 \eta_{-2,-2} &= - \frac{3}{2} \delta_{-2,-2}\, , \hspace{125.5pt} \eta_{-2,-3} = -\frac{3}{2} \delta_{-2,-3}+\frac{i}{2} [\delta_{-1,-1}, \delta_{-1,-2}]\, , \nn  \\
 \eta_{-2,-4} &= -\frac{5}{4} \delta_{-2,-4}+\frac{5i}{8}[\delta_{-1,-1}, \delta_{-1,-3}]\, , \hspace{30.5pt} \eta_{-3,-3} =  -\frac{15}{8}\delta_{-3,-3}+\frac{5i}{4}[ \delta_{-2,-1}, \delta_{-1,-2}] \, , \nn \\
 \eta_{-3,-4} &= -\frac{15}{8} \delta_{-3,-4}+\frac{3i}{8}[ \delta_{-1,-1},  \delta_{-2,-3}]  \\
 &\quad+\frac{3i}{2}[ \delta_{-2,-1},  \delta_{-1,-3}]+\frac{3i}{4}[ \delta_{-2,-2},  \delta_{-1,-2}] +\frac{1}{8} [  \delta_{-1,-1}, [ \delta_{-1,-1},  \delta_{-1,-2}]]\, , \nn  \\
   \eta_{-4,-4} &= -\frac{35}{16} \delta_{-4,-4} +\frac{63i}{32} [\delta_{-3,-1}, \delta_{-1,-3}]+\frac{21i}{16} [\delta_{-3,-2}, \delta_{-1,-2}]+\frac{21i}{16} [\delta_{-2,-1}, \delta_{-2,-3}] \nn \\
   & \quad +\frac{7}{48}[\delta_{-1,-1},  [\delta_{-1,-1}, \delta_{-2,-2}]]+\frac{7}{24}[\delta_{-2,-1},  [\delta_{-1,-1}, \delta_{-1,-2}]]+\frac{7}{24}[\delta_{-1,-2},  [\delta_{-1,-1}, \delta_{-2,-1}]]\, . \nn
\end{align}


\bibliographystyle{jhep}
\bibliography{references}


\end{document}